\documentclass[useAMS,usenatbib,fleqn]{mnras}

\usepackage{natbib}
\usepackage{subfig}
\usepackage{graphicx}
\usepackage{etex}
\usepackage{amsmath}
\usepackage{amssymb}
\usepackage[titletoc,title]{appendix}
\usepackage{hyperref}
\hypersetup{colorlinks=true,linkcolor=blue,citecolor=blue,filecolor=blue,urlcolor=blue}

\title[Bulges, Discs and Two Component Galaxies at $z<3$]{The Formation of Bulges, Discs and Two Component Galaxies in the CANDELS Survey at $z<3$}
\author[Margalef-Bentabol et al.]{Berta Margalef-Bentabol$^{1}$\thanks{Berta.Margalef@nottingham.ac.uk}, Christopher J. Conselice$^{1}$, Alice Mortlock$^{2}$, \newauthor Will Hartley$^{3}$,  Kenneth Duncan$^{4}$, Harry C. Ferguson$^{5}$, Anton M. Koekemoer$^{5}$, \newauthor Avishai Dekel$^{6}$, Joel R. Primack$^{7}$\\
$^{1}$University of Nottingham, School of Physics and Astronomy, Nottingham, NG7 2RD UK\\
$^{2}$SUPA Institute for Astronomy, University of Edinburgh, Royal Observatory, Edinburgh EH9 3HJ\\
$^{3}$ETH Zurich, Institute fur Astronomie, Konigstuhl 17, D-69117, Heidelberg, Germany\\
$^{4}$Leiden Observatory, Leiden University, 2300 RA Leiden, The Netherlands\\
$^{5}$Space Telescope Science Institute, 3700 San Martin Drive, Baltimore, MD 21218, USA\\
$^{6}$Racah Institute of Physics, The Hebrew University, Jerusalem 91904, Israel\\
$^{7}$Physics Department, University of California, Santa Cruz, CA 95064, USA
}

\date{Accepted XXX. Received YYY; in original form ZZZ}

\pubyear{2016}

\begin{document}
\label{firstpage}
\pagerange{\pageref{firstpage}--\pageref{lastpage}}
\maketitle

\begin{abstract}
We examine a sample of $1495$ galaxies in the CANDELS fields to determine the evolution of two component galaxies, including bulges and discs, within massive galaxies at the epoch $1 < z < 3$ when the Hubble sequence forms. We fit all of our galaxies' light profiles with a single S\'ersic fit, as well as with a combination of exponential and S\'ersic profiles. The latter is done in order to describe a galaxy with an inner and an outer component, or bulge and disc component. We develop and use three classification methods (visual, $F$-test and the $RF\!F$) to separate our sample into $1$-component galaxies (disc/spheroids-like galaxies) and $2$-component galaxies (galaxies formed by an `inner part' or bulge and an `outer part' or disc). We then compare the results from using these three different ways to classify our galaxies. We find that the fraction of galaxies selected as $2$-component galaxies increases on average $50$ per cent from the lowest mass bin to the most massive galaxies, and decreases with redshift by a factor of four from $z=1$ to $z=3$. We find that single S\'ersic `disc-like' galaxies have the highest relative number densities at all redshifts, and that $2$-component galaxies have the greatest increase and become at par with S\'ersic discs by $z = 1$. We also find that the systems we classify as $2$-component galaxies have an increase in the sizes of their outer components, or `discs', by about a factor of three from $z = 3$ to $z = 1.5$, while the inner components or `bulges' stay roughly the same size. This suggests that these systems are growing from the inside out, whilst the bulges or protobulges are in place early in the history of these galaxies. This is also seen to a lesser degree in the growth of single `disc-like' galaxies vs. `spheroid-like' galaxies over the same epoch.

\end{abstract}


\section{Introduction}\label{sec.intro}

Galaxy structure and morphology are important observables in order to both describe galaxies fully, as well as a critical property for understanding how galaxies form and evolve through cosmic time. We observe that in the local Universe most massive galaxies are classifiable into Hubble types, i.e.\ with a well defined structure, such as spheroids or spirals. However, at higher redshift a population of peculiar galaxies dominates in terms of number densities, \citep[e.g.][]{Conselice05,Mortlock13}. In particular, it is found that the majority of galaxies at $z>2$  are peculiar with a smaller number of spheroid-like galaxies \citep[e.g.][]{Mortlock13}, and with very few traditional disc galaxies. At lower redshifts we find  a gradual transition between peculiar and Hubble type galaxies with a $\sim50\!:\!50$ split between peculiars and Hubble types  at $z\sim 1.5$ \citep{Conselice05, Mortlock13}.

Uncovering the internal processes involved in changing the morphology and structures of galaxies is therefore a useful way to understand how galaxies evolve in terms of physical processes such as star formation and merging. One of the traditional ways of doing this is to examine the effective radius and S\'ersic index of galaxy populations to determine how they evolved \citep[e.g.][]{Ferguson04, Daddi05,Trujillo06, Trujillo07, Toft07, Buitrago08,vanDokkum10, Cassata11}. For instance, \cite{Buitrago08} and others studied the size evolution of massive galaxies showing that, while at $z>1$ these galaxies are extremely compact, in the local Universe we observe that their counterparts are larger, so there must have been a growth in physical size over cosmic time at a given mass. These findings have been confirmed and expanded upon by many others since in great detail with many explanations for the evolution \citep[e.g.][]{Barro13, vanDokkum14,vanderWel14}. However, this only tells part of the story, as at the same time these galaxies grow in size, they also become less peculiar, and develop into bulge+disc systems, something that simple S\'ersic fitting cannot fully quantify.

To comprehend how galaxies make the transition to become the galaxies we observe in the nearby Universe, it is especially important to study them at high redshift ($z>1$) when they are undergoing these transformations, and to do so in a wavelength which probes the underlying stellar mass of the system. Since one of the major hallmarks of the Hubble sequence is the bulge and disc dichotomy, a natural next step in understanding the evolution of galaxies and their structures is to determine when and how discs and bulges and especially disc+bulge systems first formed. 

These higher order structural parameters can be obtained by light decomposition, i.e. by fitting galaxy surface brightness profiles to well known functions, such as exponential plus de Vaucouleur light profiles. However, for high redshift galaxies, this is quite a difficult task, as galaxies are not resolved as well as  they are in the local Universe. It is thus critically important to understand the effects of redshifts on our measurements of the light decomposition of these galaxies, which we also examine.

Due to the advent of the WFC3 camera on \textit{Hubble}, we can take advantage of high quality and high resolution images of high-$z$ galaxies, and instead of just studying them as a whole, we can  perform bulge to disc decomposition with unprecedented accuracy. In fact, there have been studies at high redshift using light decomposition in two dimensions using different codes and methods \citep[e.g.][]{Buitrago08,vanderWel12,Bruce12,Lang14} with a variety of results suggesting that galaxies indeed become more `disky' at high redshift, i.e. high redshift massive galaxies contain on average lower S\'ersic indices at high redshift than at lower redshifts \citep{vanderWel11}. The bulge-disc decomposition allows us to study properties of these two fundamental components separately. In these works galaxies are typically fitted using a combination of a de Vaucouleurs and an exponential profile to describe, respectively, an assumed bulge and disc component in each galaxy.

Previously, using bulge and disc decompositions, \cite{Bruce12} claim that at low redshift, massive galaxies are bulge-dominated. While at redshifts $1<z<2$, galaxies are a mix of bulge+disc systems, and by $z>2$ they are mostly disc-dominated. Up to $z=3$ there are other results showing that stellar mass correlates with the redshift at which Hubble type galaxies start to dominate over peculiar \citep{Mortlock13}. Nevertheless, it remains unclear what causes this transition and when the dominant structures of the local Universe (bulges and discs) appear as well as if these are related events. In this paper we investigate the structures of these distant galaxies to determine when, and in what way, discs and spheroids first appear in the massive galaxy population.

We perform one and two component light decompositions using \textsc{galfit} \citep{Peng02} and \textsc{galapagos} \citep{Barden12} to a mass selected sample of galaxies at $1<z<3$. We fit the observed two-dimensional surface brightness profiles of galaxies with several models, the first one being a single S\'ersic profile (with free $n$ and $R_e$), and the second one a combination of S\'ersic profile (again with free $n$) and an exponential profile. The latter combination describes, respectively, a bulge and a disc. However, it is important to notice that we do not assume that this dichotomy translates directly and simply to high redshift systems, where something more complicated, or a transition phase are potentially present between peculiar systems and the classic bulge+disc systems we see in today's Universe. By allowing the S\'ersic index to vary we are considering more general bulges, in comparison with previous work where bulges are assumed to be the classical bulge described by a S\'ersic law with $n=4$. In this work, we also study a larger sample of galaxies at high redshift than previous works and a wider range in masses. This can lead to a better interpretation of the role that total stellar mass plays in the evolution of bulges and discs. 

By fitting the surface brightness to such models, we obtain raw structural parameters for both one and two dimensional fits. However, it is important to know whether an individual galaxy is better fit by a two-component profile (bulge+disc) rather than a single S\'ersic profile, as in the case for pure spheroid-like galaxies and disc-like galaxies. This is a difficult task and there have been attempts using different methods: \cite{Simard11} use the $F$-test probability to determine the most appropriate model, while  \cite{Lang14} use both the reduced $\chi^2$ of the model fits and the Akaike information criterion (AIC). In this work we study and combine three different methods: visual classification, $F$-test and a method based on the Residual Flux Fraction, $RF\!F$ \citep{Hoyos12}, and explore how each method affects the results.

The structure of this paper is as follow. Section \ref{sec.data} is devoted to describing the data we use. In Section \ref{sec.method} we describe how the structural parameters of the galaxies in our sample are obtained, and explain the different methods used to classify them. In Section \ref{sec.results} the main results of the paper are gathered, and in Section \ref{sec.conclusions} we discuss and summarize the results. Finally, an 
Appendix is included with some simulations to better understand our results. Throughout this paper we use $AB$ magnitude units and assume the following cosmology: $H_0=70\mathrm{\ Km s}^{-1}\mathrm{ Mpc}^{-1}$, $\Omega_{\lambda}=0.7$, and  $\Omega_m=0.3$.

\section{Data}\label{sec.data}

\subsection{Imaging}

\begin{figure*}
  \includegraphics[width=1\linewidth]{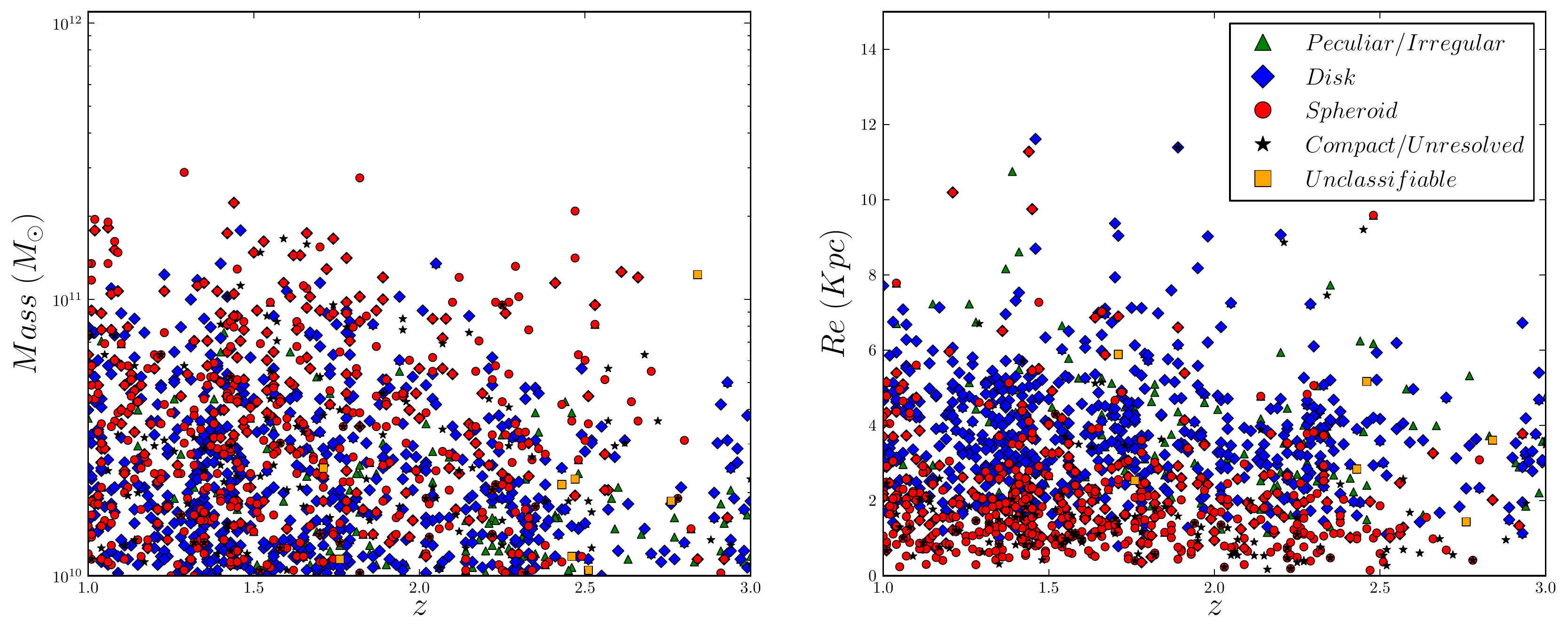}
  \caption{The stellar masses (left) and effective radius, $R_{e}$, from our $1$-component fit (right) vs. redshift for galaxies used in this work. Colours represents the classification from \protect\cite{Kartaltepe15}. Green triangles are peculiar or irregular galaxies, blue diamonds are disc galaxies, red circles are spheroids, black stars represent compact or unresolved sources, and yellow squares, unclassifiable objects.} \label{Figure1}
\end{figure*}

For this work we examine a sample of $1495$ galaxies at redshifts $1< z < 3$ with stellar masses $M_{\ast}\geq10^{10}\textrm{M}_{\odot}$ (see Figure \ref{Figure1}) from the CANDELS UDS field. CANDELS \citep{Grogin11,Koekemoer11} is a Multi Cycle Treasury Program which images the distant Universe with both the near-infrared Wide Field Camera $3$ (WFC3) and the visible-light Advanced Camera for Surveys (ACS). In total, CANDELS consists of $902$ orbits with the \textit{Hubble Space Telescope} (\textit{HST}) and covers $800\ \textrm{arcmin}^2$. The survey targets five distinct fields (GOODS-N, GOODS-S, EGS, UDS and COSMOS) at two distinct depths. The deep portion of the survey is referred to as `CANDELS/Deep', with exposures in GOODS-N and GOODS-S. `CANDELS/Wide' is the shallow portion and images all five CANDELS fields. We have used the WFC3 data from the UDS which comprises $4\times 11$ tiles and covers an area of $187\ \textrm{arcmin}^2$ in the $F160$ ($H$-band) filter. The $5\!\,\sigma$ point-source depth for this filter is $H=27.1$ ($AB$ mag).

The CANDELS UDS field is a subset of the larger UDS area which contains data from the \textit{U}-Band CHFT, \textit{B}, \textit{V}, \textit{R}, \textit{i}, \textit{z}-band SXDS data and \textit{J}, \textit{H} and \textit{K}-band data from UKIDSS. This includes $F606W$ and $F814W$ imaging ACS, $H160$, and CANDELS  $J125$-band \textit{HST} WFC3 data, \textit{Y} and \textit{Ks} bands taken as part of the HAWK-I UDS and GOODS-S survey (HUGS; VLT large programme ID 186.A-0898, PI: Fontana; \citealt{Fontana14}). For the CANDELS UDS, the $3.6$ and $4.5$ $\mu \textrm{m}$ data are taken as part of the \textit{Spitzer} Extended Deep Survey (SEDS; PI: Fazio; \citealt{Ashby13}). SEDS is deeper than SpUDS, which is used in the UDS data set, but is only available over a $0.17$ deg$^{2}$ region. Therefore, SEDS is a more appropriate choice for the smaller CANDELS UDS region. For a detailed discussion of the CANDELS UDS region photometry see \cite{Galametz13}.

In Figure \ref{Figure1} we show how the stellar mass and effective radii of our galaxies are distributed with redshift, along with the morphological classification from \cite{Kartaltepe15}, where galaxies are visually classified into five main morphology classes. Such classes are based on the typical Hubble sequence types: discs, spheroids, irregular/peculiar, compact/unresolved and unclassifiable (more than one of these options can be selected for each galaxy). We divide our sample into star forming and passive galaxies using the rest-frame \textit{UVJ} colours \citep[see][]{Mortlock13}, where a galaxy is classified as red/passive if it satisfies the following criteria

\begin{equation}\label{eq_uvj}
\begin{cases}
 (U-V)>1.3 \\
 (V-J)<1.6 \\
 (U-V)>0.88\cdot (V-J)+0.49
\end{cases} 
\end{equation}

\noindent and as blue otherwise (see Ownsworth et al. submitted, for more details).

\subsection{Redshifts and Stellar Masses}

We use a combination of photometric and spectroscopic redshifts as described in \cite{Mortlock15} and \cite{Hartley13}. The photometric redshifts and stellar masses we use are described in \cite{Mortlock13}. The photometric redshifts were computed by fitting template SEDs to the photometric data points described in the previous Section using the \textsc{eazy} code \citep{Brammer08}. The photometry was fit to the linear combinations of the six default \textsc{eazy} templates, and an additional template which is the bluest \textsc{eazy} template with a small amount of Small Magellanic Cloud-like extinction added ($A_V = 0.1$). The redshifts are retrieved from a maximum likelihood analysis. For full details of the fitting procedure and resulting photometric redshifts see \cite{Hartley13} and \cite{Mortlock15}.

A comparison of the photometric redshifts used in this work to spectroscopic redshifts which are available in the UDS was carried out in \cite{Mortlock15} where it is discussed the spectroscopic redshifts versus the photometric redshifts for the $285$ CANDELS galaxies with spectroscopic redshifts \citep[see also][for details]{Galametz13}. The dispersion of $z_\mathrm{photo.}$ vs. $z_\mathrm{spec.}$ is $\delta z /(1 + z) = 0.026$ for the photometric redshifts, after removing the $2$ per cent of catastrophic outliers. However, note that we have only a small sample of spectroscopic redshifts to compare to within the CANDELS UDS region.

Stellar masses are obtained by creating a large grid of synthetic SEDs from the stellar population of \cite{Bruzual03}, using a Chabrier Initial Mass Function (IMF) \citep{Chabrier03}. And the UDS sample is complete down to $10^{9.5} M_\odot$ at $2.5<z<32$ \citep[see][]{Mortlock15}, therefore, our sample of massive galaxies is mass complete. We use as a $PSF$ the combination of the TinyTim simulated $PSF$ and a stacked star empirical $PSF$. The reason for using this $PSF$ is that the TinyTim $PSFs$ are better in the core region (where empirical $PSFs$ tend to broaden), while empirical $PSFs$ appear to fit real stars better in the wings.

\section{Method}\label{sec.method}

We have used \textsc{galfit} and \textsc{galapagos} to perform our morphological analysis on our sample. \textsc{galfit} is a two-dimensional fitting code used to model the surface-brightness of an object with predefined functions. This program allows the user to fit any number of components and different light profiles (e.g. S\'ersic, Exponential disc, Gaussian, Moffat, Nuker, etc.) The most used and useful functions to describe galaxies are the Exponential disc profile and the S\'ersic profile \citep{Sersic} for fitting, respectively, disc and bulges/spheroids.

The S\'ersic profile has the following functional form given by,
\begin{equation}\label{eq_sersic}
 \Sigma(R)=\Sigma_{e}\exp\left\{-\kappa_n\left[\left(\frac{R}{R_e}\right)^{1/n}-1\right]\right\},
\end{equation}

\noindent where the parameter $R_e$ is the effective radius, such that half of the total flux is within $R_e$. $\Sigma_e$ is the surface brightness at the effective radius $R_e$. The parameter $n$ is the S\'ersic index, and it determines the shape of the light profile. Finally $\kappa_n$ is a positive parameter that for a given $n$, can be determined from the definition of $R_e$ and $\Sigma_e$. It satisfies the equation $\Gamma(2n)=2\gamma(2n,\kappa_n)$, a non-linear equation which can be solved numerically, where $\Gamma$ is the gamma function and $\gamma$ is the incomplete gamma function \citep[see][]{Graham05}. The classic de Vaucouleurs profile that describes spheroids and massive galaxy bulges is a special case of the S\'ersic profile with $n=4$ and $\kappa_4=7.67$. The Exponential disc profile is also a special case of the S\'ersic function when $n=1$ and $\kappa_1=1.68$. The best fit model is obtained by $\chi^2$-minimisation using a Levenberg-Marquardt algorithm in \textsc{galfit}.

We carry out our fitting with \textsc{galapagos} and \textsc{galfit}. \textsc{galapagos} is a software that uses \textsc{se}xtractor \citep{Bertin96} to detect and extract sources and performs an automated S\'ersic profile fit using \textsc{galfit}. It is divided into four main stages: the first one detects sources by running \textsc{se}xtractor, the second one cuts out postage stamps for all detected objects, the third block estimates the sky background, prepares and runs \textsc{galfit}, and the last stage compiles a catalogue of all galaxies. We then fit all of our sample galaxies with both $1$ and $2$ dimensional profiles. 

\subsection{One Component Model}\label{one_comp_model}

We run \textsc{galapagos} on all of our $H$-band galaxy images to fit our sample galaxies with a single S\'ersic profile, with $n$ as a free parameter as in equation \eqref{eq_sersic}. \textsc{galapagos} creates a mask for each individual postage stamp and decides whether a neighbouring object is masked or fit simultaneously, taking into account the distance and relative brightness to the main object. It also calculates the sky value to be used in the fit. As a result we obtain, for all sources, the following parameters: position of the galaxy within the stamp $(x,y)$, effective radius $R_e$, S\'ersic index $n$, $A\!B$-magnitude $m$, axis ratio $q$ and position angle $P\!A$. We discard any fitting with unphysical parameters: effective radius smaller than $0.5$ pixels, or larger than the size of the image stamp, $q<0.1$, and $n<0.5$ or $n>8$ ($\sim 7$ per cent of the objects).

\subsection{Two Component Model}

After the previous procedure, we then run \textsc{galfit} on the same postage stamps and use the sky value obtained by \textsc{galapagos} in Section \ref{one_comp_model} for the single component fit. We fit the surface brightness of the main galaxy to a S\'ersic (free $n$) plus an exponential profile (S\'ersic profile with $n$ fixed to $n=1$), where the total light distribution ($\Sigma$) is the sum of these two models:

\begin{align}\label{eq_model2}
\Sigma(R)&=\Sigma_{e}\exp\left\{-\kappa_n\left[\left(\frac{R}{R_e}\right)^{\frac{1}{n}}-1\right]\right\} \nonumber \\
 &\qquad + \Sigma_{e}\exp\left[-1.68\left(\frac{R}{R_e}-1\right)\right],  
\end{align}
\noindent fitting simultaneously or masking neighbour objects in the same way \textsc{galapagos} does for the one component model. We constrain the centre of both components to be the same. The result is a list of structural parameters for all the sample galaxies: position in the stamp $(x,y)$, effective radius of bulge and disc components $(R_{e\,\mathrm{B}},R_{e\,\mathrm{D}})$, S\'ersic index of the bulge $n_\mathrm{B}$, $AB$-magnitude for bulge and disc $(m_\mathrm{B},m_\mathrm{D})$, axis ratio of bulge and disc $(q_\mathrm{B},q_\mathrm{D})$ and position angle of both components $(P\!A_\mathrm{B},P\!A_\mathrm{D})$. As in the previous model, we exclude any fitting with unphysical parameters in any component for the effective radius, axial ratio or S\'ersic index $(R_{e\,\mathrm{B}},R_{e\,\mathrm{D}},q_\mathrm{B},q_\mathrm{D},n_\mathrm{B})$.

Adding an extra component increases the degrees of freedom, hence it is more likely that the fitting gets trapped in a local minimum of $\chi^2$ in the minimisation process. To ensure that the $\chi^2$ obtained from the fitting is the global minimum, we have run \textsc{galfit} starting with different initial values of magnitudes, effective radius and S\'ersic index. For the S\'ersic index, we choose alternatively as initial values $n=1,2.5,4$. The starting values of the magnitudes of each components are: both equal to a magnitude that corresponds to half of the total flux obtained from the one component model, one magnitude which correspond to $80$ per cent of the total flux while the other is $20$ per cent and vice versa. The starting values for the effective radius are: both equal to the effective radius obtained from the one component model, one of the components half the size of that radius while the other is $20$ per cent times larger, and vice versa. We therefore run \textsc{galfit} for the $3^3=27$ possibilities. We choose the model that delivers the smallest $\chi^2$ and does not have any unphysical parameters.

We first try to fit all the central components of our galaxies with a free $n$ for the S\'ersic profile (first term of eq. \ref{eq_model2}), but in some cases ($\sim 40$ per cent) the fitting results in an unrealistic S\'ersic index (either too small or too big). In such cases, we redo the fitting with the S\'ersic index fixed first at $n=4$ and then at $n=1$ in eq. \ref{eq_model2}, and choose the fitting with the smallest $\chi^2$. In $\sim 73$ per cent of these cases, the model prefers $n=1$. There are still some objects ($\sim 20$ per cent) that do not have any realistic result with two components, those will be directly classified as $1$-component galaxies (if the fitting in this case is considered good) in all methods. In the end, only about $\sim 8$ per cent of the galaxies are not well represented with either the one or the two component model. These galaxies are either very compact objects, or considerably faint/small, and have an average redshift of $z=2$.

We later discuss in Section \ref{subsec.size} how the ratio of the fluxes in the two components changes with redshift. Overall we find that there is a fairly broad distribution of the ratio between the fluxes of the two components. Only about $6$ per cent of the galaxies have a second component which is less than $10$ per cent of the total flux. Otherwise, $70$ per cent of the sample of two component galaxies are disc-dominated, with $B/T < 0.5$.

\subsection{Morphological K-correction}\label{subsec.kcorrect}

\begin{center}
   \begin{figure}
      \includegraphics[width=1\linewidth]{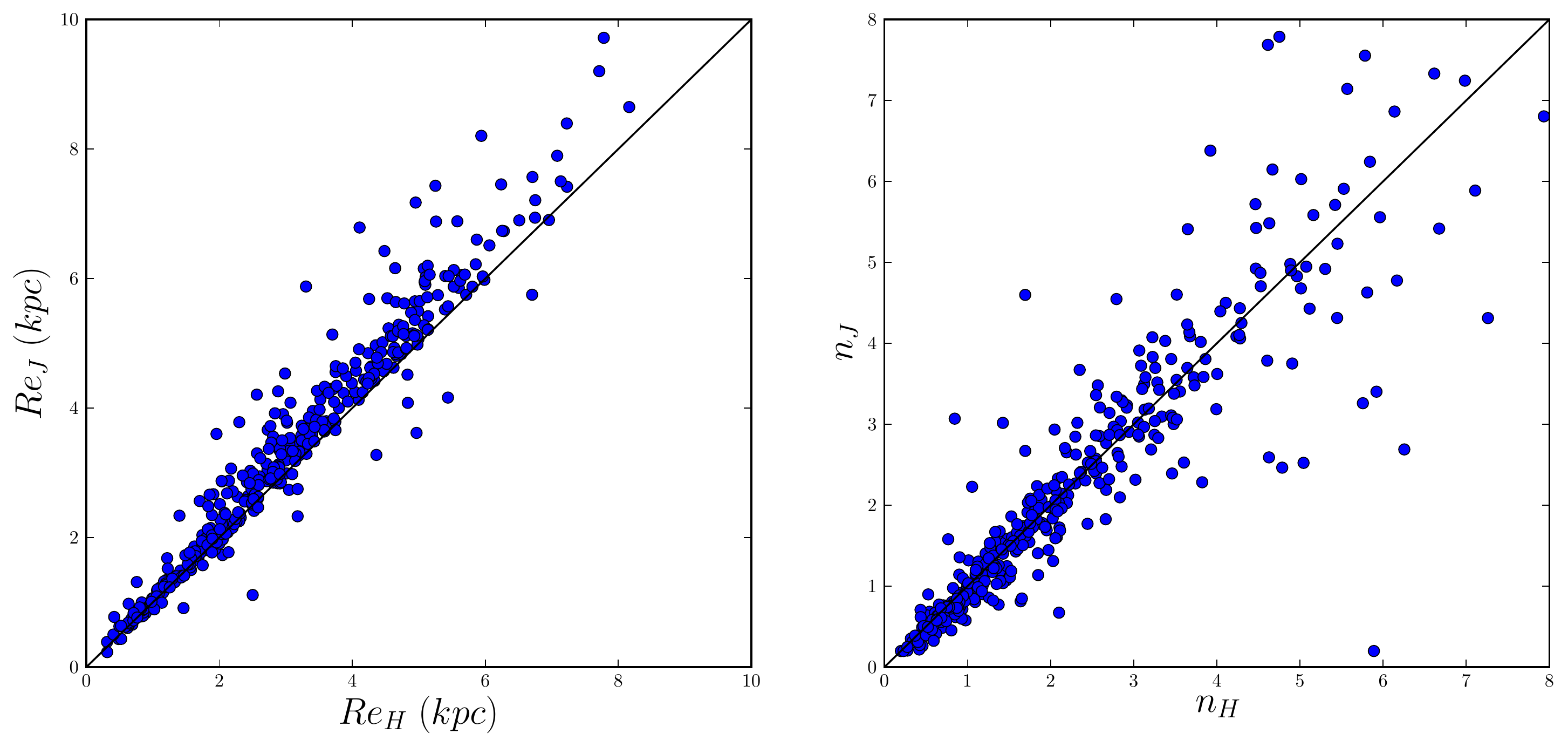}
      \caption{Comparison between effective radius (left) and S\'ersic index (right) obtained from fitting the surface brightness in the $J$-band and $H$-band with a single S\'ersic profile at $z=1$ with a $\Delta\lambda=200\ \textrm{nm}$ difference.}  \label{Figure2}
  \end{figure}
\end{center}

We also investigate whether we should consider the effect of the morphological K-correction in our study, as the quantitative structure of galaxies changes as a function of wavelength \citep[e.g.][]{TaylorMager07}. Using the $H$-band in the redshift range of $1<z<3$ means that we are observing and comparing galaxies at a rest-frame wavelength from visible to near-IR \citep{Conselice11}. Therefore, the difference in rest-frame wavelength is $\leq 350\ \textrm{nm}$ at the highest and lowest redshifts. To test whether this difference can have an effect on the structure and morphology of our galaxies, we select a subsample with $z\sim1$ and fit their surface brightness to a single S\'ersic profile in the $J$-band. The observed rest-frame in this case is $\sim 600\ \textrm{nm}$ and by comparing with the same galaxies in the $H$-band, we see the effect caused by a difference in rest-frame wavelength of $\sim 200\ \textrm{nm}$ which is in a similar range to that of our whole sample of galaxies. In Figure \ref{Figure2} we see that we recover the same structural parameters (effective radius and S\'ersic index) whether we use the $J$- or $H$-band. This means that the spanning in redshift for our sample of galaxies does not affect the observed structure and morphology. Therefore, we can continue our study without having to consider the morphological K-correction.

\subsection{Classification}\label{subsec.classification}
Once we have the two models for each of our galaxy profiles, we need a method to choose whether to use $1$ or $2$ component fits for each galaxy. This is critical for both determining the evolution of $1$-component galaxies, as well as for how multiple component galaxies form and evolve over the epoch $1 < z < 3$. In this paper we investigate three different methods of deciding whether a galaxy is better `fit' as a $1$ or $2$ component system, and compare the results of these methods to see how internally consistent they are. Our first method consists in visually classifying galaxies into $1$- or $2$-component systems. Our second method is based on an index called the Residual Flux Fraction ($RF\!F$), and the final method is based on an statistical test ($F$-test). All of these methods, are explained below.

\subsubsection{Visual Inspection}\label{subsec.visual}

\begin{figure*}
  \centering
  \subfloat[][$1$-component model]{\label{Figure3a}
  \includegraphics[width=0.45\linewidth]{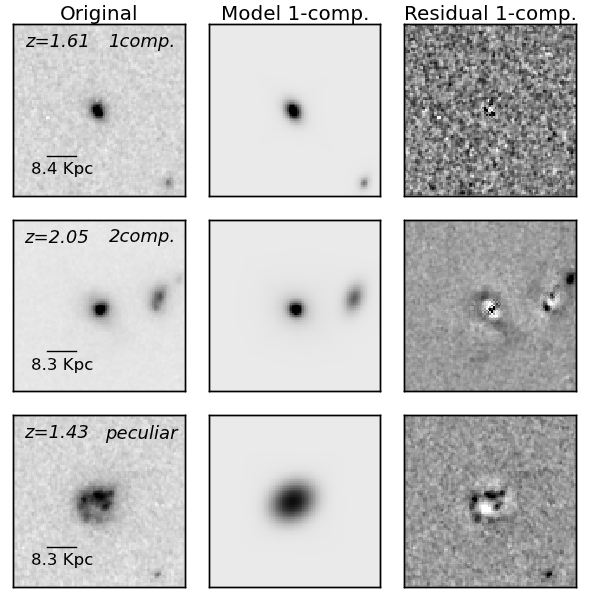}}
  \subfloat[][$2$-component model]{\label{Figure3b}  
  \includegraphics[width=0.45\linewidth]{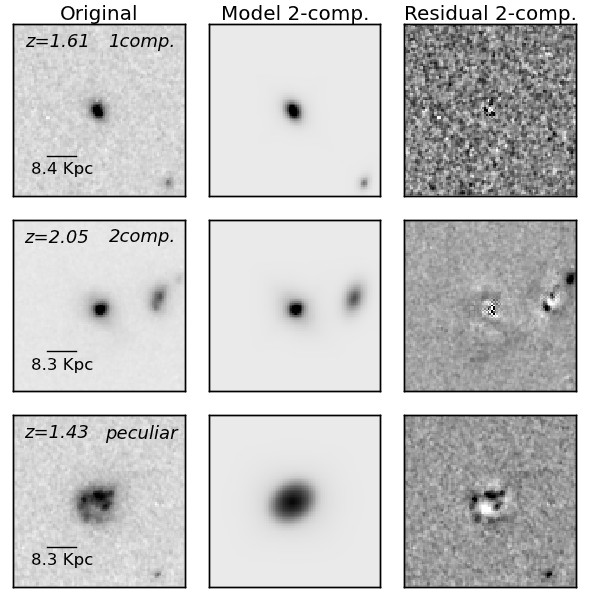}}
   \caption{Visual classification. Left: $1$-component model (original image, model and residual). Right: $2$-component model (original image, model and residual). Top row: example of $1$-component best fit. Middle row: example of $2$-component best fit. Bottom row: example of a peculiar galaxy}
\end{figure*}

The first method of determining whether a galaxy has one or two components consists of visually inspecting all the sample galaxies, and their correspondent residual images from both $1$ and $2$ component best fit models. In Figure \ref{Figure3a} we show examples of the fitting using one component for three different types of galaxies. For each model we show the original image (left), the model image (middle), and the residual image (right) which is obtained by subtracting the model to the original image. Analogously, in Figure \ref{Figure3b} we show the fitting using two components. 

\begin{figure*}
  \includegraphics[width=0.8\linewidth]{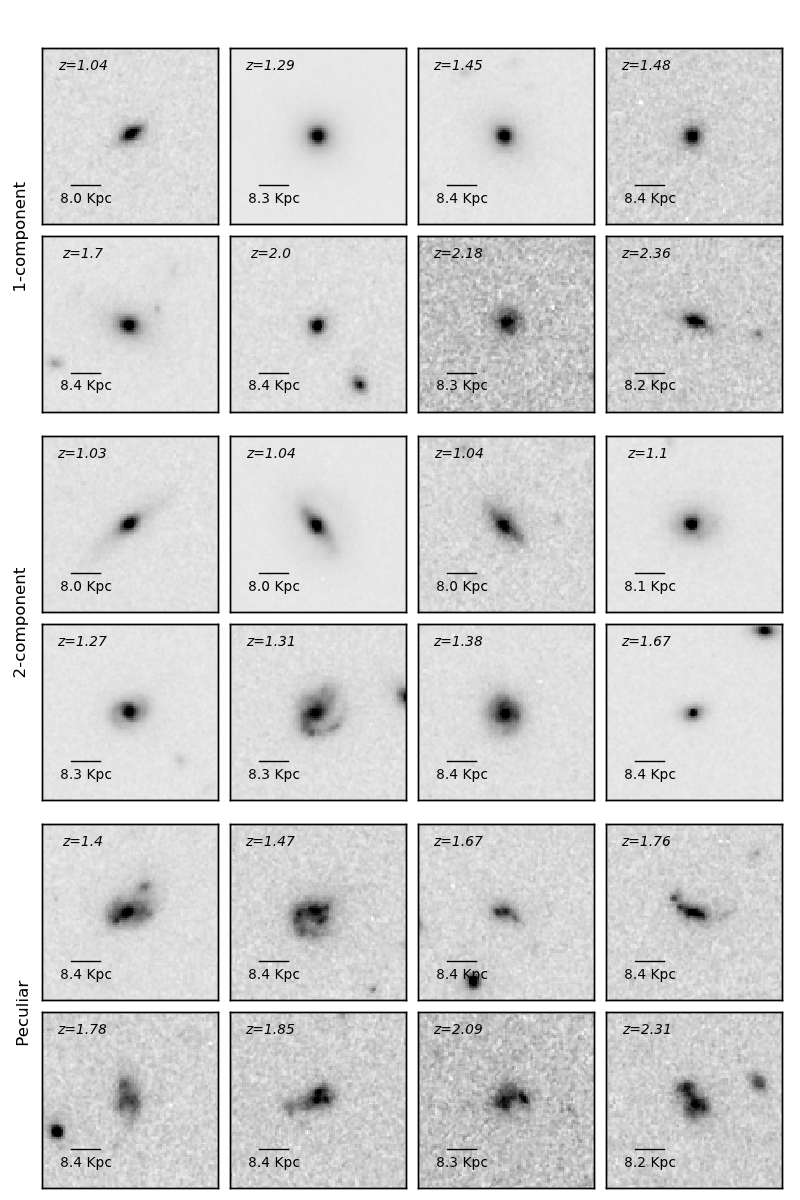}
  \caption{Examples of galaxies visually classify as $1$-component galaxies (top), $2$-component galaxies (middle) and peculiar galaxies (bottom). The postage stamps are $6\times 6$ arcseconds in size.}\label{Figure4}
\end{figure*}

We have visually classified all the galaxies in our sample into one of three types (examples in Figure \ref{Figure4}) based on both the visual appearance of the galaxy and also the residuals left over from the galaxy after the best fitting one and two component profiles are fit.

\begin{description}
 \item[\textbf{One component galaxies}.] These are disc-like or spheroid-like galaxies, which show no evidence of needing a second component. Indeed, a single S\'ersic profile fitting is able to reproduce well the surface brightness of the galaxy as shown by the lack of structures left in the residual image. 
 \item[\textbf{Two component galaxies}.] These are sometimes disc galaxies with a bulge component. They are better fit with a composition of a S\'ersic profile plus exponential profile. They show less residual light from the $2$-component models than with the single one, although a significant amount of residual can be left due to spiral arms in the disc. 
 \item[\textbf{Peculiar galaxies}.] These are disturbed galaxies or mergers. They show residuals from both models, and the addition of another component does not improve the fitting. 
\end{description}

There is a very small fraction of galaxies ($\sim 3$ per cent) that are removed from our sample: unresolved or unclassifiable (due to problems with the image) galaxies. Note that galaxies classified as one or two component galaxies can display irregular or merger features, but unlike peculiar galaxies, they are still well represented by either the single S\'ersic model or the S\'ersic plus exponential model respectively, as these features do not dominate the structure of the galaxy.

\subsubsection{Residual Flux Fraction}\label{subsec.rff}

The Residual Flux Fraction, or $RF\!F$ \citep{Hoyos11} is defined as the fraction of the signal contained in the residual image that cannot be explained by fluctuations of the background. Hence, the smaller the $RF\!F$ value the better the fitting. This index is defined as follows

\begin{equation}\label{eq_rff}
 RF\!F=\frac{\displaystyle\sum_{(j,k)\in A}|I_{j,k}-I^{\mathrm{GALFIT}}_{j,k}|-0.8\cdot{}\hspace*{-1.8ex}\sum_{(j,k)\in A}\sigma_{B\, j,k}}{FLUX\_ AUTO},
\end{equation}

\noindent where $I$ is the actual galaxy image, $I^\mathrm{GALFIT}$ is the model image created by \textsc{galfit}, $\sigma_{B}$ is the background RMS image and $FLUX\_ AUTO$ is the total flux of the galaxy calculated by \textsc{se}xtractor. Finally, $A$ represents the area in which we calculate this index. The $0.8$ factor in the numerator guarantees that for a Gaussian noise error image, the expected value of the $RF\!F$ is $0$ \citep[see][for details]{Hoyos11}. It is important to note that the $RF\!F$ diagnosis does not work well for large areas \citep{Hoyos11}. In those cases, as the galaxy decays towards zero flux at large radius, outer areas will dominate the $RF\!F$ computation, making it small even when the residual is not good at the centre. Taking this into consideration we have decided to use the area inside the $2.5\cdot{}R_{kron}$ radius of each galaxy to calculate the $RF\!F$, where $R_{kron}$ is the Kron radius obtained from \textsc{se}xtractor \citep[e.g.][]{Zhao15}.

To calculate the first term of the numerator in equation \eqref{eq_rff}, we sum the absolute value of the pixels inside the chosen area from the residual image (original image subtracted by the model). If there is a nearby, but different, object inside this area we do not take into account the pixels corresponding to that object, in order to reduce as much as possible bad fittings from nearby objects affecting the $RF\!F$ of the main object. To compute the second term of the numerator we assume
\begin{equation}\label{eq_sigma}
\sum_{(j,k)\in A}\sigma_{B\, j,k}=N\cdot{}\langle\sigma_{B}\rangle,
\end{equation}

\noindent where $\langle\sigma_{B}\rangle$ is the mean value of the background sigma for the whole image, and $N$ the number of pixels in the area we are considering in the calculation of the $RF\!F$ (excluding those pixels belonging to nearby objects). We obtain the value $\langle\sigma_{B}\rangle$ directly from the sky measures from \textsc{se}xtractor.

We compute the $RF\!F$ for both the one and two component models (denoted as $RF\!F_{1}$ and $RF\!F_{2}$ respectively) for all the objects in our sample. Peculiar and spiral galaxies have similar $RF\!F$ values, namely the average value of $RF\!F$ for the $1$-component model in spirals is $RF\!F_1=0.07$ while for peculiars it is $RF\!F_1=0.08$, making it difficult to distinguish these two populations using just the $RF\!F$. To solve this problem we use our visual classification (\ref{subsec.visual}) to separate these two populations.

\begin{center}
    \begin{figure}
      \includegraphics[width=1\linewidth, trim=2.3cm 0cm 2.3cm 0cm, clip=true]{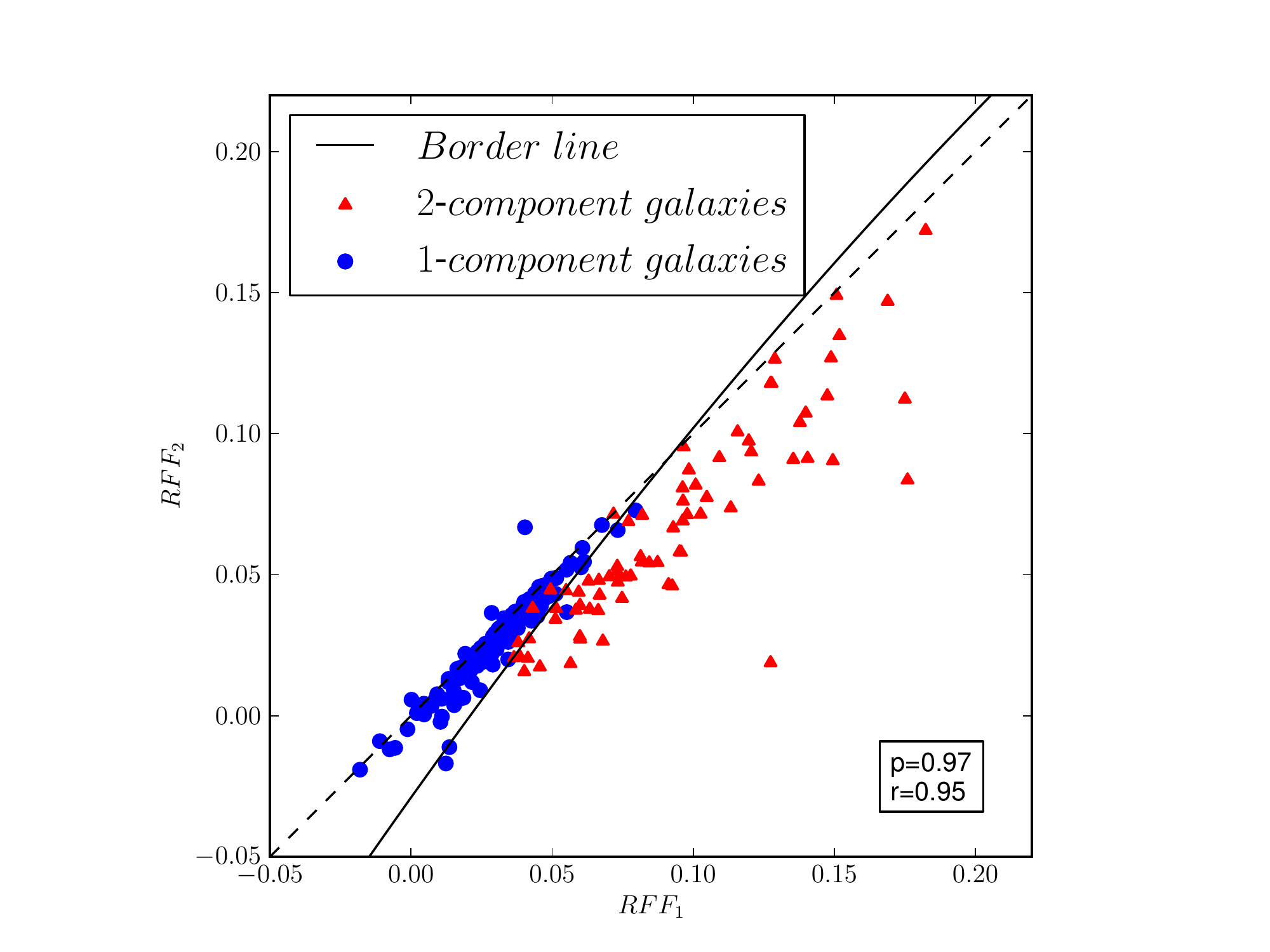}
      \caption{$F$-score technique using $RF\!F$ measures (see Section \ref{subsec.rff} for more details). Using a training sample obtained by visual classification (red triangles: $2$-component galaxies, blue circles: $1$-component galaxies) we obtain the line (black solid line) that separates these two subsamples given by equation \eqref{eq_best_line}. $p$ and $r$ are the completeness of the two subsamples. The dash line is the identity function.}  \label{Figure5}
    \end{figure}
\end{center}

Spheroid-like galaxies have small $RF\!F_1$ ($RF\!F_1$ $\lesssim$ $0.5$) and $RF\!F_2\cong RF\!F_1$, as they are well fit by a single S\'ersic profile model. Meanwhile, galaxies that contain a bulge and a disc will generally have a larger $RF\!F_1$ ($RF\!F_1 \gtrsim 0.5$), due the spiral arms and $RF\!F_2<RF\!F_1$, as the two-component fitting will be better than the single-component model. Therefore they will occupy a different region in the plane of $RF\!F_2$ vs. $RF\!F_1$ (see Figure \ref{Figure5}).

We have also used the $F$-score technique \citep{Hoyos12} to find the border in the $RF\!F_2$ vs. $RF\!F_1$ diagram that best separates these two populations ($1$-component and $2$-component galaxies). This method consists in finding the parameters of a function (the border, that in our case will be a second order polynomial) that maximise the $F$-score, $F_\beta$ \citep{Rijsbergen79}, defined as
\begin{equation}\label{eq_F}
 F_\beta=\frac{1+\beta^2\cdot{}p\cdot{}r}{\beta^2\cdot{}p+r},
\end{equation}

\noindent where $r$ and $p$ are the sensitivity or completeness of both populations and are given by the equations
\begin{align}
 r&=\frac{\#\{True\ Population_1\}}{\#\{True\ Population_1\}+\#\{False\ Population_2\}}\label{eq_r}\\
 p&=\frac{\#\{True\ Population_2\}}{\#\{True\ Population_2\}+\#\{False\ Population_1\}}\label{eq_p}
\end{align}

\noindent In these definitions $\#\{True\ Population_1\}$ is the number of objects correctly classified as $Population_1$ by the method while $\#\{False\ Population_1\}$ is the number of those objects of the $Population_2$ misclassified as belonging to $Population_1$. We define analogously $\#\{True\ Population_2\}$ and $\#\{False\ Population_2\}$. Hence $r$ measures the fraction of the actual elements in $Population_1$ correctly classified as $Population_1$. Finally $\beta$ is a control parameter, specified by the user, that determines the relative importance of $r$ and $p$. We have chosen $\beta=1.0$ because having a complete sample of the $Population_1$ is as important as having a complete sample of $Population_2$. 

To apply this method to our sample of galaxies we have performed the $F$-score technique in a training sample. This training sample is formed by the galaxies that have been visually classified most confidently as either $1$-component ($Population_1$) or $2$-component systems ($Population_2$). This allows us to obtain the (second order polynomial) line that best separates these two populations, from which we can then classify the rest of the galaxies in our sample according to this criterion. It is interesting to note that in our case, changing the value of $\beta$ does not significantly change the border line, as both populations of the training sample are clearly separated in the $RF\!F_2$ vs. $RF\!F_1$ plane. This maximisation has been performed with the Amoeba algorithm \citep{Press}, using a second order polynomial as the border line. The result of this maximization is shown in Figure \ref{Figure5} and can be expressed as

\begin{equation}
 RF\!F_2=-0.023+1.40\cdot{} RF\!F_1-0.94\cdot{} RF\!F_1^2\label{eq_best_line}.
\end{equation}

\noindent This line gives the following values for the completeness of the two populations, for the training sample: $r=0.95$, $p=0.97$.

\begin{figure}
    \centering
        \includegraphics[width=1\linewidth, trim=2.3cm 0cm 2.3cm 0cm, clip=true]{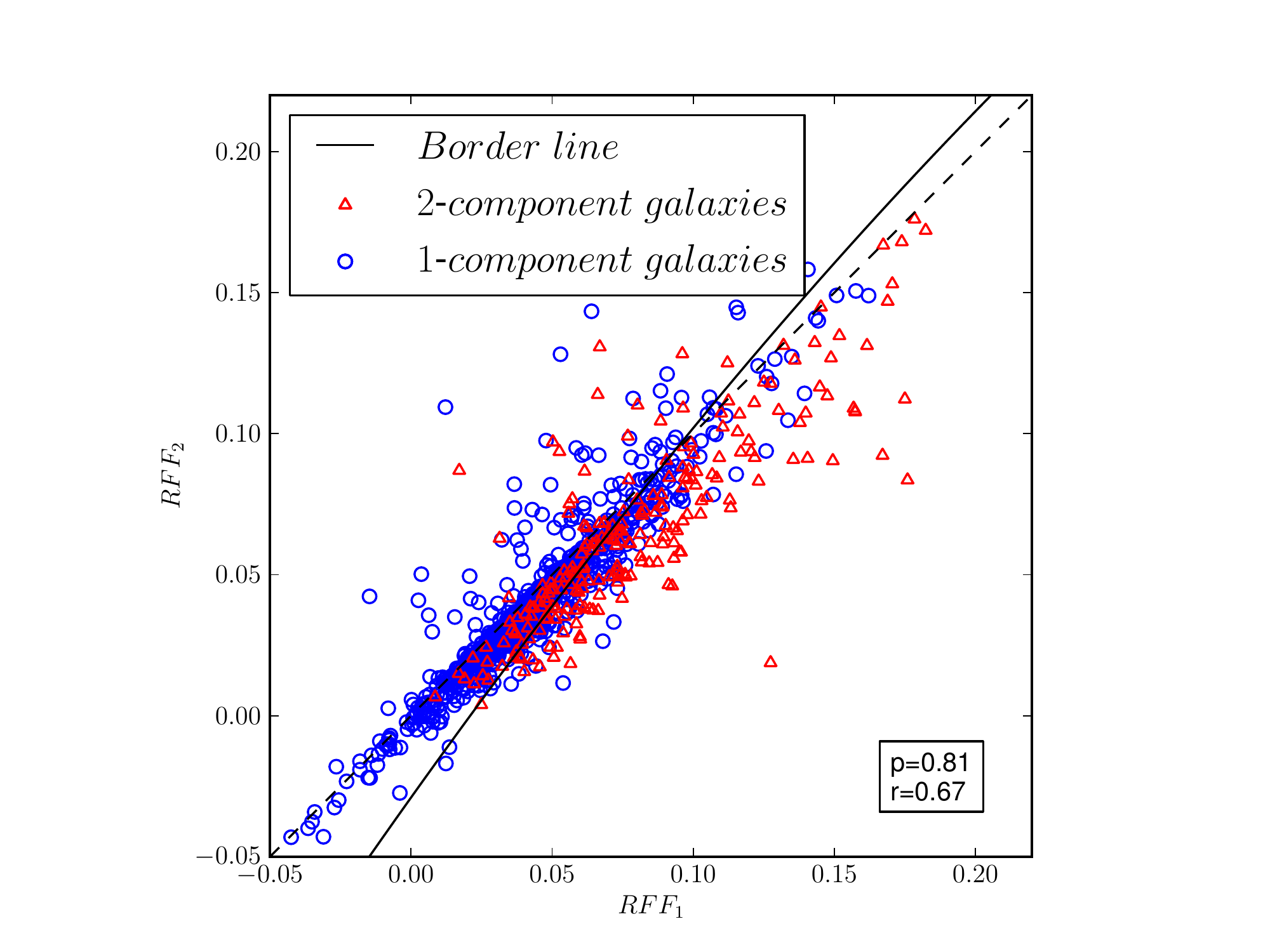}
        \includegraphics[width=1\linewidth, trim=2.3cm 0cm 2.3cm 0cm, clip=true]{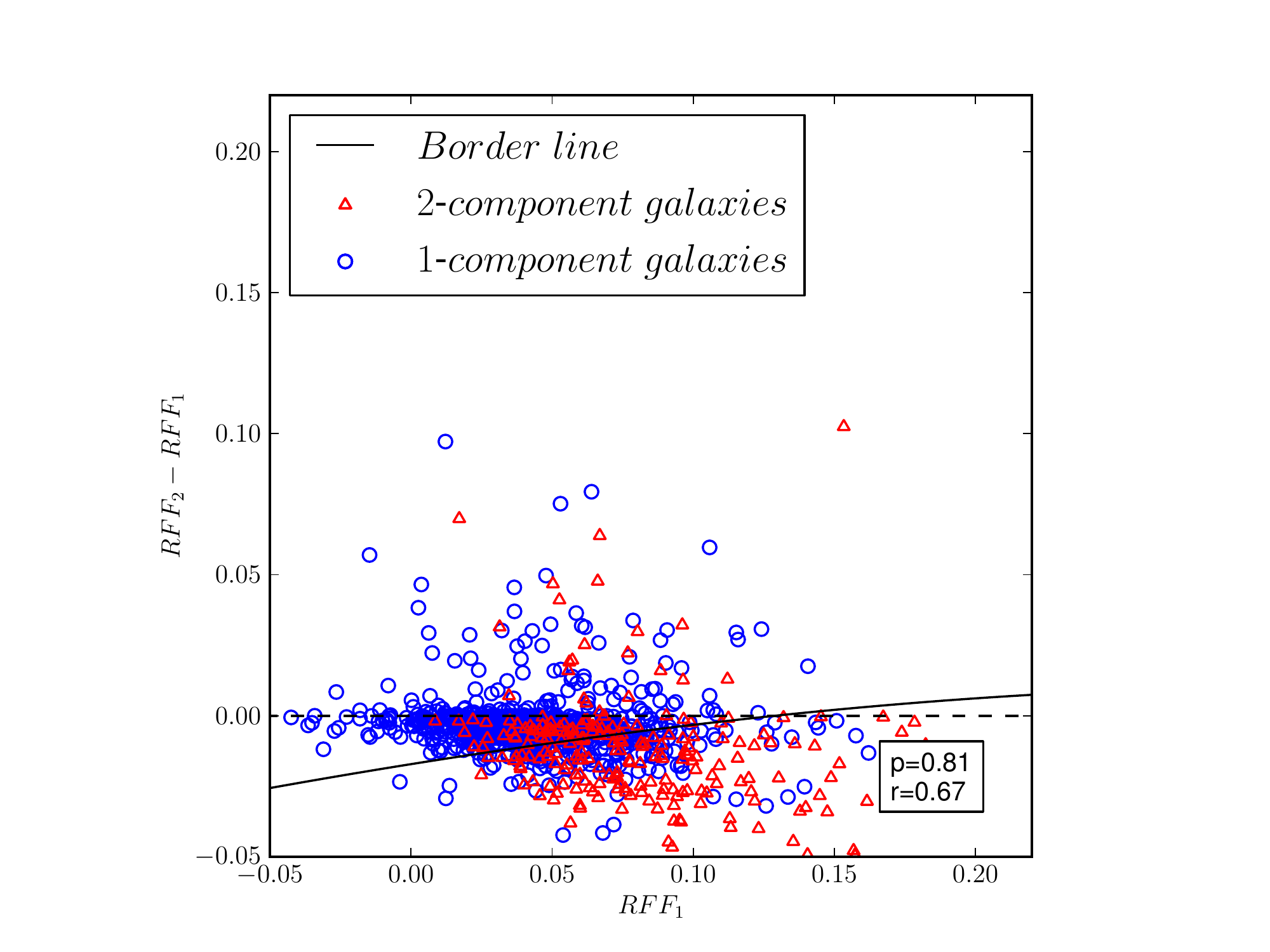}
    \caption{$RF\!F_2$ (top) and difference in $RF\!F=RF\!F_2-RF\!F_1$ (bottom) versus $RF\!F_1$, with the classification of all the galaxies using equation \eqref{eq_best_line} (black solid line), below which galaxies are considered as having $2$ components, and above the line, they are classified as $1$-component galaxies. Red triangles are galaxies visually classified as $2$-component galaxies and blue circles as $1$-component ones. }  \label{Figure6}
\end{figure}

As mentioned earlier, once we know this line, we can plot in the $RF\!F_2$ vs. $RF\!F_1$ plane all our objects (see Figure \ref{Figure6}) and classify them according to their position with respect to the equation of line \eqref{eq_best_line}: as $1$-component galaxies if they lie above the line or as $2$-component galaxies if they are under the line. In Figure \ref{Figure6} we plot the entire sample and, just for comparison with the visual classification, we have plotted in blue circles those objects that have been visually classified as $1$-component galaxies and in red triangles those classified as $2$-component galaxies. There is overall a good agreement between the two methods.

\subsubsection{$F$-test}\label{sugbsec.f_test}

The $F$-test is a statistical test in which the statistic has an $F$-distribution under the null hypothesis. An $F$-distribution is formed by the ratio of two independent $\chi^2$ variables divided by their respective degrees of freedom.

We have performed the $F$-test following the method described in \cite{Simard11}, who also use it for $1$ vs. $2$ component separation within SDSS data. In our study we have two models: S\'ersic profile (model $1$) and a S\'ersic+exponential profile (model $2$). We consider the $\chi^2$ for each model from the residual image, and take as degrees of freedom the number of resolution elements $n_{res}$ minus the number of free parameters in the model. The number of resolution elements can be calculated as follow
\begin{equation}\label{n_res}
n_\mathrm{res}=\frac{n_\mathrm{pixels}}{\pi\theta^2},
\end{equation}

\noindent where $n_\mathrm{pixels}$ is the number of unmasked object pixels used in the fitting, and $\theta=1.38\ \mathrm{pix}$ is the $H$-band seeing half-width half maximum, in units of pixels. As in the $RF\!F$ calculation, we compute the $\chi^2$ in the area inside the $2.5\cdot{}R_{kron}$ of each galaxy.

To know whether the $\chi^2$ from model $2$ is significantly smaller than the $\chi^2$ from model $1$, we have to perform a one tailed test. The hypothesis of such test can be formulated as follows
\begin{itemize}
    \item Null hypothesis $H_{0}$: $\chi^2_1 \leq \chi^2_2$, (the simpler model is correct).
    \item Alternative hypothesis $H_1$: $\chi^2_1 > \chi^2_2$.
\end{itemize}

From the statistic of the test $F=\chi^2_{\mathrm{red},2}/\chi^2_{\mathrm{red},1}$ the probability $P$ of accepting the null hypothesis (i.e.\ the probability that the more complex model is not required) can be calculated. Following \cite{Simard11} we set a $1\!\,\sigma$ threshold value $P_0=0.32$ below which we consider galaxies to be better fit by model $2$ (S\'ersic+exponential), and therefore, classified as $2$-component galaxies. Meanwhile those with $P > P_0$ are classified as $1$-component galaxies.

Notice however that with this method, we cannot distinguish peculiars from $1$-component galaxies, as in both cases the more complex model is not required. As in the $RF\!F$ method, we have used the visual classification to separate the peculiar galaxies. Those can also be separated by using the asymmetry index \citep[e.g.][]{Conselice03}, finding similar galaxies \citep{Mortlock13}.

\section{Results}\label{sec.results}

In this Section we compare how the selection of our galaxy sample into one or two component types varies from one method to another. For our final results, we average the properties for the three methods, to take into account the strengths and weaknesses of each method. These final results include examining the fraction of $1$- or $2$-component galaxies as a function of redshift and stellar mass, as well as the evolution of the sizes of these components with redshift. In the Appendix \ref{appendix} we present simulations to test the robustness of our conclusions.

\subsection{Method Comparison and Basic Trends}\label{subsec.comparison}

\begin{center}
    \begin{figure}
      \includegraphics[width=1\linewidth]{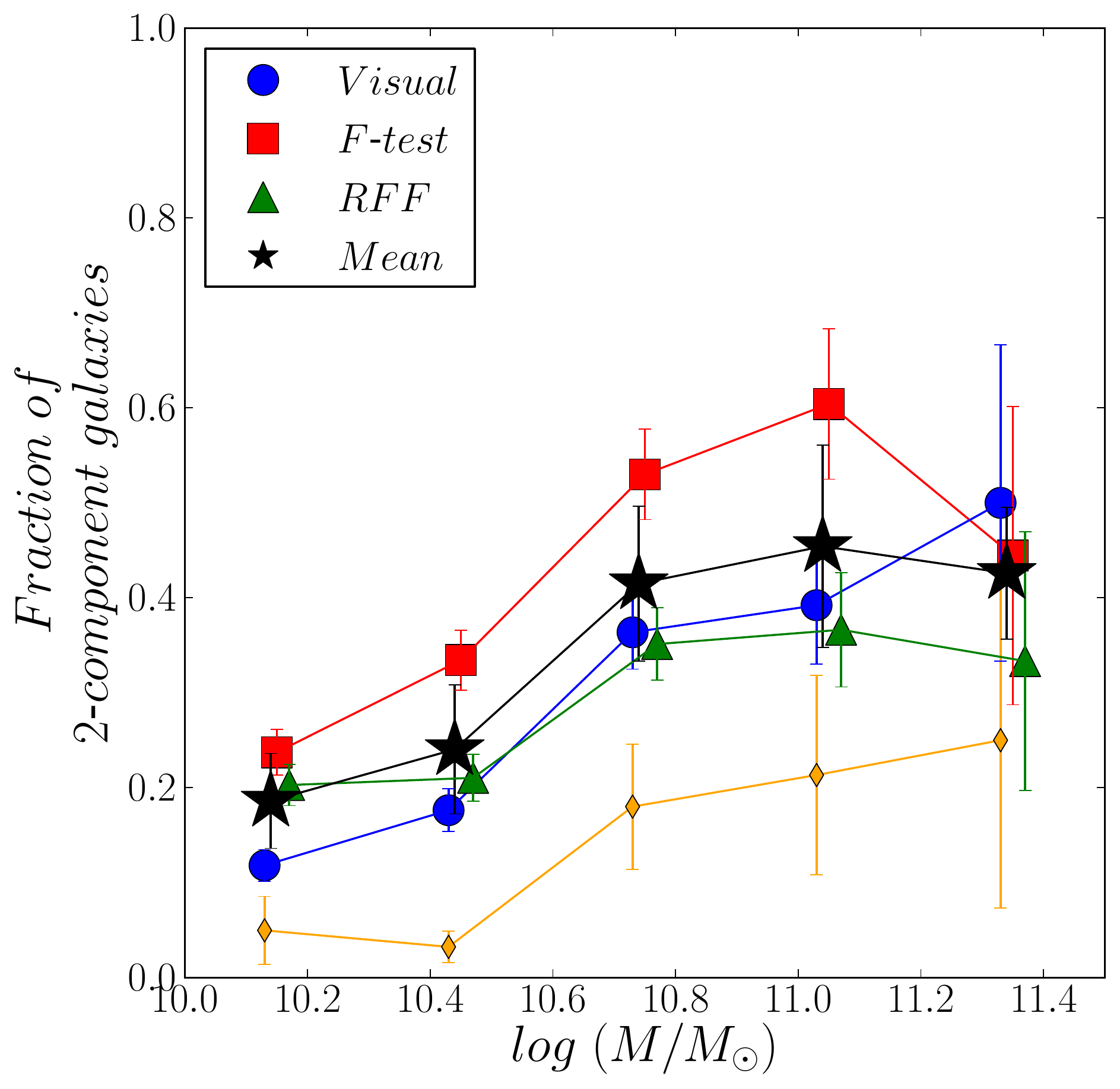}
      \caption{Fraction of $2$-component galaxies as a function of stellar mass at redshifts $1<z<3$, selected by the three selection methods (visual classification: blue circles, $F$-test: red squares, and $RF\!F$ method: green triangles) as a function of the stellar mass. The black stars represent the mean fraction in each mass bin. The yellow diamonds are the fraction of $2$-component galaxies from the high redshift simulated galaxies (see Appendix \ref{appendix}). Error bars represent the random errors.}\label{Figure7}
    \end{figure}
\end{center}

\begin{center}
    \begin{figure}
      \includegraphics[width=1\linewidth]{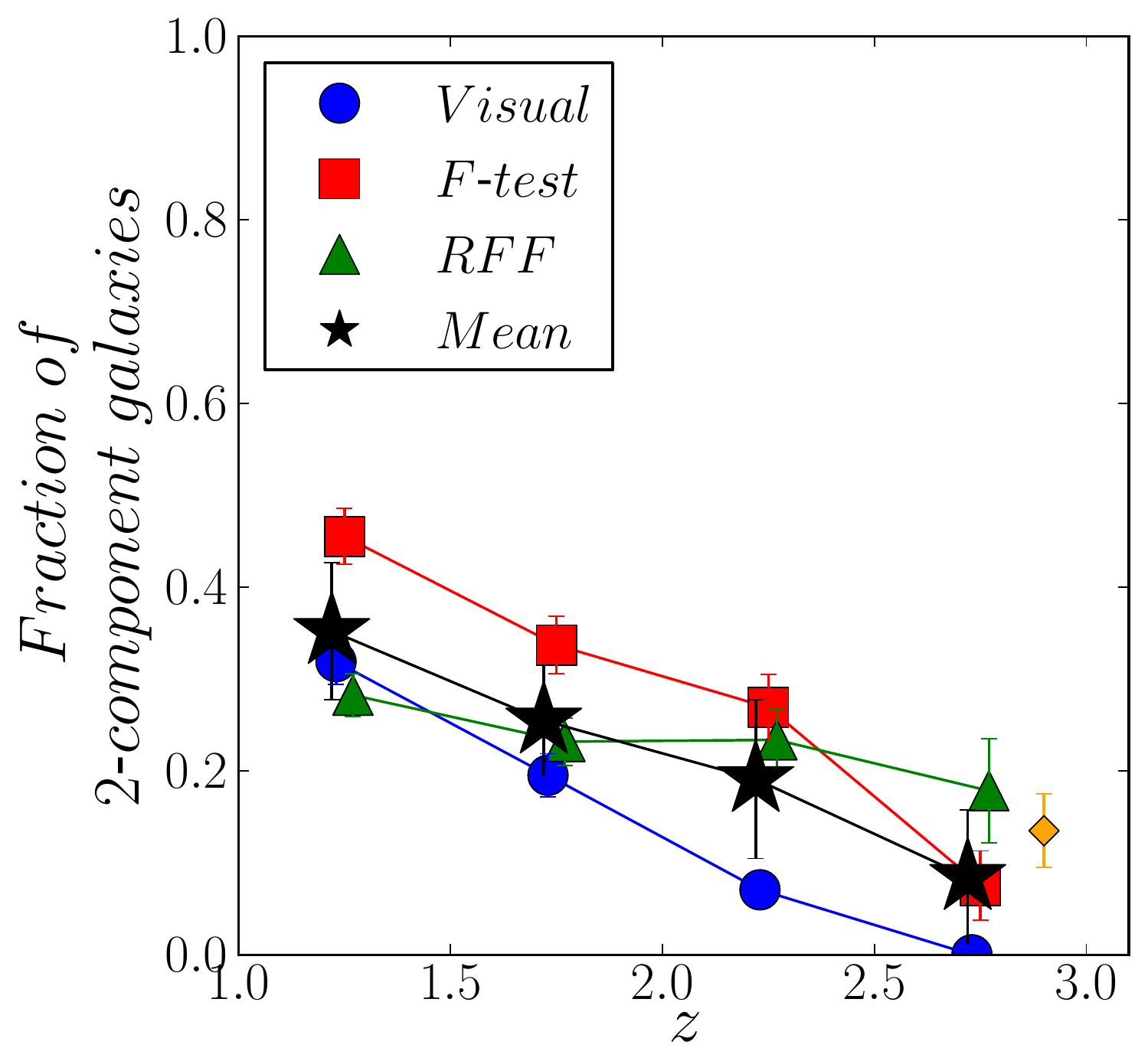}
      \caption{Fraction of $2$-component galaxies selected by the three selection methods (visual classification: blue circles, $F$-test: red squares, and $RF\!F$ method: green triangles) as a function of redshift. The black stars represent the mean fraction in each redshift bin. The yellow diamond is the fraction of $2$-component galaxies from the high redshift simulated galaxies (see Appendix \ref{appendix}). Error bars represent the random errors.}\label{Figure8}
     \end{figure}
\end{center}

We first explore how the three methods select different galaxies as being $1$- or $2$-component systems. We demonstrate this in Figures \ref{Figure7} and \ref{Figure8} which show the fraction of galaxies selected as $2$-component galaxies by each method, as a function of mass and redshift respectively, normalized by the total number of galaxies in each bin. The first thing to take away from these figures is that the agreement between the three methods is good, with the average of the three methods shown as the black stars in both figures.  

In more detail, we see that the fraction of $2$-component galaxies increases with stellar mass (Figure \ref{Figure7}) by a factor of $\sim 2$ from the lowest mass ($19\pm5$ per cent) to the highest mass bin ($43\pm6$ per cent). We explored the possibility that this trend was due to $S/N$ instead of stellar mass, but we observed that regardless of the $S/N$ the trend of $2$-component galaxies with stellar mass was preserved, so we believe that this trend is real. This trend is also not a result of redshift effects as we show in the Appendix \ref{appendix}.

In a similar way, we see a trend in terms of the fraction of $2$-component galaxies within our $M_{*} > 10^{10} \textrm{M}_{\odot}$ sample at $z < 3$ (Figure \ref{Figure8}). The fraction of $2$-component galaxies decreases with higher redshift for all the methods, from $35\pm6$ per cent at $z\sim 1$ to $8\pm6$ per cent at $z\sim3$. This is a significant change over a relatively quick $\sim 2$ Gyr time-span. However, some of this evolution could be due to redshift effects (see Appendix \ref{appendix}), so this must be a tentative conclusion at present.

Another thing to notice is that in terms of the fraction of $2$-component galaxies evolving with redshift, the $RF\!F$ method appears to be roughly constant while the other two methods decline with higher redshift. This is likely due to our galaxies having some residual light even at high redshift. This can arise from having two components at lower redshifts, where indeed the $RF\!F$ agrees with the $F$-test and visual methods. However, at higher redshifts, what we are likely seeing is higher $RF\!F$ values which are due to galaxy formation processes such as residuals from merging or star formation \citep[e.g.][]{Conselice03, Conselice08,Bluck12}. These signatures would not as easily be seen in the other two methods.


\subsection{Number Density Evolution of Galaxy Components}\label{subsec.ndensity}

Investigating the number density (number of galaxies normalized by the co-moving volume) of different galaxy selections allows us to determine at which epoch different types of galaxies dominate, and how they evolve throughout the history of the Universe. We can also compare the rate at which the number density grows between different kind of systems. In this paper we explore how the number density of galaxies best fit with either $1$ and $2$ components evolve in terms of their number densities during the epoch $1 < z < 3$.

\begin{figure*}
    \centering
    \subfloat[][$1$-component disc-like galaxies.]{
      \includegraphics[width=0.33\linewidth]{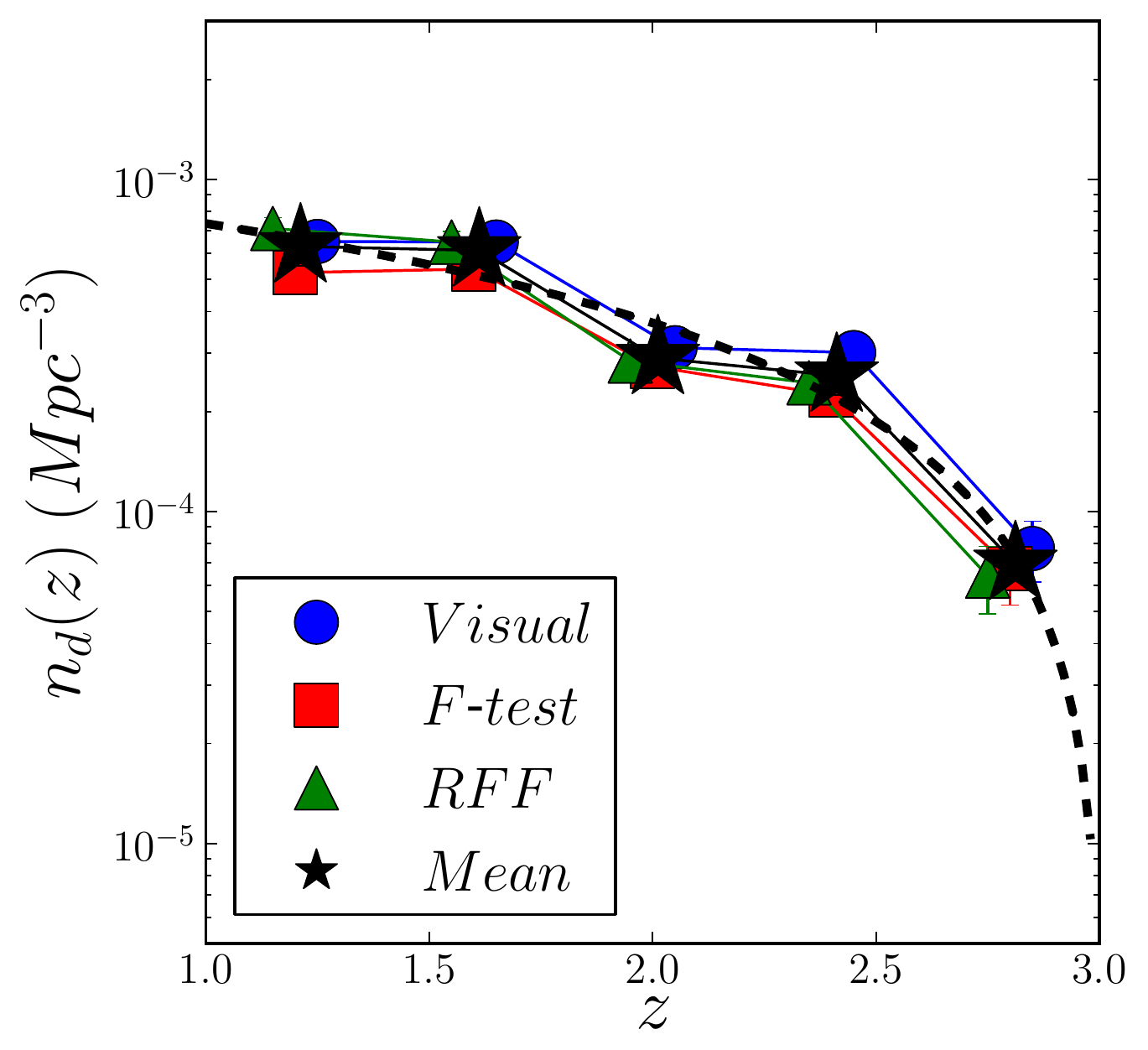}\label{Figure9a}\hspace{-1em}
    }
    \subfloat[][$1$-component spheroid-like galaxies.]{
      \includegraphics[width=0.33\linewidth]{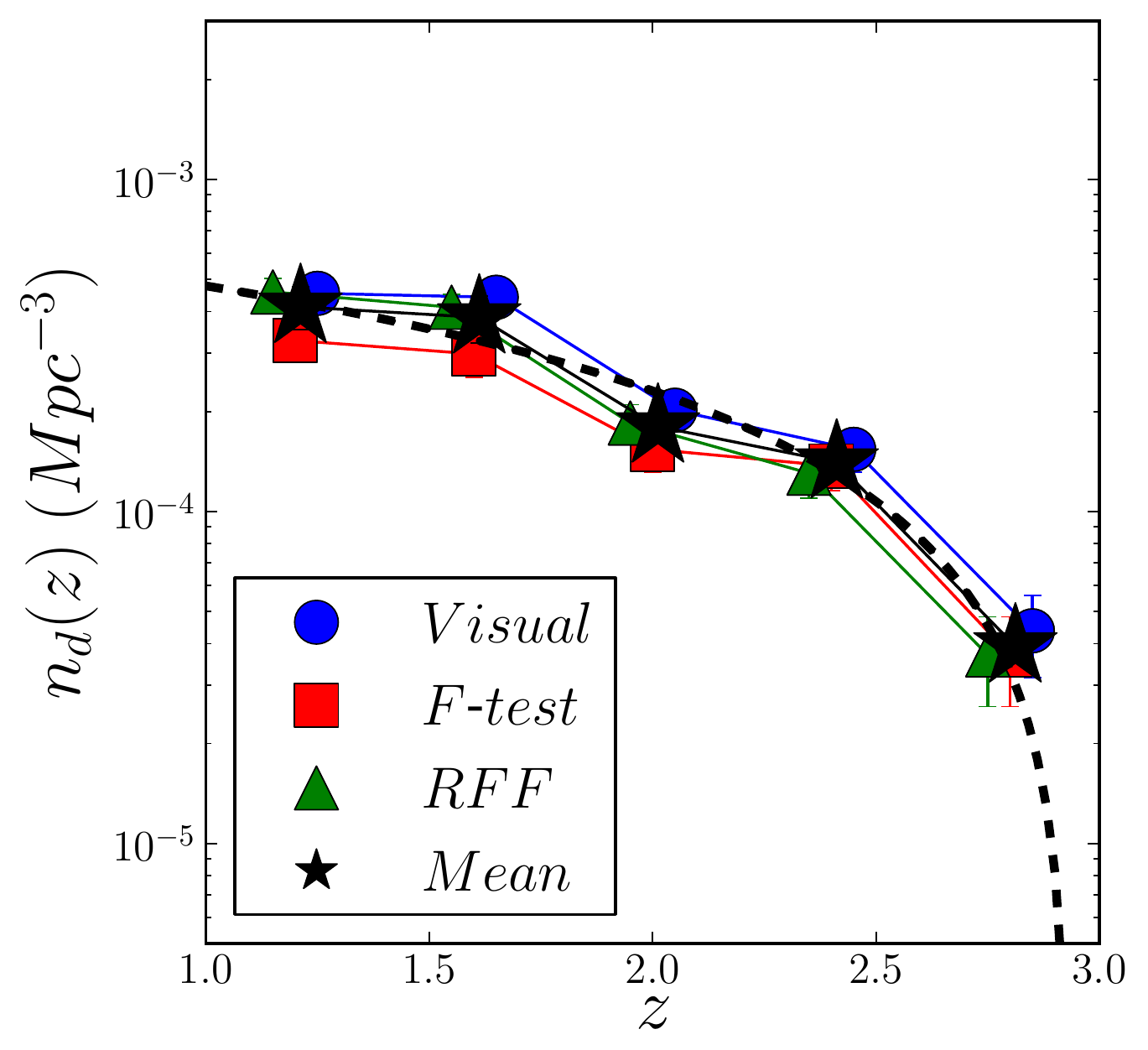}\label{Figure9b}\hspace{-1em}
    }
    \subfloat[][$2$-component galaxies.]{
      \includegraphics[width=0.33\linewidth]{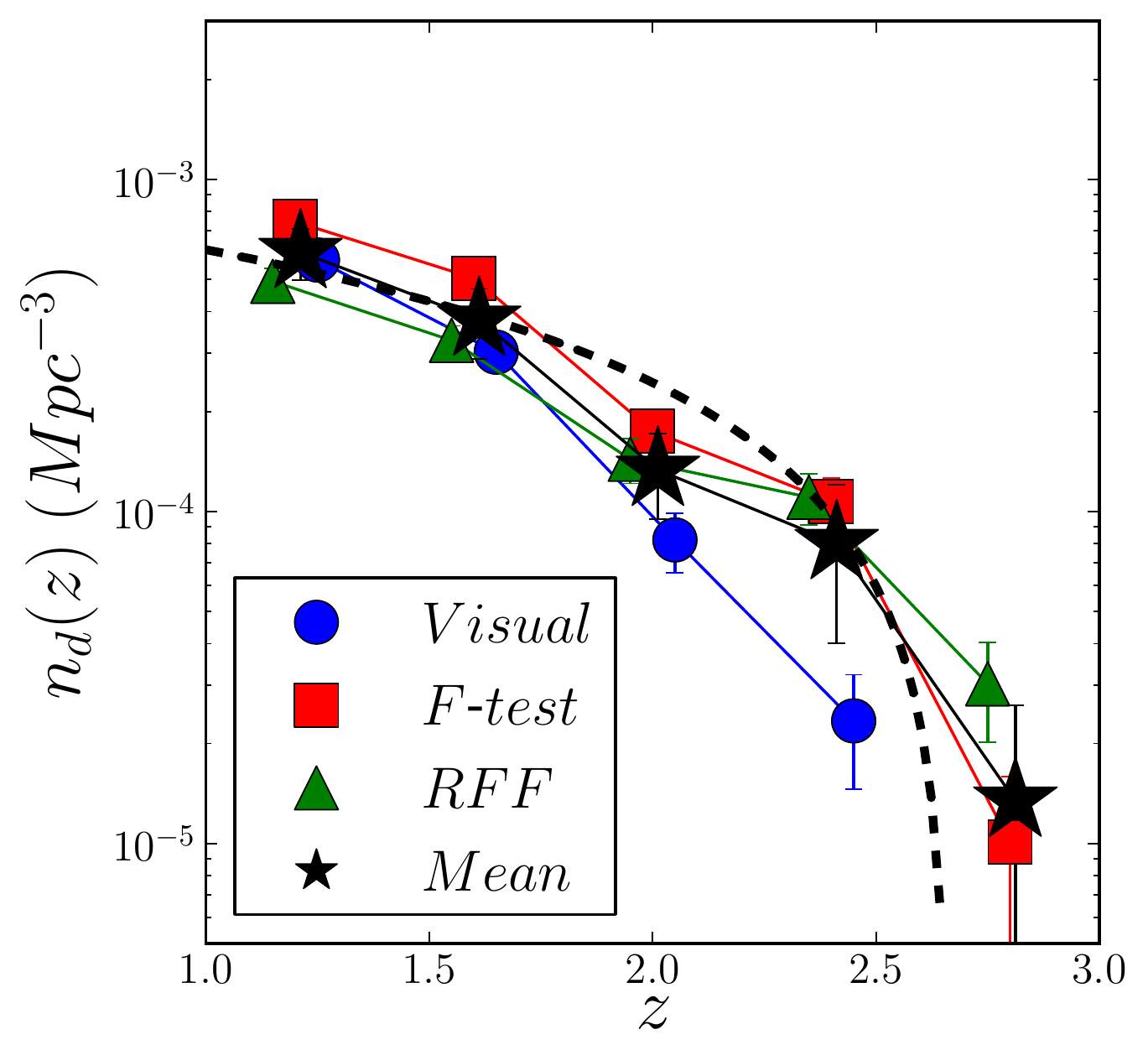}\label{Figure9c}}
  \caption[]{Number density $n_\mathrm{d}(z)$ as a function of redshift for each type of galaxy (disc-like \subref{Figure9a}, spheroid-like \subref{Figure9b} and $2$-component galaxies \subref{Figure9c}) for the three selection methods (visual classification, $F$-test and $RF\!F$ method). The points are coloured as in Figure \ref{Figure7}. The dashed black lines are the lines that best fit the mean values (note the log scale).}  \label{Figure9}
\end{figure*}

We have split our galaxies into four categories: $1$-component discs or disc-like galaxies ($n<2.5$), $1$-component spheroids or spheroid-like galaxies ($n>2.5$), two component galaxies, and peculiar galaxies. In Figure \ref{Figure9} we plot the number density of galaxies in five redshift bins for the different types of $1$ and $2$ component galaxies (number of galaxies normalized by the co-moving volume in $\mathrm{Mpc}^3$ corresponding to that bin). In black we plot the mean number density of the three methods.

We fit the mean number density $n_\mathrm{d}(z)$ of each type of galaxy using three different functions. Firstly we fit a linear function
\begin{equation}\label{eq_linear}
f_1(z)=a\cdot{}z+b,
\end{equation}

\noindent secondly we also fit a power-law function
\begin{equation}\label{eq_power}
f_2(z)=\gamma\cdot{}(1+z)^{\alpha},
\end{equation}

\noindent and lastly an exponential function
\begin{equation}\label{eq_expo}
f_3(z)=n_0\cdot{}e^{(-z/z_0)}.
\end{equation}

\begin{figure}
   \centering
   \includegraphics[width=1\linewidth]{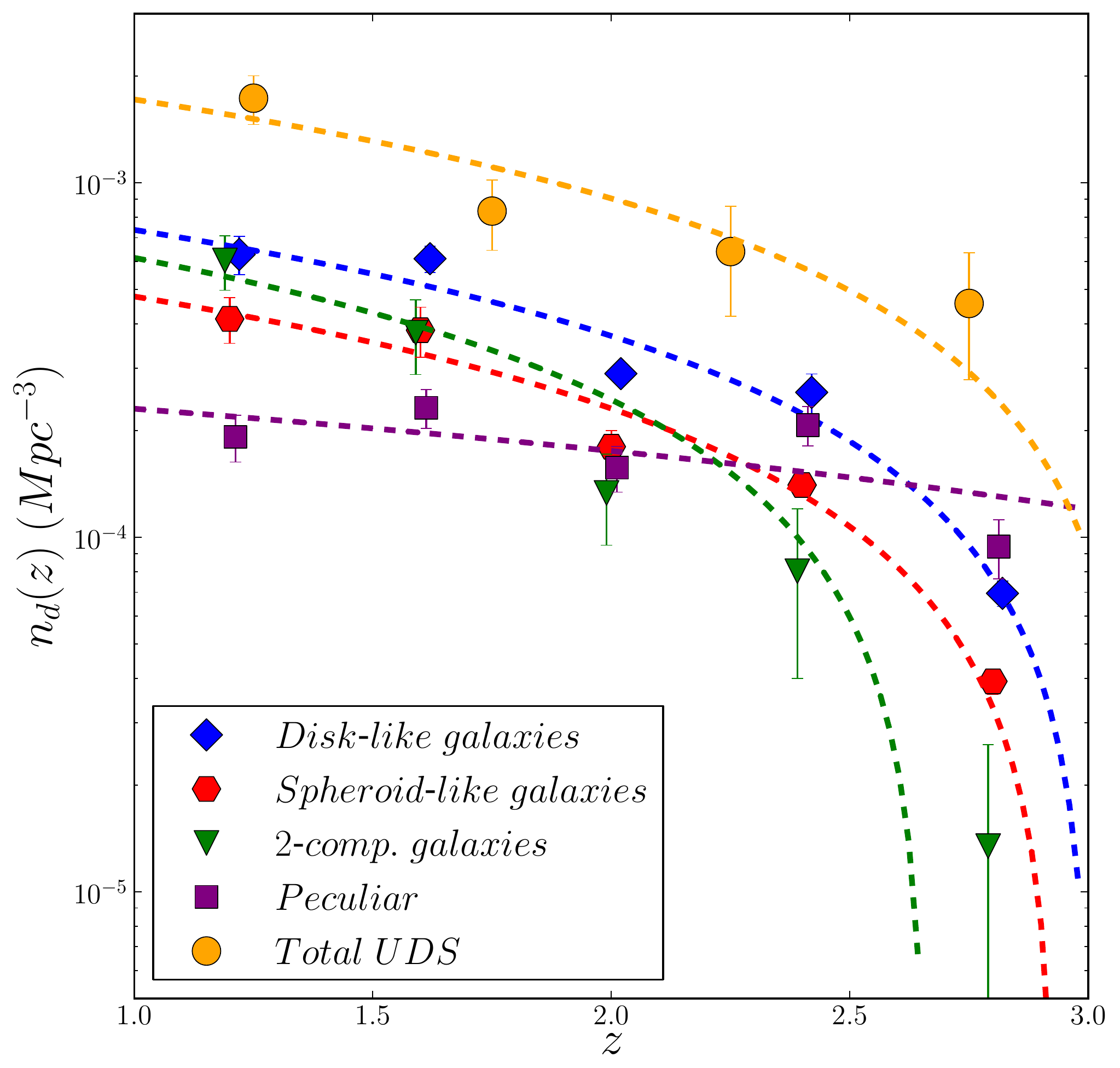}
   \caption{Total number density evolution of the different types of galaxies in our sample (disc-like galaxies: blue diamonds, spheroid-like galaxies: red hexagons, $2$-component galaxies: green triangles, and peculiar galaxies: purple squares). The yellow circles are the total number density of galaxies in UDS calculated by equation \eqref{eq_phi_tot}. The dashed lines are the straight lines that better fit the mean values (in log scale).}\label{Figure10}
\end{figure}

\noindent We show in Table \ref{Table1} the result for all three fits, noting that the function that best fits the data is the linear function. To compare the number density of the different types of galaxies, we plot in Figure \ref{Figure10} all four number densities, and the total number density of galaxies in the CANDELS-UDS \citep{Mortlock15}. The dashed lines show the best fit of the linear function, equation \eqref{eq_linear}. 

Already there are several trends which are visible on Figure \ref{Figure10}. The first is the rise of the $2$-component galaxies. While they make up a small fraction of the galaxy population at $z \sim 2.5$, they rise by a factor of $30$ in number density to become just as abundant as the $1$-component galaxies. This reveals that this epoch of $1 < z < 3$ is when two component galaxies form and dominate the abundances of massive galaxies.  

\begin{table*}
\caption{Parameters of the fittings to the mean number density of the different types of galaxies and to the total density of UDS.} 
\centering  
\begin{tabular}{c c c c c c} 
\hline                      
  Function & & Disc-like galaxies & Spheroid-like galaxies & $2$-comp. galaxies & Total UDS \\ [0.5ex] 
\hline \hline                 
Linear      & $a     $ & $-0.00038 \pm 0.00004$ & $-0.00025 \pm 0.00004$ & $-0.00036 \pm 0.00006$ & $-0.0008 \pm 0.0002$ \\ 
    & $b     $ & $0.0011   \pm 0.0001 $ & $0.0007   \pm 0.0001 $ & $0.0010   \pm 0.0001 $ & $0.0025  \pm 0.0004$ \\
\hline                      
Power-law    & $\alpha$ & $-2.5     \pm 0.3    $ & $-2.7     \pm 0.3    $ & $-4.2     \pm 0.8    $ & $-2.9    \pm 0.8   $ \\
    & $\gamma$ & $0.005    \pm 0.001  $ & $0.004    \pm 0.002  $ & $0.02     \pm 0.02   $ & $0.02    \pm 0.02  $ \\
\hline                      
Exponential & $z_0   $ & $1.1      \pm 0.1    $ & $1.0      \pm 0.1    $ & $0.6      \pm 0.1    $ & $1.0     \pm 0.3   $ \\
    & $n_0   $ & $0.0021   \pm 0.0005 $ & $0.0015   \pm 0.0004 $ & $0.004    \pm 0.004  $ & $0.006   \pm 0.004 $ \\ [1ex]      
\hline 
\end{tabular}\label{Table1} 
\end{table*}

We also compare in Figure \ref{Figure10} our number densities for the individual galaxy types and the total number density of all galaxies.  
The total number density of galaxies within the UDS \citep{Mortlock15} is calculated as:
\begin{equation}\label{eq_phi_tot}
 \phi_\mathrm{Total}=\int_{M_1}^{M_2}\phi(M;\phi^{\ast},M^{\ast},\alpha)\mathrm{d}M,
\end{equation}

\noindent where $M_1=10^{10} \textrm{M}_{\odot}$ and $M_2=10^{12} \textrm{M}_{\odot}$. The map $\phi(M)$ is the Schechter function \citep{Schechter76} given by
\begin{equation}\label{eq_phi_m}
 \phi(M;\phi^{\ast},M^{\ast},\alpha)=\phi^\ast \mathrm{ln}10 \left(10^{M-M^{\ast}}\right)^{1+\alpha}\exp\left(-10^{M-M^{\ast}}\right),
\end{equation}

\noindent where $\phi^\ast$ is the normalisation of the Schechter function, $M^{\ast}$ is the turn over mass in units of dex, $\alpha$ is the faint end slope of the Schechter function and the variable $M$ is the stellar mass in units of dex. The parameters $\phi^\ast$, $M^{\ast}$ and $\alpha$ depend on the redshift range, and the values we use to obtain $\phi_\mathrm{Total}$ in each redshift bin are calculated in \cite{Mortlock15}.

From Figure \ref{Figure10} and Table \ref{Table1} we see that the total number density of $1$-component disc-like galaxies evolves at a similar rate to that of the $2$-component galaxies, but its value is about $1.5$ times larger. The number density of spheroid-like galaxies increases slower than those of the other types of galaxies. We observe that for the whole redshift range the number density of disc-like galaxies is the highest of the four types of galaxies. The $z \gtrsim 2$ spheroid-like galaxies have a higher number density than the $2$-component galaxies, but at $z \lesssim 2$ the number density of $2$-component galaxies become greater than the spheroid-like galaxies. 

\begin{figure}
   \centering
    \includegraphics[width=1\linewidth]{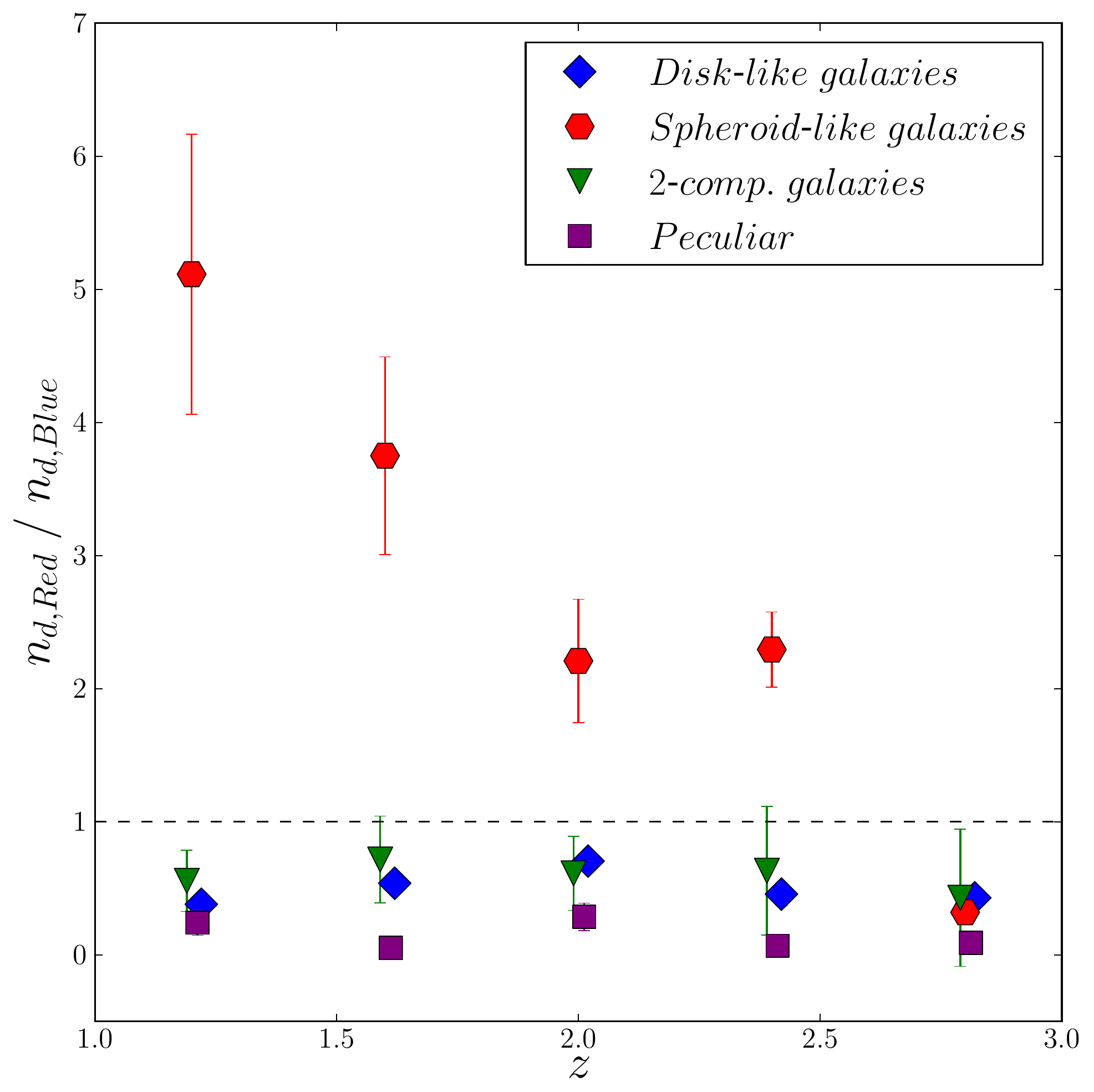}
    \caption{Ratio of the number density of red galaxies and blue galaxies for the different types of systems (disc-like galaxies: blue diamonds, spheroids-like galaxies: red hexagons, $2$-component galaxies: green triangles, and peculiar galaxies: purple squares).}\label{Figure11}
\end{figure}

In Figure \ref{Figure11} we compare blue and red galaxies according to the $UV\!J$ selection (eq. \ref{eq_uvj}) for the four different types of galaxies. Peculiar galaxies are mostly blue at all redshifts, with just a small number of them being red. For disc-like and two component galaxies, the fraction of blue galaxies is greater than half. Interestingly, at redshift $z\sim3$ most of the spheroid-like galaxies are blue, but as redshift decreases, the fraction of red spheroids rapidly increases while blue spheroids decrease. By redshift $z\sim1$, the vast majority ($85$ per cent) of spheroid-like galaxies are red.

\subsection{Total Stellar Mass Density}\label{subsec.density}

By investigating the stellar mass density, we can determine which types of galaxies dominates the stellar mass in the Universe at which epoch. This is closely related to the number density evolution, but here we are investigating essentially whether the galaxies in each selection are more massive on average than in the other selections.  

\begin{figure*}\centering
    \subfloat[][$1$-component disc-like galaxies.]{
      \includegraphics[width=0.33\linewidth]{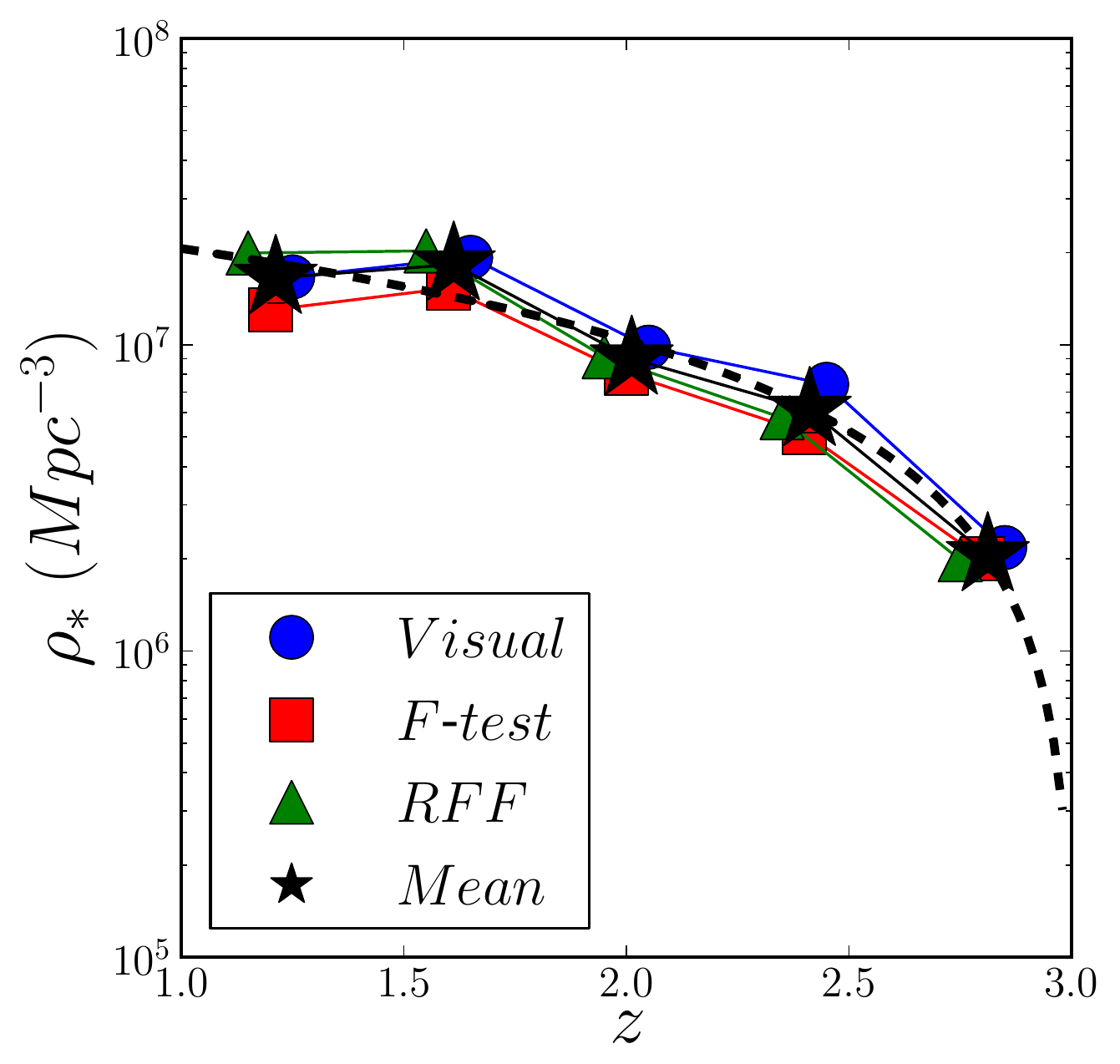}\label{Figure12a}\hspace{-1em}
    }
    \subfloat[][$1$-component spheroid-like galaxies.]{
      \includegraphics[width=0.33\linewidth]{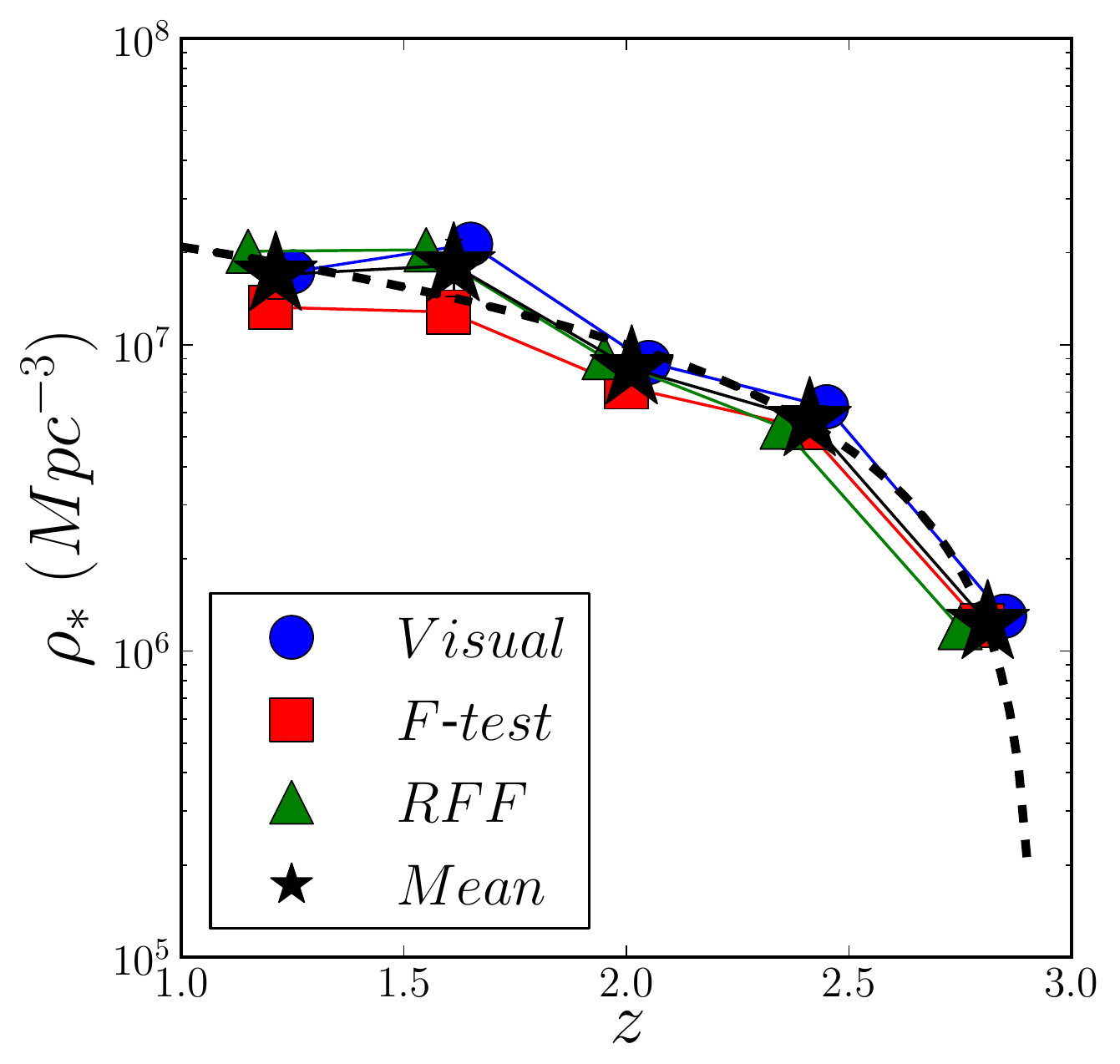}\label{Figure12b}\hspace{-1em}
    }
    \subfloat[][$2$-component galaxies.]{
      \includegraphics[width=0.33\linewidth]{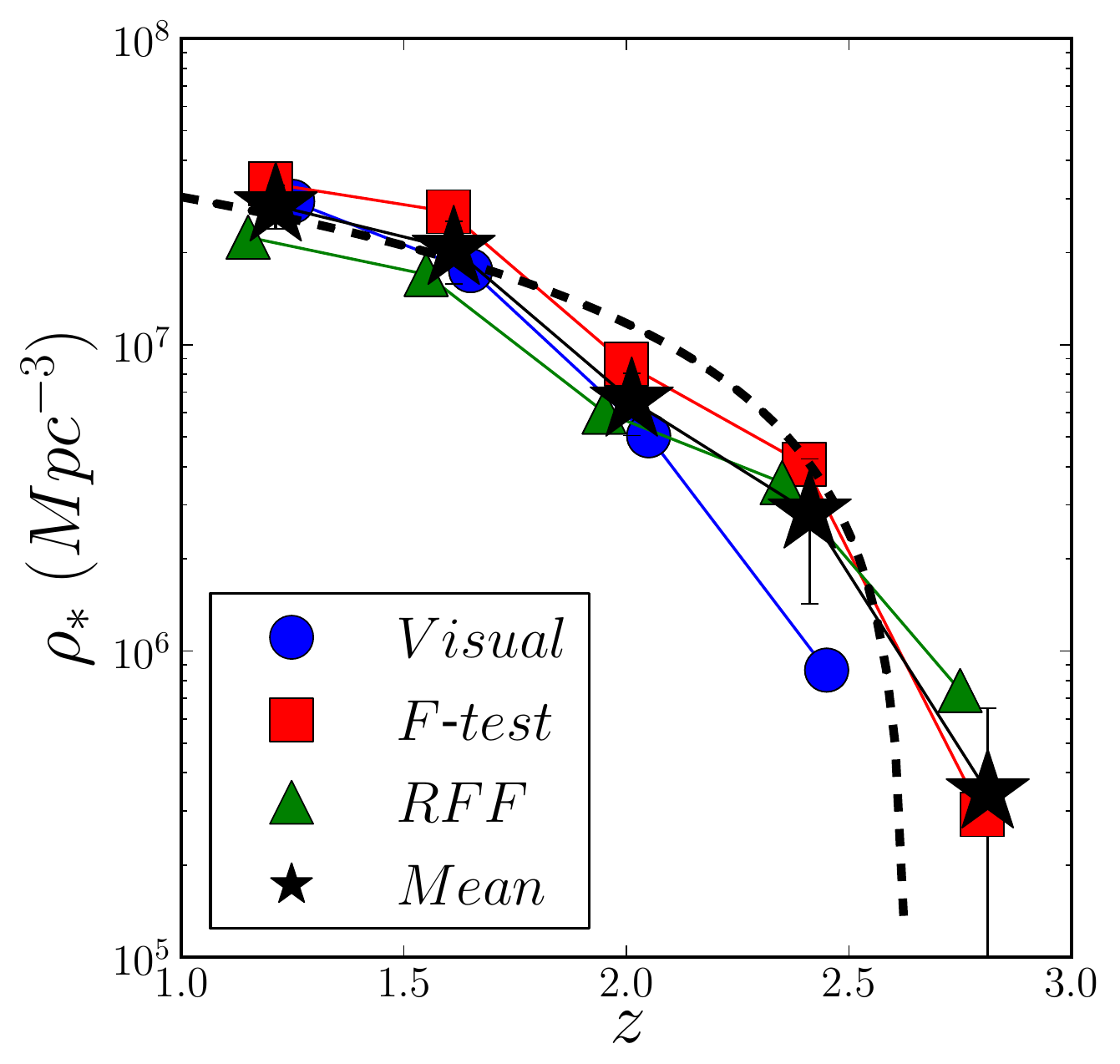}\label{Figure12c}
    }
  \caption[]{Mass density as a function of redshift for each type of galaxy in our sample (disc-like \subref{Figure12a}, spheroid-like \subref{Figure12b} and $2$-component galaxies \subref{Figure12c}) for the three selection methods (visual classification, $F$-test and $RF\!F$ method). The points are coloured as in Figure \ref{Figure7}. The dashed black line are the straight lines that best fit the mean values (black stars) (in log scale).}\label{Figure12}
\end{figure*}

In Figure \ref{Figure12} we plot the mass density $\rho_{\ast}$ of the different types of galaxies as a function of redshift, and in black we plot the mean mass density of the three methods. 

\begin{figure}\centering
        \includegraphics[width=1\linewidth]{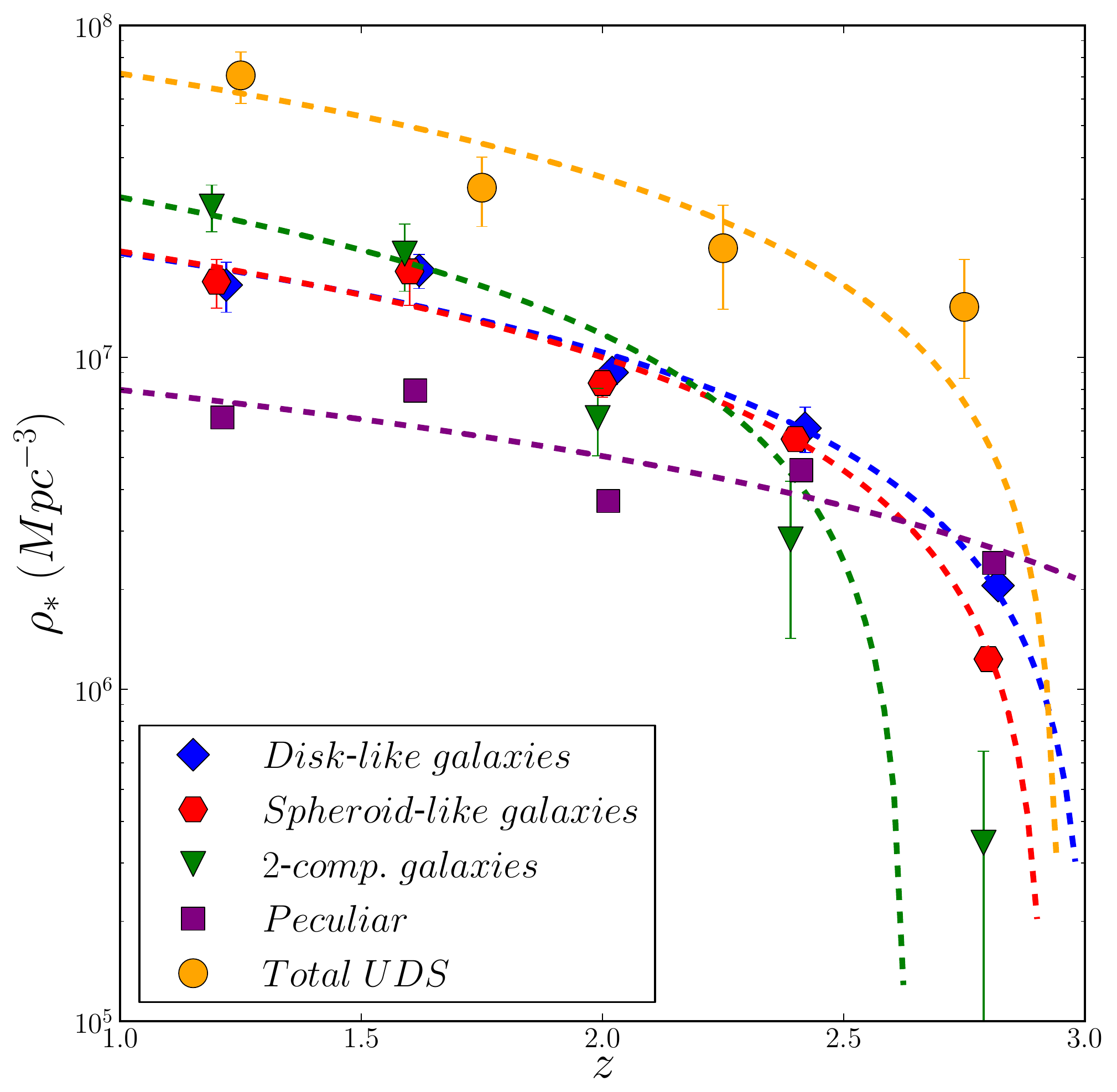}
    \caption{Mean mass density evolution of the different types of galaxies (disc-like galaxies: blue diamonds, spheroids-like galaxies: red hexagons, $2$-component galaxies: green triangles, and peculiar galaxies: purple squares). The yellow circles are the total mass density of galaxies in UDS calculated by equation \eqref{eq_phi_mass}.}\label{Figure13}
\end{figure}

We fit the mean mass density of each type of galaxy using the same functions as in Section \ref{subsec.ndensity} (linear function, power-law and exponential function). We show the result of the fits in Table \ref{Table2} for all three fits. In Figure \ref{Figure13} we plot the mass density for all four classes of galaxies using the average selection, as well as the total mass density from the UDS. The dashed lines show the best fit of the linear function, equation \eqref{eq_linear}.

\begin{table*}
\caption{Parameters of the fittings to the mean mass density of the different types of galaxies and to the total density of UDS} 
\centering  
\begin{tabular}{c c c c c c} 
\hline                      
  Function & & Disc-like galaxies & Spheroid-like galaxies & $2$-comp. galaxies & Total UDS \\ [0.5ex] 
\hline \hline                 
Linear      & $a\ (10^7)    $ & $-1.0 \pm 0.2 $ & $-1.1 \pm 0.02$ & $-1.8 \pm 0.03$ & $-3.6 \pm 0.8$ \\ 
    & $b\ (10^8)    $ & $0.31  \pm 0.04 $ & $0.32   \pm 0.05 $ & $0.48   \pm 0.06 $ & $1.1  \pm 0.2$ \\
\hline                      
Power-law    & $\alpha$ & $-2.5     \pm 0.3    $ & $-2.7     \pm 0.3    $ & $-4.2     \pm 0.8    $ & $-2.9    \pm 0.8   $ \\
    & $\gamma\ (10^8)$ & $1.2    \pm 0.5  $ & $1.6    \pm 0.8  $ & $8.1     \pm 0.4   $ & $11    \pm 3  $ \\
\hline                      
Exponential & $z_0   $ & $1.1      \pm 0.2    $ & $1.0      \pm 0.1    $ & $0.6      \pm 0.1    $ & $0.8     \pm 0.3   $ \\
    & $n_0\ (10^8)  $ & $0.5   \pm 0.2 $ & $0.7   \pm 0.2 $ & $ 2    \pm 1  $ & $3   \pm 1 $ \\ [1ex]      
\hline 
\end{tabular}\label{Table2} 
\end{table*}

The total mass density of galaxies in the UDS is calculated as
\begin{equation}\label{eq_phi_mass}
 \rho_{\ast,\mathrm{Total}}=\int_{M_1}^{M_2} M_{*}\cdot \phi(M;\phi^{\ast},M^{\ast},\alpha)\mathrm{d}M,
\end{equation}

\noindent where $M_1=10^{10} \textrm{M}_{\odot}$, $M_2=10^{12} \textrm{M}_{\odot}$, and $\phi(M)$ is the Schechter function defined in equation \eqref{eq_phi_m}.

The mass density of the $1$-component galaxies evolves at similar rates independently of the S\'ersic index selection (i.e.\ being discs or spheroids). The mass density of $2$-component galaxies have the highest increase over the whole redshift range, and its contribution to the total mass density is smaller than that of $1$-component galaxies at $z=1.75-3$, but for $z=1-1.75$ the mass density of this type of galaxy becomes dominant. The $2$-component galaxies in fact dominate the mass density of massive galaxies at these lower redshifts.

Overall, we find that the mass density for $2$-component galaxies increases by a factor of $\sim 100$, which is roughly a factor of $3$ higher than for its number density increase. Therefore, we see a larger effect in the integrated mass density for our galaxies than the increase in the number density. This implies that the galaxies which are driving this increase are more massive at the lower end of the redshift range around $z \sim 1$ than at higher redshifts, relative to the $1$-component galaxies. This implies that the most massive galaxies preferentially become the two component systems at lower redshifts, while the $1$-component systems are relatively lower mass.

\subsection{The Size Evolution of Components}\label{subsec.size}

\begin{figure*}\centering
    \subfloat[][$1$-component disc-like galaxies]{
      \includegraphics[width=0.45\linewidth]{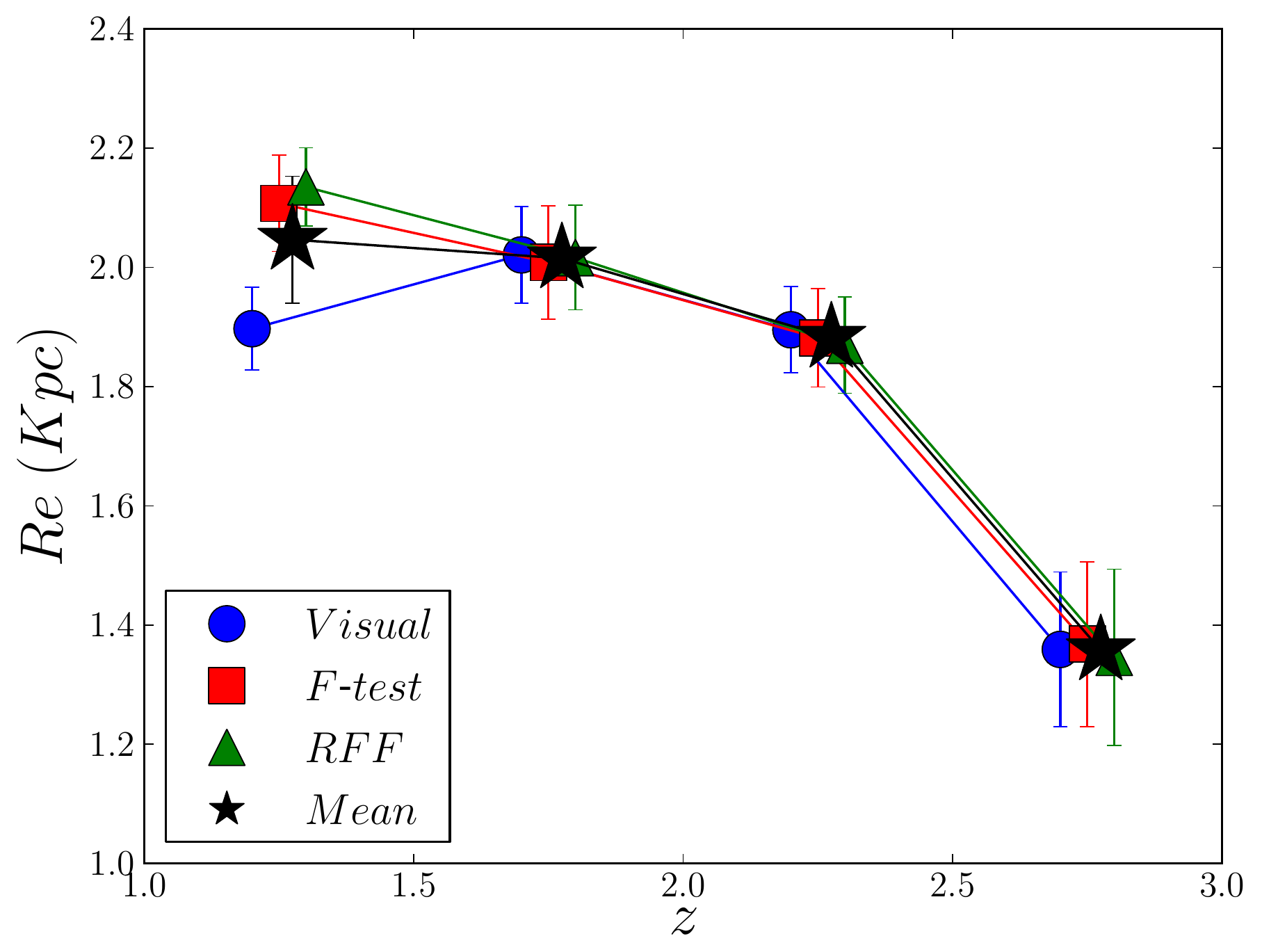}\label{Figure14a}
    }
    \subfloat[][$1$-component spheroid-like galaxies]{
      \includegraphics[width=0.45\linewidth]{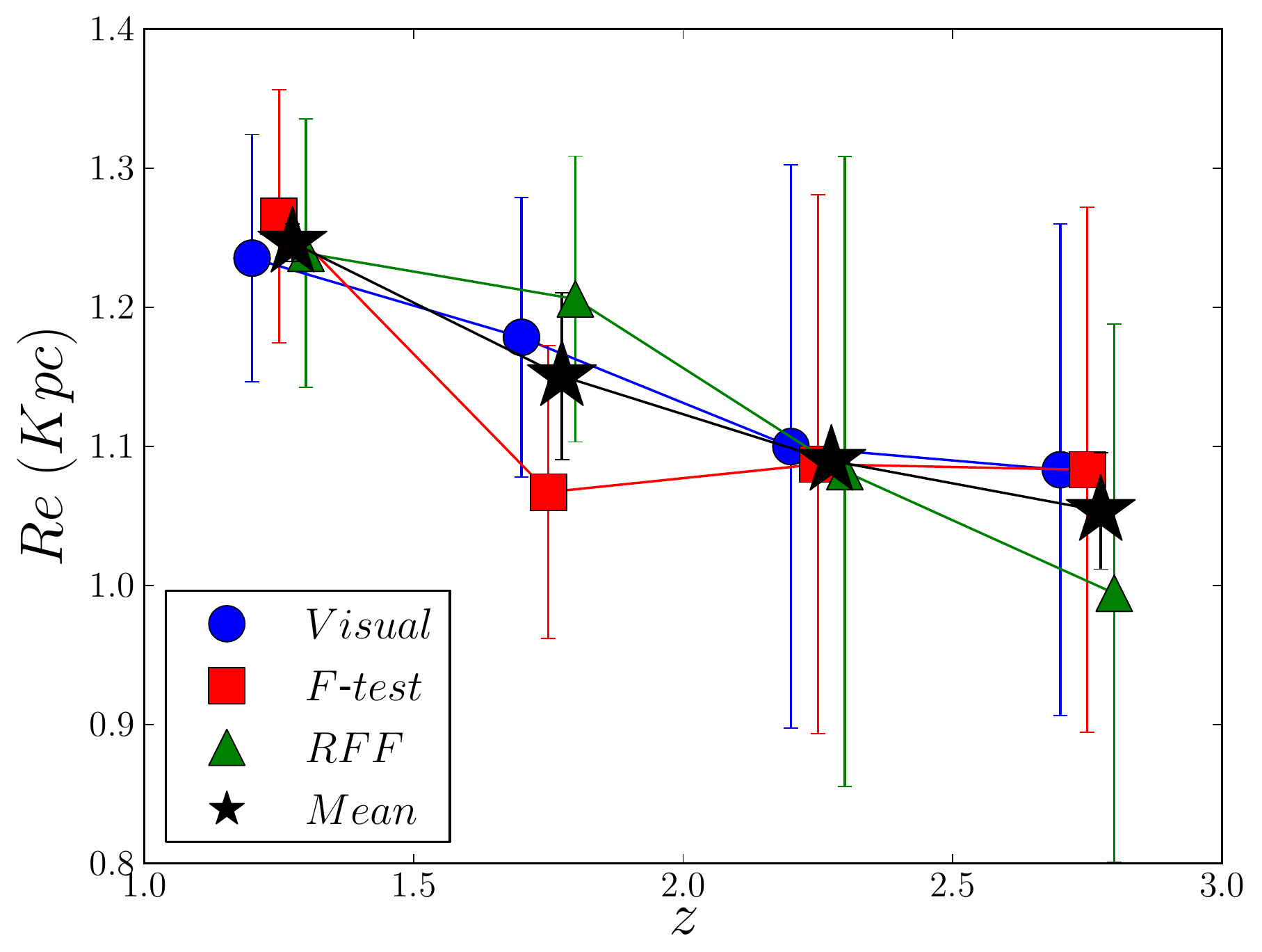}\label{Figure14b}
    }\\
    \subfloat[][Discs of $2$-component galaxies]{
      \includegraphics[width=0.45\linewidth]{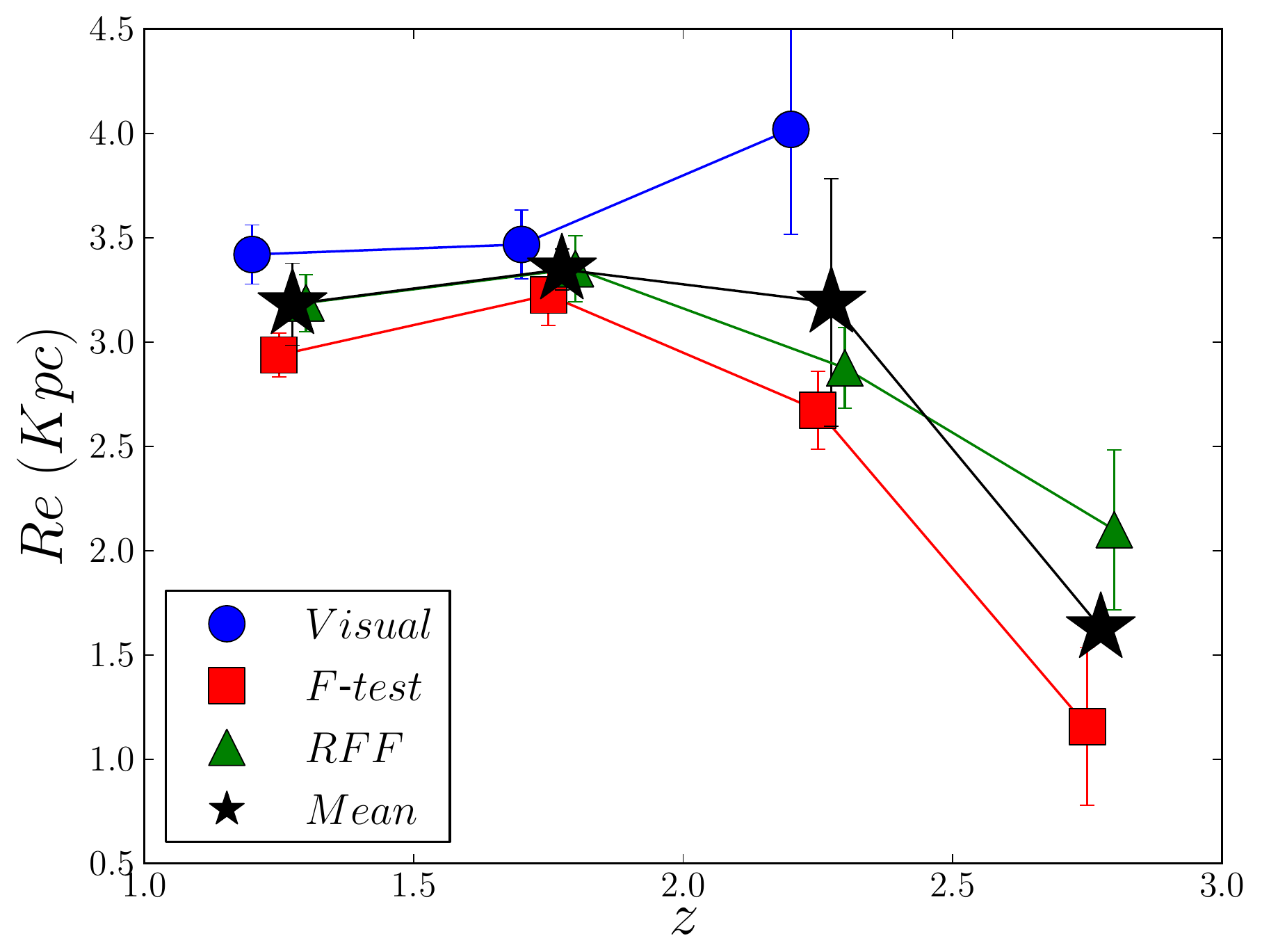}\label{Figure14c}
    }
     \subfloat[][Bulges of $2$-component galaxies]{
      \includegraphics[width=0.45\linewidth]{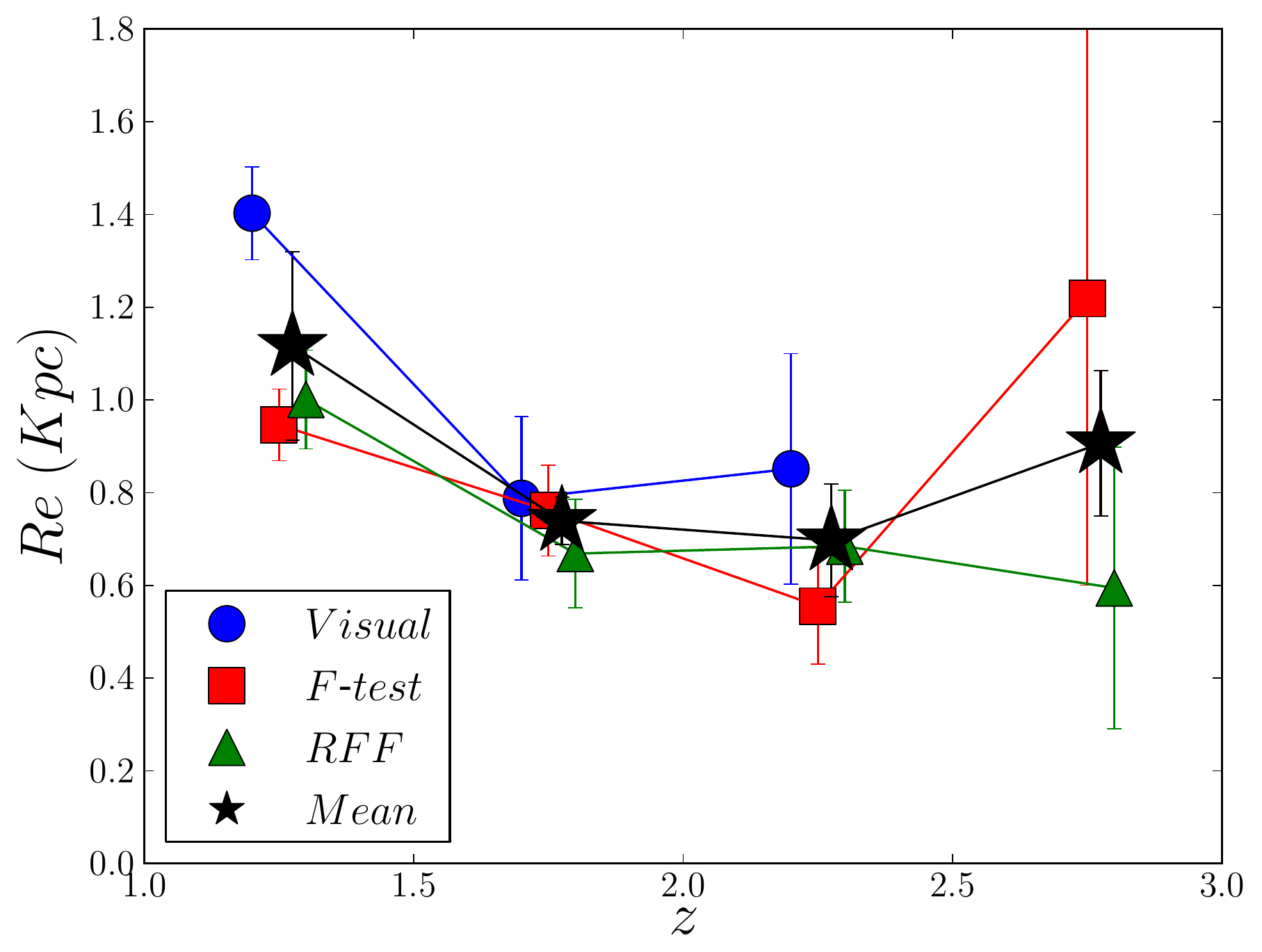}\label{Figure14d}
    }
  \caption[]{Observed size evolution with redshift. Plotted are the median sizes of galaxies in our sample: disc-like \subref{Figure14a}, spheroid-like \subref{Figure14b}, discs of $2$-component galaxies \subref{Figure14c} and bulges of $2$-component galaxies \subref{Figure14d} for the three different selection methods (visual classification, $F$-test and $RF\!F$ method). The points are coloured as in Figure \ref{Figure7}. In the $y$-axis we plot the median sizes of the galaxies. Notice that the visual classification does not select any $2$-component galaxies at the highest redshift bin, and therefore, in the bottom plots there is not a blue point at high redshift. The error bars on the points represent the standard deviation}  \label{Figure14}
\end{figure*}

We explore the evolution in size of our galaxy sample, and their components, to determine if the inner and outer components grow together or not. As the effective radius $R_e$ calculated from \textsc{galfit} corresponds to the major axis of an ellipse containing half of the light, in order to compare with other results, we have calculated a circularized radius $R_{e,\mathrm{circ}}=R_e\cdot{}\sqrt{b/a}$, where $b/a$ is the axis ratio. In Figure \ref{Figure14} we plot the median circularized effective radius of the $1$-component galaxies, as well as the discs and bulges of $2$-component galaxies, as a function of redshift (for the three different methods as well as the average of the three methods). 

First, we observe that there is a trend for the $1$-component galaxies: on average they grow in size at lower redshifts, although the evolution appears stronger in disc-like galaxies than in spheroid-like galaxies.Between $1 < z < 3$, the disc-like $1$-component galaxies grow on average from $1.3$ kpc to $2.1$ kpc i.e. an increase of $60$ per cent. On the other hand, the $1$-component spheroid-like galaxies grow on average from $1.05$ kpc to $1.24$ kpc i.e. an increase of only $18$ per cent. Thus during this epoch the disc-like systems dominate the growth. Note that our simulations of $z=1$ galaxies to $z=2.75$ show if anything this increase in size may be more dramatic (see Appendix \ref{appendix}). These results also show that the $1$-component disc-like galaxies are larger on average than $1$-component spheroid-like galaxies.

We find a very interesting trend when we examine the evolution of the inner `bulge' component and outer `disc' component of $2$-component galaxies. We first note that the discs in $2$-component galaxies are larger in size than disc-like galaxies at all redshifts. The discs of the $2$-component galaxies increases in size on average from $1.6$ kpc to $3.2$ kpc i.e. an increase of a factor of $2$. On the other hand, we find that the bulge components of the $2$-component systems increases very slightly from $0.9$ kpc to $1.1$ kpc on average.  

\begin{figure}\centering
        \includegraphics[width=1\linewidth]{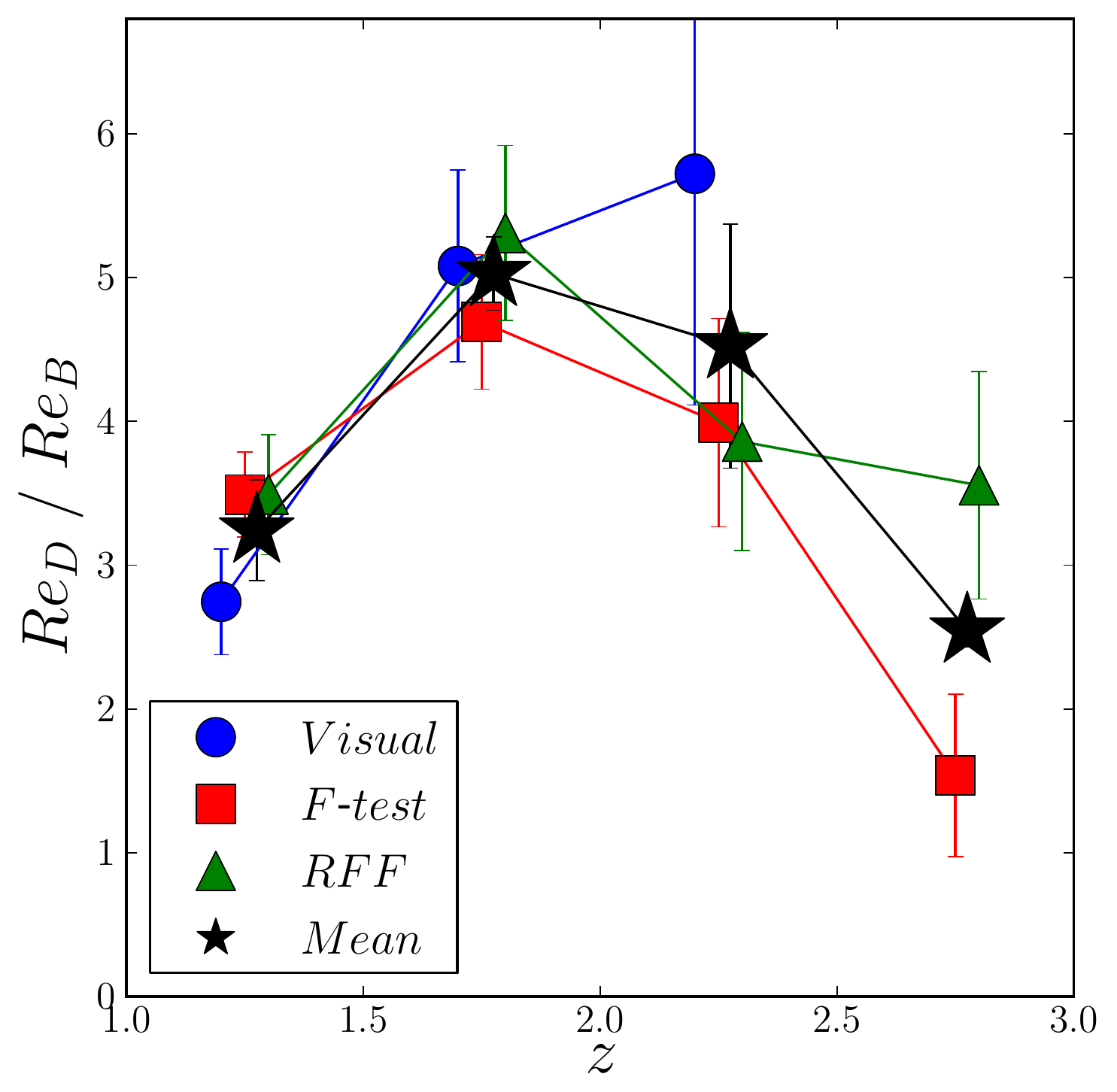}
    \caption{Evolution of the ratio between the sizes of discs and bulges of $2$-component galaxies. Notice that the visual classification does not select any $2$-component galaxies at the highest redshift bin, and therefore, there is not a blue point at high redshift.}  \label{Figure15}
\end{figure}

\begin{figure}\centering
        \includegraphics[width=1\linewidth]{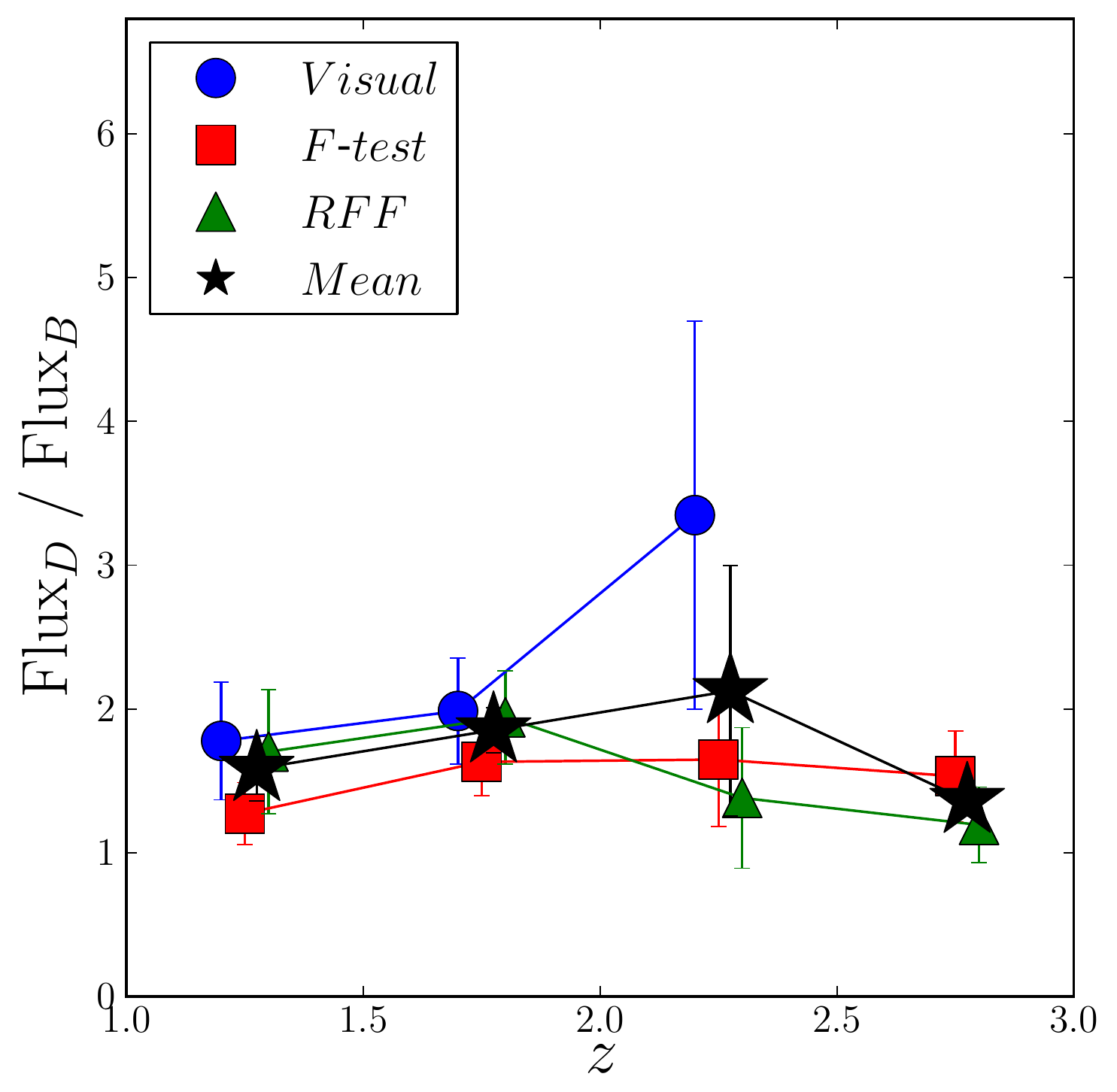}
    \caption{Evolution of the ratio between the flux of discs ($\mathrm{Flux}_D$) and bulges ($\mathrm{Flux}_B$) of $2$-component galaxies. Notice that the visual classification does not select any $2$-component galaxies at the highest redshift bin, and therefore, there is not a blue point at the highest redshift bin.}  \label{Figure16}
\end{figure}

What we are likely seeing therefore is an inside-out formation of two component galaxies such that the inner component is in place before the outer component. This is in agreement with previous studies such as \cite{vanDokkum10} and \cite{Carrasco10} which study the evolution of massive galaxies since $z \sim 2$ at fixed aperture. Likewise, we find the inner component of our sample does not grow as significantly as the outer component does. This is seen more clearly in Figure \ref{Figure15}, where we plot the ratio between the mean effective radius of the inner and outer component as a function of redshift. Discs seem to grow earlier, and they increase at a higher rate relative to the growth of the bulges until $z\sim1.5$, when bulges appear to rapidly grow in size. We only have, however, one point showing this, which will need to be further confirmed. Nevertheless, in Figure \ref{Figure16} we observe that the ratio between the flux of these two components remains fairly constant with redshift.

Overall, what we find is that the stellar masses, as traced by the light distribution (assuming a similar mass to light ratio), of these components are increasing at a similar rate, while at the same time the sizes of the outer components are growing faster than their inner components. There are several ways to interpret this. One possibility to explain these observations is through how the additional mass is distributed in the two components. For the outer components, this new mass is added to the outer parts, increasing the size of the disc component, but within the bulges the additional mass is still centrally concentrated.  This is one way in which the mass ratio can remain constant while the size ratio increases with time. We will investigate this in more detail in a future paper (Margalef-Bentabol et al. 2016, in preparation).

It may be argued that the circularized effective radius $R_{e,circ}$ is not the most appropriate quantity to measure the size of discs, as for flat disky objects, $R_{e,circ}$ will depend on the inclination. In which case, the size could be better quantify by the effective radius $R_e$ (major axis of an ellipse containing half of the light). However, we find that using $R_e$ instead of $R_{e,circ}$ for our disky objects does not change our results. In the case of $1$-component discs we observe the same growth of $60$ per cent over the redshift range, but the sizes are about $\sim 1.4$ times larger than the circularized radius. For discs of $2$-component galaxies, we observe an increase of a factor of $1.7$ over the redshift range, which is still much stronger than the growth of the bulges at the same redshifts. In this case, the sizes are on average about $1.6$ times greater than using the circularized values.

\begin{figure}\centering
        \includegraphics[width=1\linewidth]{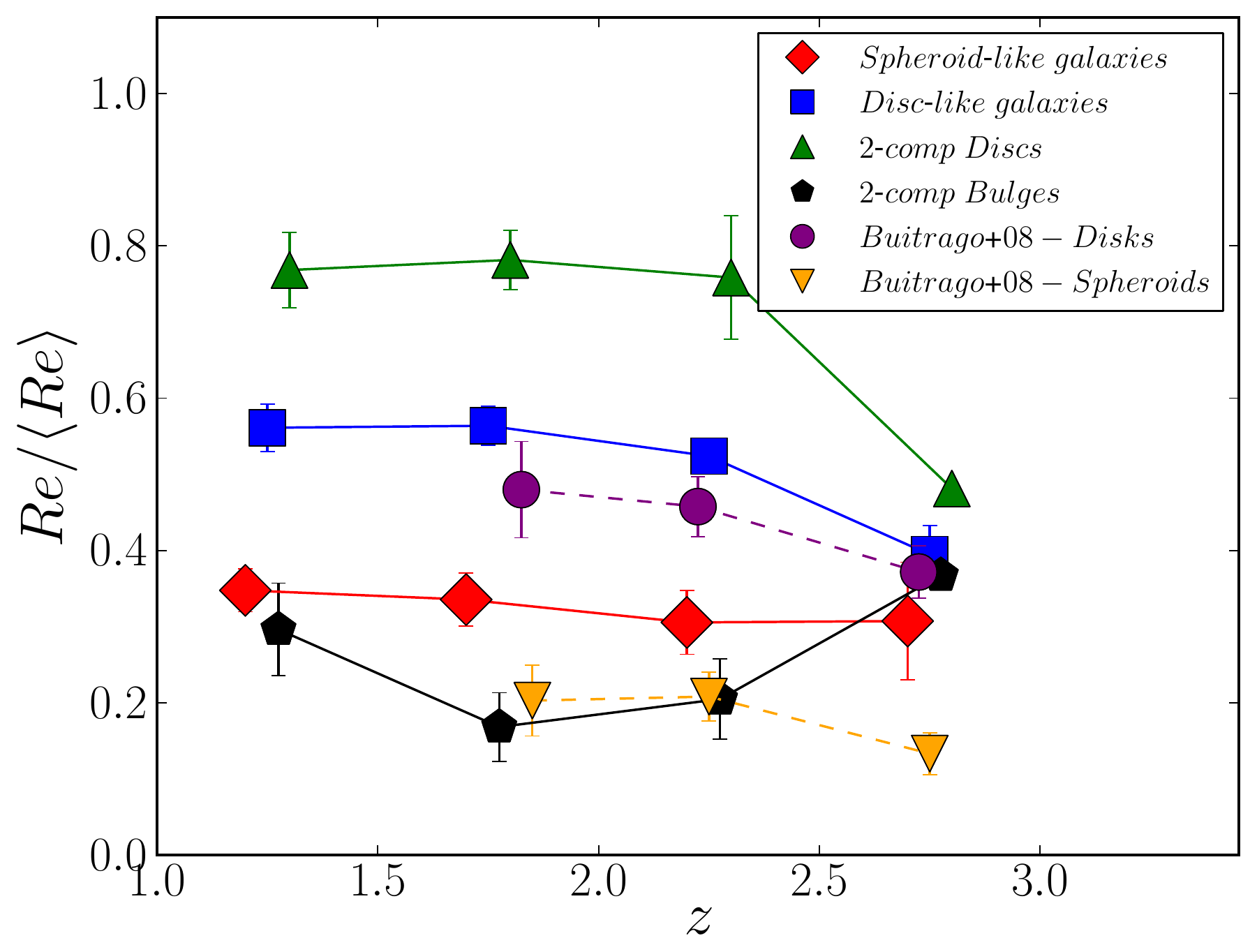}
    \caption{Evolution of the ratio between the sizes of the galaxy components (disc-like galaxies, spheroid-like galaxies, disc of $2$-component galaxies and bulges of $2$-component galaxies), and the sizes of galaxies from the nearby Universe at the same mass $\langle R_e \rangle$.}  \label{Figure17}
\end{figure}

In Figure \ref{Figure17} we plot the ratio between the sizes of the galaxy components (disc-like galaxies, spheroid-like galaxies, discs of $2$-component galaxies and bulges of $2$-component galaxies), and the sizes of galaxies within the nearby Universe at the same mass. These nearby galaxy sizes were obtained by the size--mass relation from GAMA results \citep{Lange15}. We use the early-type relation to compare with spheroid-like galaxies, and the bulges of $2$-component galaxies, and the late-type relation for comparison with disc-like galaxies and discs in $2$-component galaxies.

\begin{itemize}
   \item Early type $\ R_e(\mathrm{Kpc})=c\left(\dfrac{M_{\ast}}{\textrm{M}_{\odot}}\right)^{\!\eta}$\\[1ex]
   \item Late type$\ R_e(\mathrm{Kpc})=d\left(\dfrac{M_{\ast}}{\textrm{M}_{\odot}}\right)^{\!\lambda}$
\end{itemize}

\noindent where $c=36.04\cdot 10^{-5} \mathrm{\ Kpc}$, $\eta=0.38$, $d=25.26\cdot 10^{-3}\mathrm{\ Kpc}$ and $\lambda=0.21$.

We have over-plotted the data from \citet{Buitrago08} (galaxies in the same redshift range as our sample but with higher masses at $M_{\ast}>10^{11}\textrm{M}_{\odot}$) calculating the ratios in the same manner. $2$-component discs seem to remain the same over redshift while compared with their low redshift counterparts, there may be a slight growth before $z=2.5$, but they remain constant at lower redshifts. Disc-like galaxies are smaller at high redshift compared with disc-like galaxies of the same mass at the present time, in agreement with \citet{Buitrago08}. Spheroid-like galaxies also seem to have grown in size over cosmic time, but not as much as disc-like galaxies. The bulges of disc-like galaxies seem to have grown with time, particularly after $z=2$. We observe that size of spheroid-like galaxies and bulge of $2$-component galaxies is less than $40$ per cent of that of early-type galaxies in the local universe. This implies a significant growth from $z=1$ to present day for spheroid dominated galaxies. Disc-like galaxies undergo a less dramatic growth over the same epoch. This results are in agreement with previous studies, such as \cite{vanderWel14}, where they observe a growth by a factor of $\sim 2$ since $z=1$ for early-type galaxies of similar masses, and only a moderate growth is seen for late-type galaxies.

\section{Summary and Conclusions}\label{sec.conclusions}1
We have carried out a detailed investigation of the light decomposition of galaxies within the CANDELS UDS field using $1495$ massive galaxies with $M_{\ast}>10^{11}\textrm{M}_{\odot}$ at $1 < z < 3$. In this paper we set out a new methodology for deciding whether a galaxy should be considered a single or two component system using the observed $H$-band imaging, and then we examined the evolution of the individual component's sizes and mass as a function of redshift and stellar mass.

We have used three different methods to determine if a galaxy is better fit by a $1$ or $2$ component model for their surface brightness distributions. The three methods are: visual morphology, the $F$-test, and by examining the residual flux fraction ($RF\!F$). We find that all three methods are largely in agreement with each other, and we calculate the mean value of the various parameters we study derived within the $3$ methods.  

One major result is that the fraction of $2$-component galaxies increases with higher stellar mass for all three methods. In fact, on average there are $\sim2$ times more galaxies selected as $2$-component for the most massive galaxies than in the lowest mass bin. We also find an evolution with redshift, such that the fraction of $2$-component systems decreases from about $35$ per cent to $8$ per cent from $z=1$ to $z=3$. However, this decrease with redshift might be partially due to the degraded data at higher redshifts.

We find that disc-like galaxies have the highest relative number density at all redshifts, while spheroid-like galaxies have the lowest increase in that epoch, and by $z\sim2$, $2$-component galaxies exceed both of them in number density. The contribution to the total density due to $2$-component galaxies becomes dominant at $z<1.75$ in spite of their being the lowest at $z>1.75$. At redshift $z\sim3$ the majority of spheroid-like galaxies are blue, but as redshift decreases, the number of red spheroid-like galaxies rapidly increases. The other populations of galaxies remain mostly blue for all redshifts.

We also find that for $2$-component galaxies there is an increase in the sizes of their outer components, or `discs' by about a factor of three from $z = 3$ to $z = 1.5$, while the inner components or `bulges' stays roughly the same size. This suggests that these systems are growing from the inside out, whilst the bulges are in place early in the history of these galaxies. This is also seen to a lesser degree in the growth of single `disc-like' galaxies vs. `spheroid-like' galaxies over the same epoch.

We also carry out image simulations to determine how reliable our results are. We do this by reproducing how our galaxies would look at higher redshifts (for more details see Appendix \ref{appendix}), we conclude that the decrease in size we observe within discs and bulges for $2$-component galaxies in Figure \ref{Figure14} must be real, as the simulations show that we can accurately recover the size of the bulges in the simulated redshifted galaxies, including the smallest ones. Discs are also well recovered except for the smallest ones ($Re_\mathrm{D} < 2\ Kpc$), where we recover larger discs than the originals. However, this may hint that we are indeed observing small discs at the highest redshift bin. The simulations where the $F$-test is used to find $2$-component galaxies may induce us to think that the decreasing in the fraction of $2$-component galaxies is due to simply redshift. However, the visual classification, which is a reliable tool to distinguish patterns, only accounts for half of the decreasing, suggesting that the observed reduction is real (see Figure \ref{Figure22}). It thus seems more reasonable to think that the $F$-test is not as reliable using the same threshold $P_0$ for all redshifts. This might be due to the fact that for high redshift, galaxies features are blended with the noise.

\section*{Acknowledgements}

We thank the CANDELS team for their support in making this paper possible, as well as STFC and the University of Nottingham for financial support. A.M. acknowledges funding from the STFC and a European Research Council Consolidator Grant (P.I. R. McLure).

\bibliographystyle{mnras}
\bibliography{references}
\clearpage

\begin{appendices}
\section{Appendix} \label{appendix}
\subsection{Details of the Simulations}

Many of the results presented in the main part of this paper originate from some mixture of redshift effects as well as real changes to the $1$- and $2$-dimensional structure of massive galaxies. It is important to separate these two effects to determine the real evolution of galaxies. As such, we artificially simulate redshifted galaxies using the Ferengi code from \cite{Barden08}. From our whole galaxy sample, we take a subsample of $357$ objects with redshifts $z<1.4$ that we denote $L\!RO$ (low-redshift original galaxies). For those $L\!RO$ galaxies, we create new images of how we would observe the same galaxies if they were at redshift $z=2.75$ in the same CANDELS field, to determine the maximal effects of redshift. We denote this new sample $H\!RS$ (high-redshift simulated galaxies). This procedure modifies the angular size and the surface brightness (dimming) due to cosmological effects, but also takes into account the brightness increase of high-redshift objects. We now compare the structures of the $H\!RS$ galaxies with their corresponding $L\!RO$ ones, as well as with the actual galaxies we examine in the main paper in the redshift range of $2.5<z<3$, that we denote $H\!RO$ (high-redshift original galaxies). 

\begin{figure}\centering
        \includegraphics[width=1\linewidth]{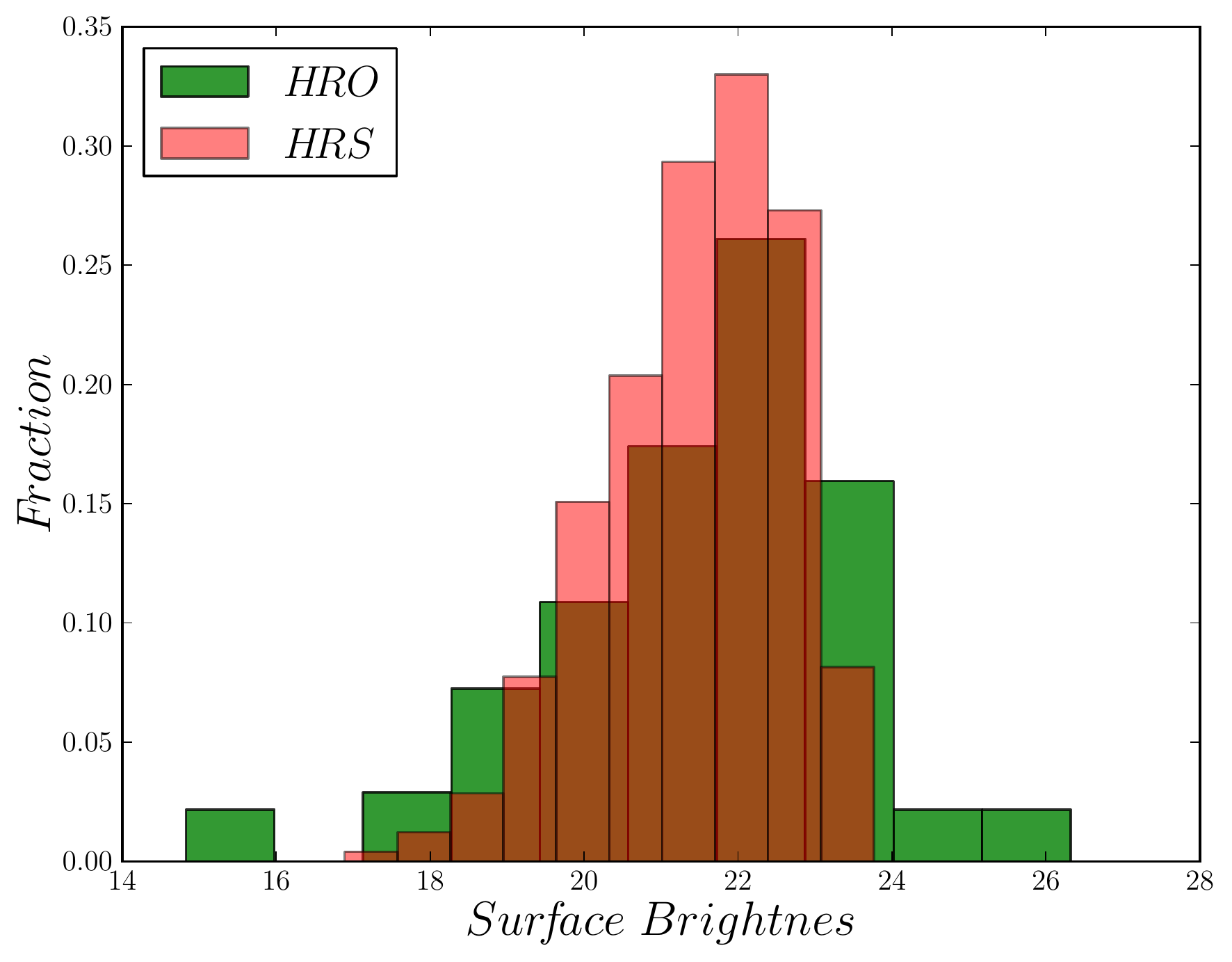}
    \caption{Comparison of the surface brightness of the $H\!RS$ (high-redshift simulated) galaxies (red) with the $H\!RO$ (high-redshift original) galaxies at redshift $2.5<z<3$ (green).}\label{Figure18}
\end{figure}

To demonstrate that our simulated galaxies are a fair comparison to the actual high redshift galaxies we first compare the surface brightness distribution for both samples. In Figure \ref{Figure18} we show the surface brightness of the $H\!RS$ sample compared with the $H\!RO$ sample, which shows that there is a good overlap between the two and therefore that, at least in the respect of magnitude and size distributions, these two samples can be compared.  

\subsection{Results}

Once we have our simulated subsample of $H\!RS$ galaxies we apply the same method described through the paper to determine whether each galaxy should be classified as $1$- or $2$-component system. The process for this can be summarized as:

\begin{itemize}
 \item We fit the surface brightness with a S\'ersic function with a free $n$ for a $1$-component fit.
 \item We fit the surface brightness with a S\'ersic + Exponential function as a $2$-component fit.
 \item We use our three methods to classify the galaxies as $1$- or $2$-components: the visual classification, $F$-test and $RF\!F$ method.
\end{itemize}

\begin{figure}\centering
	\includegraphics[width=0.7\linewidth]{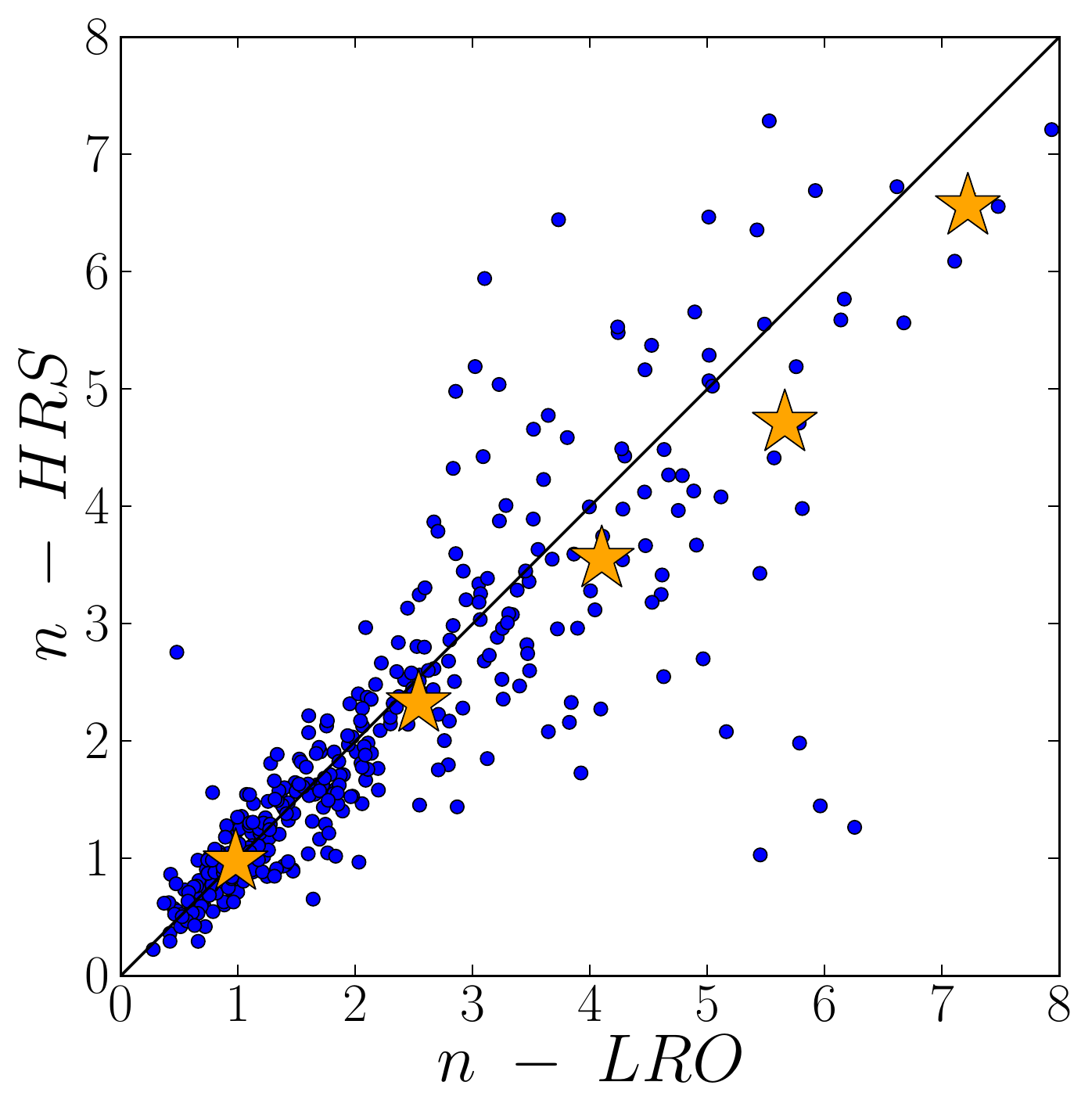}
	\includegraphics[width=0.7\linewidth]{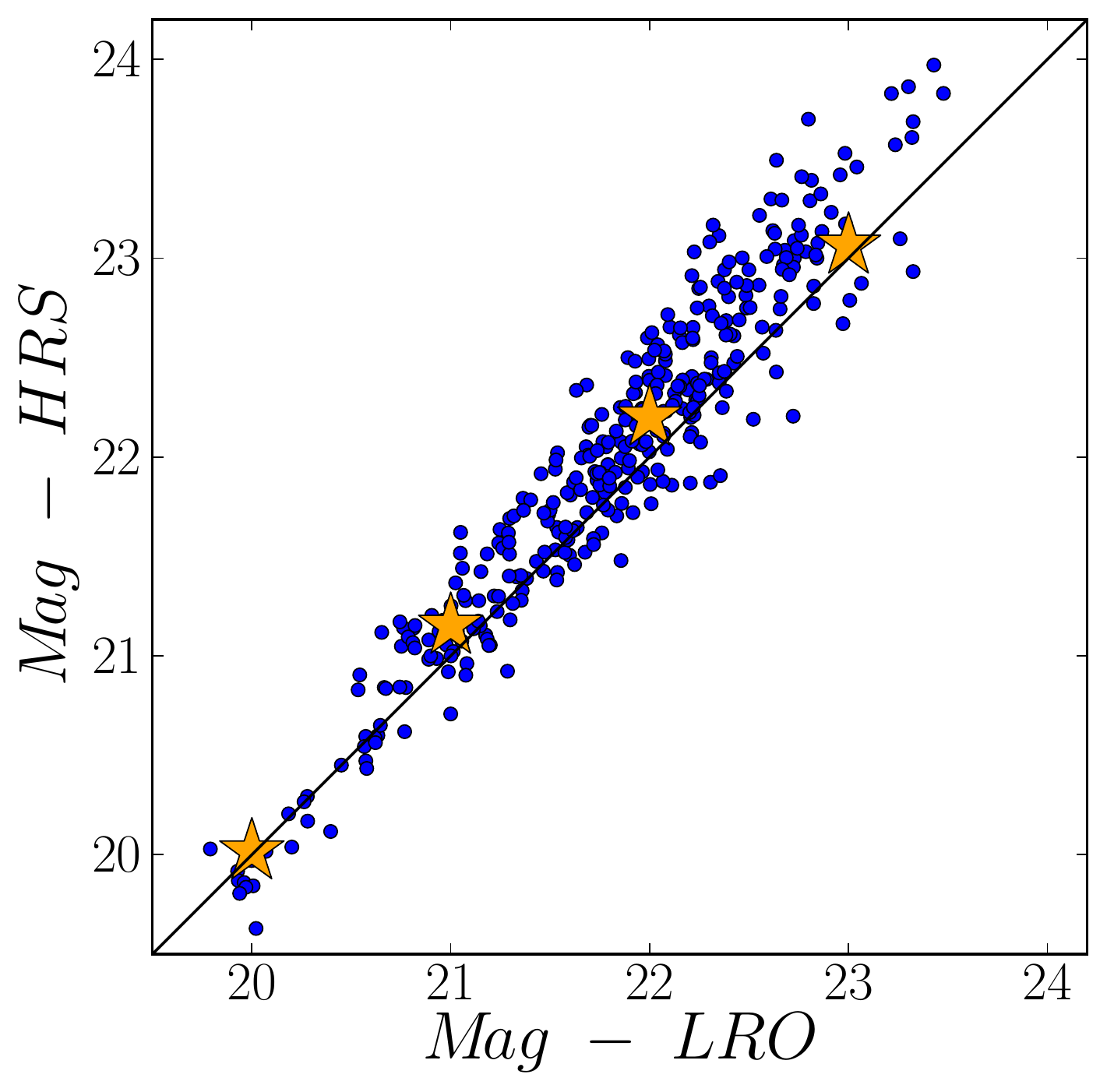}
	\includegraphics[width=0.7\linewidth]{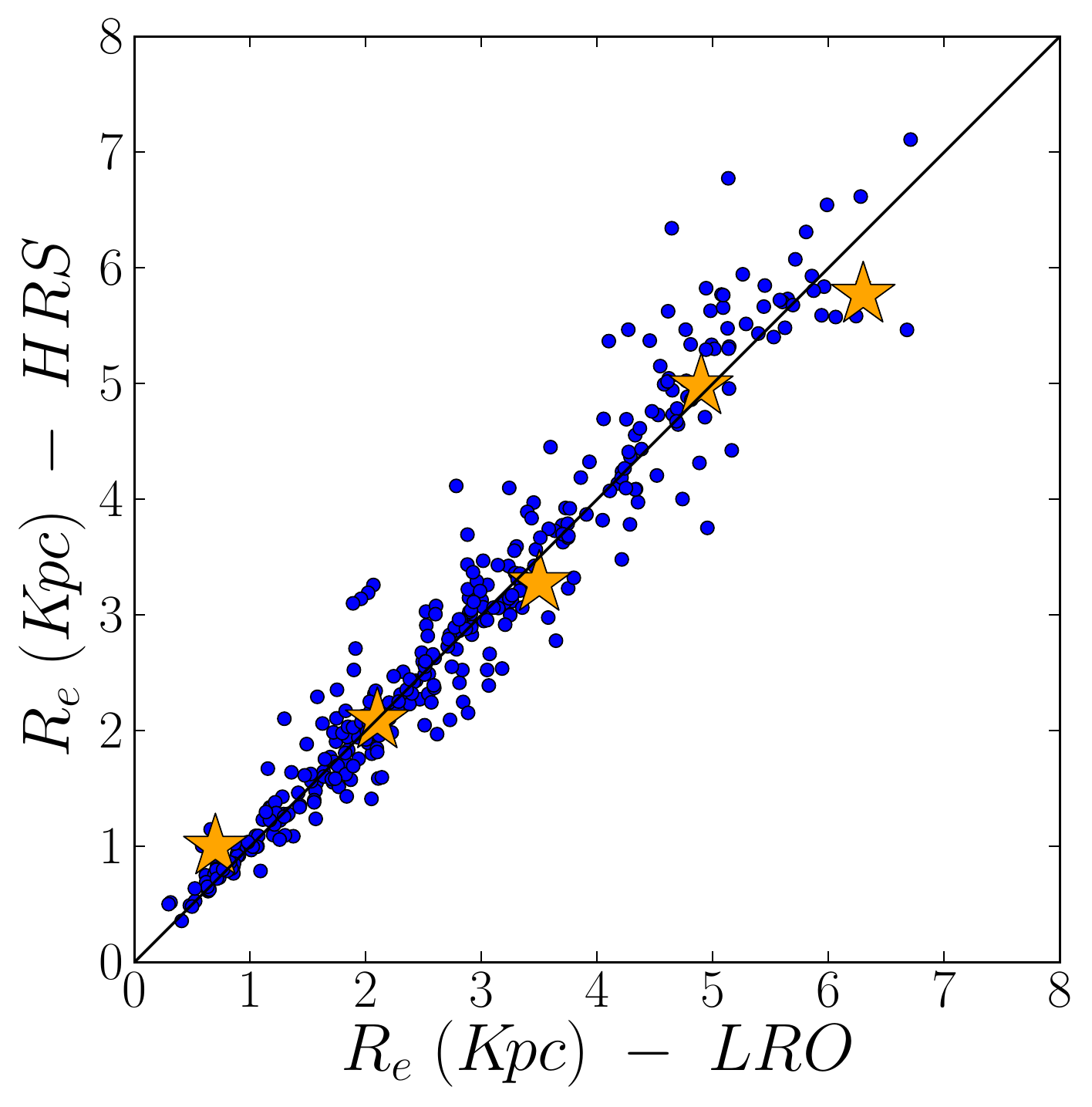}
    \caption{Comparison between the results from the fitting for the $H\!RS$ galaxies versus the $L\!RO$ sample, for the S\'ersic index (top), magnitude (middle), and effective radii (bottom). The effect on the $H\!RS$ magnitudes due to luminosity evolution and surface brightness dimming, introduced in the simulations, is removed to better compare with the $L\!RO$.  The yellow stars represent the mean values for the $H\!RS$ galaxies.}\label{Figure19}
\end{figure}

A comparison of the S\'ersic index, magnitude and effective radii between the $H\!RS$ and $L\!RO$ for $1$-component galaxies is shown in Figure \ref{Figure19}. We observe that these quantities behave as expected after the simulation. We recover on average the values of S\'ersic index and effective radius, as well as the apparent magnitude. Notice for instance that although the S\'ersic index is not as well preserved for high values, the classification due to such an index is preserved, namely $80$ per cent of galaxies with $n>2.5$ for $L\!RO$ still have $n>2.5$ after the simulation.

After applying the three methods to classify the galaxies as $1$- or $2$-components, in Table \ref{Table3} we summarize the difference in the fraction of $2$-component galaxies in the $H\!RS$ with respect to the $L\!RO$ and the $H\!RO$.

\begin{table}
\caption{Fraction of $2$-component galaxies selected by the three methods for different subsample of galaxies.} 
\centering  
\begin{tabular}{c c c c} 
\hline                      
   & $L\!RO$ galaxies & $H\!RS$ galaxies &  $H\!RO$ galaxies \\[0.5ex] 
\hline \hline                 
 Visual    & $0.38$ & $0.17$ & $0.00$ \\ 
\hline                      
 $F$-test  & $0.50$ & $0.04$ & $0.21$ \\
\hline                      
 $RF\!F$     & $0.36$ & $0.10$ & $0.31$ \\
\hline 
\end{tabular}\label{Table3} 
\end{table}

\begin{figure*}\centering
    \includegraphics[width=0.42\linewidth]{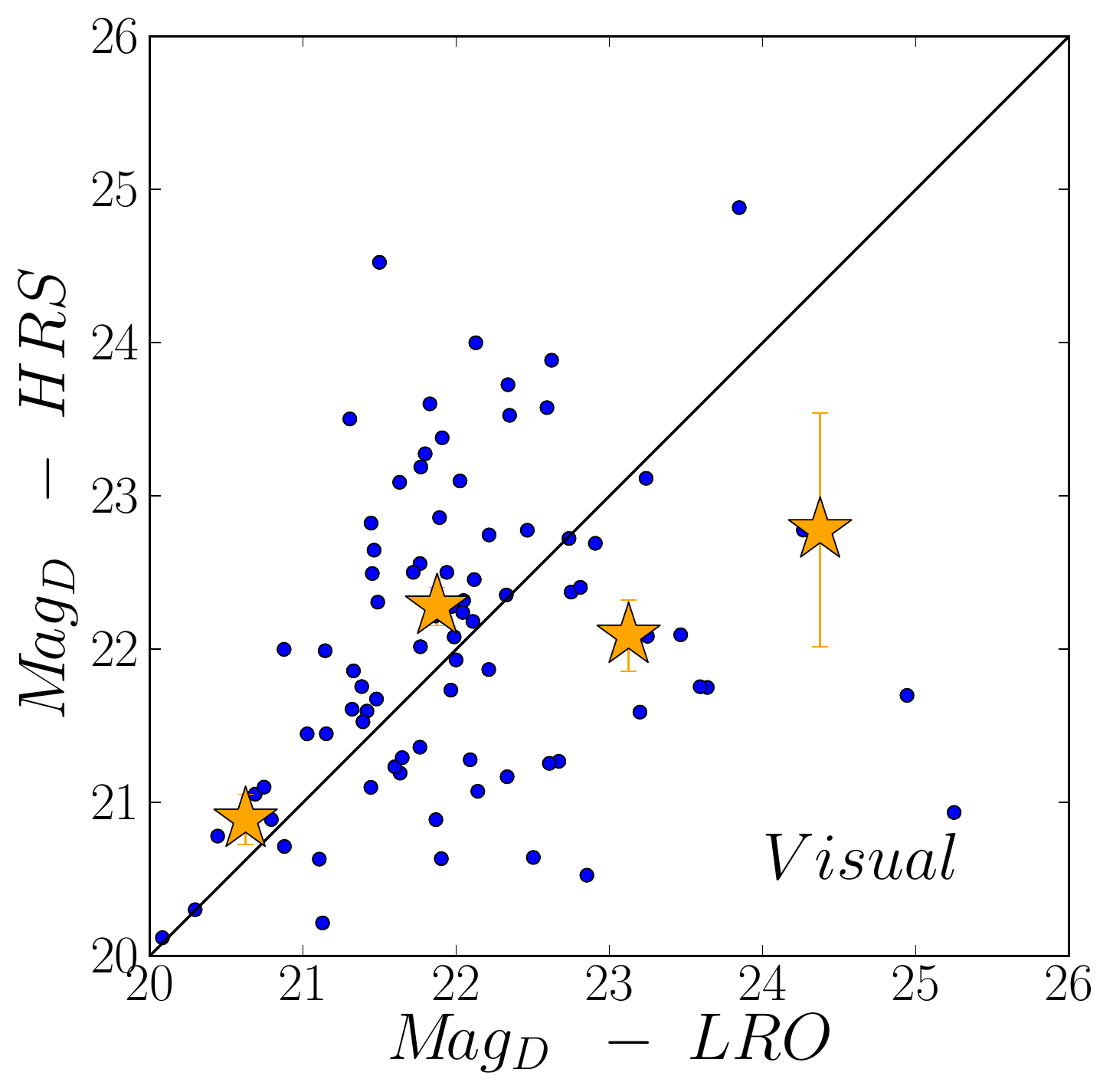}
    \includegraphics[width=0.42\linewidth]{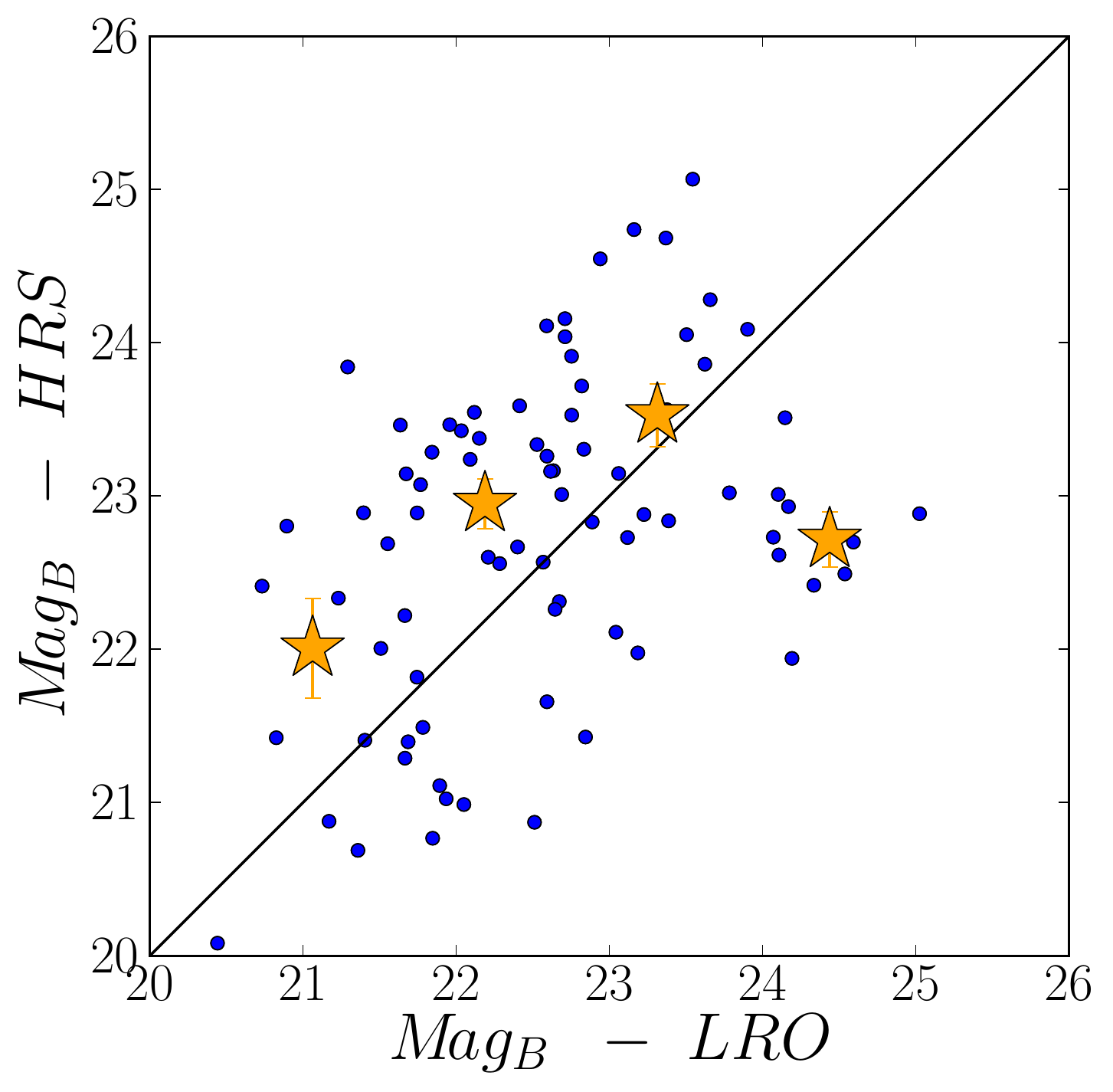} \\
    \includegraphics[width=0.42\linewidth]{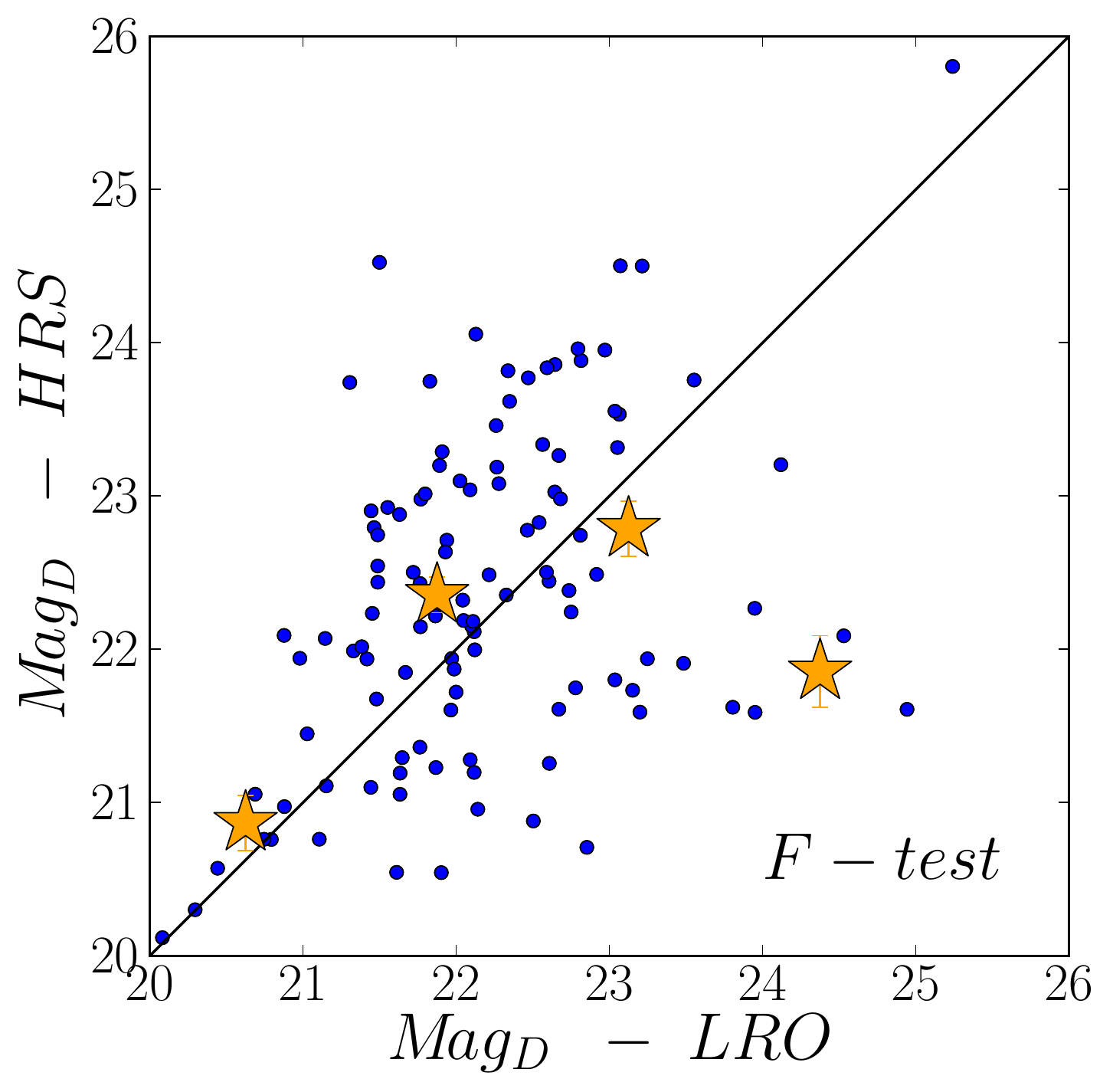}
    \includegraphics[width=0.42\linewidth]{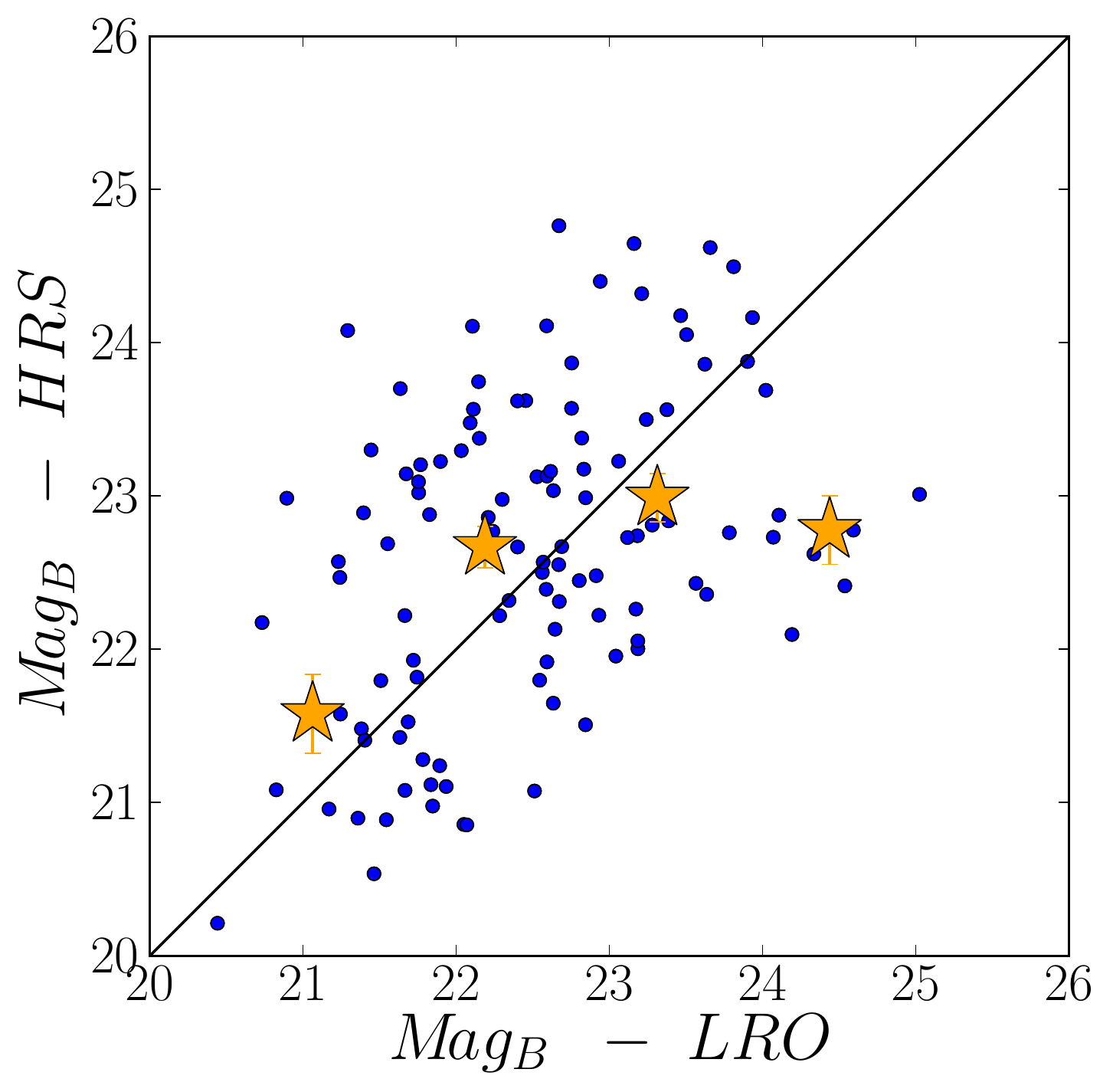}\\
    \includegraphics[width=0.42\linewidth]{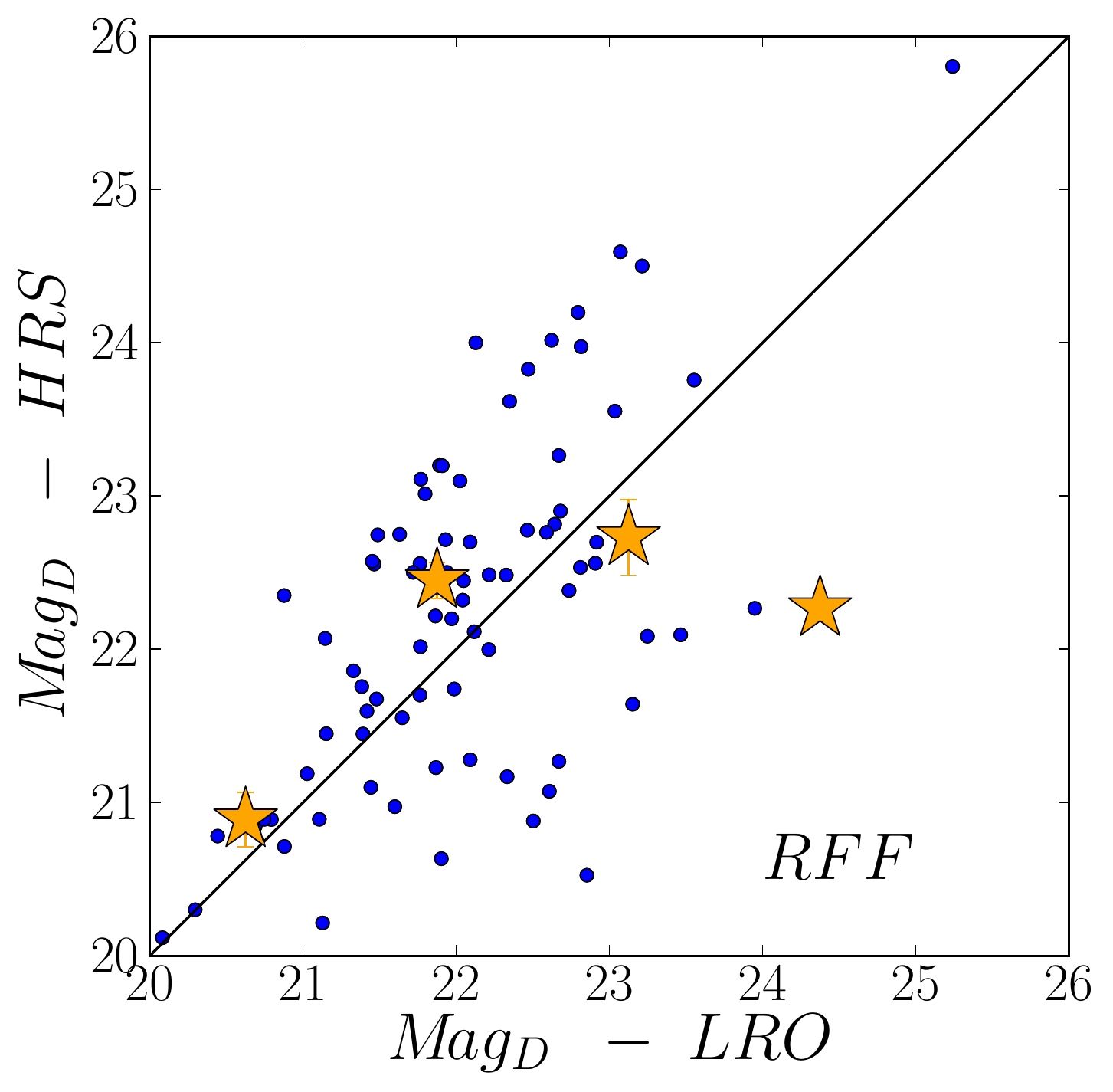}
    \includegraphics[width=0.42\linewidth]{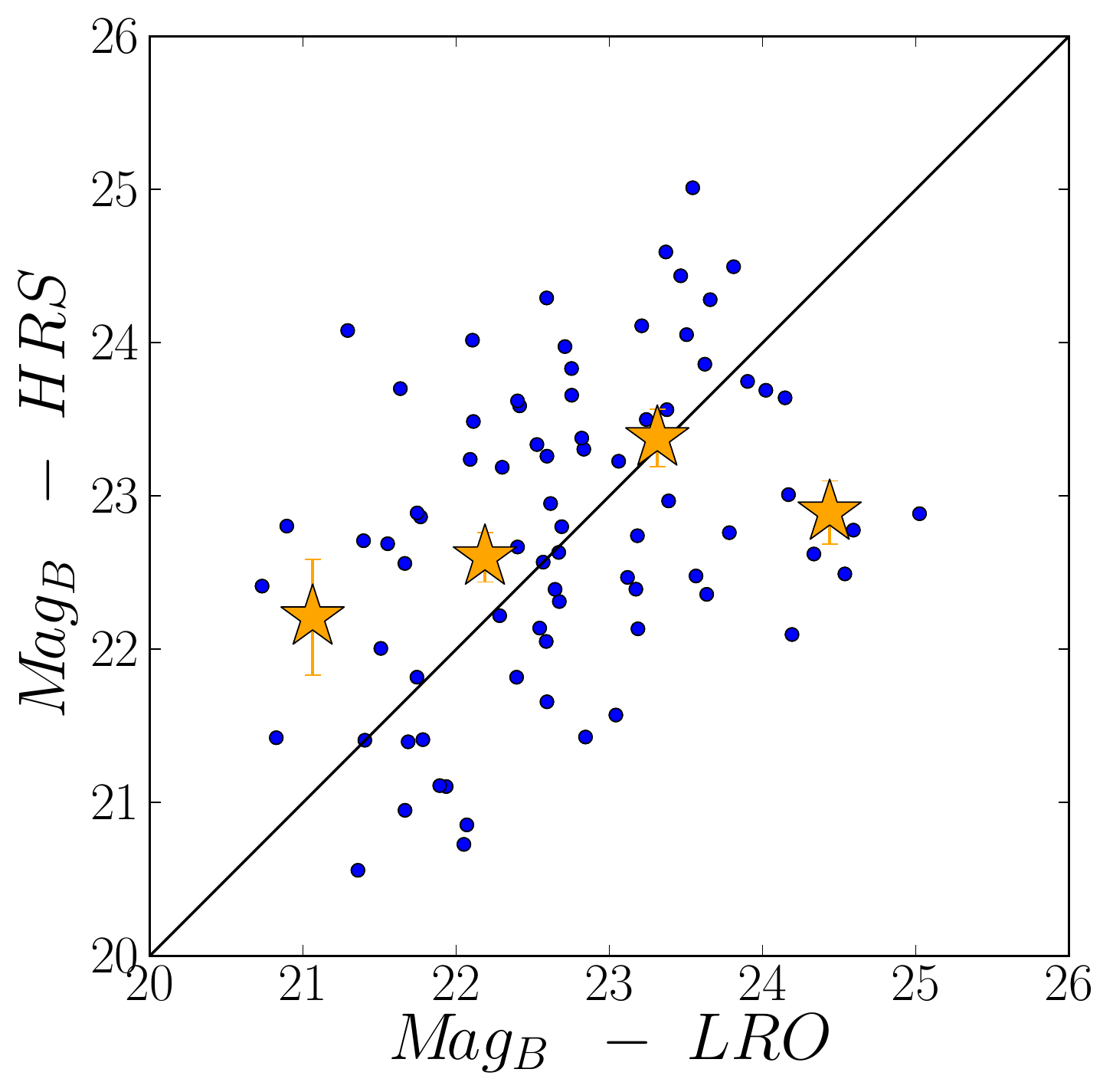}
    \caption{Magnitude of the disc-like outer (left) and bulge-like inner (right) component of high-redshift simulated  versus low-redshift original galaxies for each method (top panel: visual classification, middle panel: $F$-test and bottom panel: $RF\!F$ method). The effect on the $H\!RS$ magnitudes due to luminosity evolution and surface brightness dimming, introduced in the simulations, is removed to better compare with the $L\!RO$. The yellow stars represent the mean value of the magnitudes for the $H\!RS$ galaxies.} \label{Figure20}
\end{figure*}

\begin{figure*}\centering
    \includegraphics[width=0.41\linewidth]{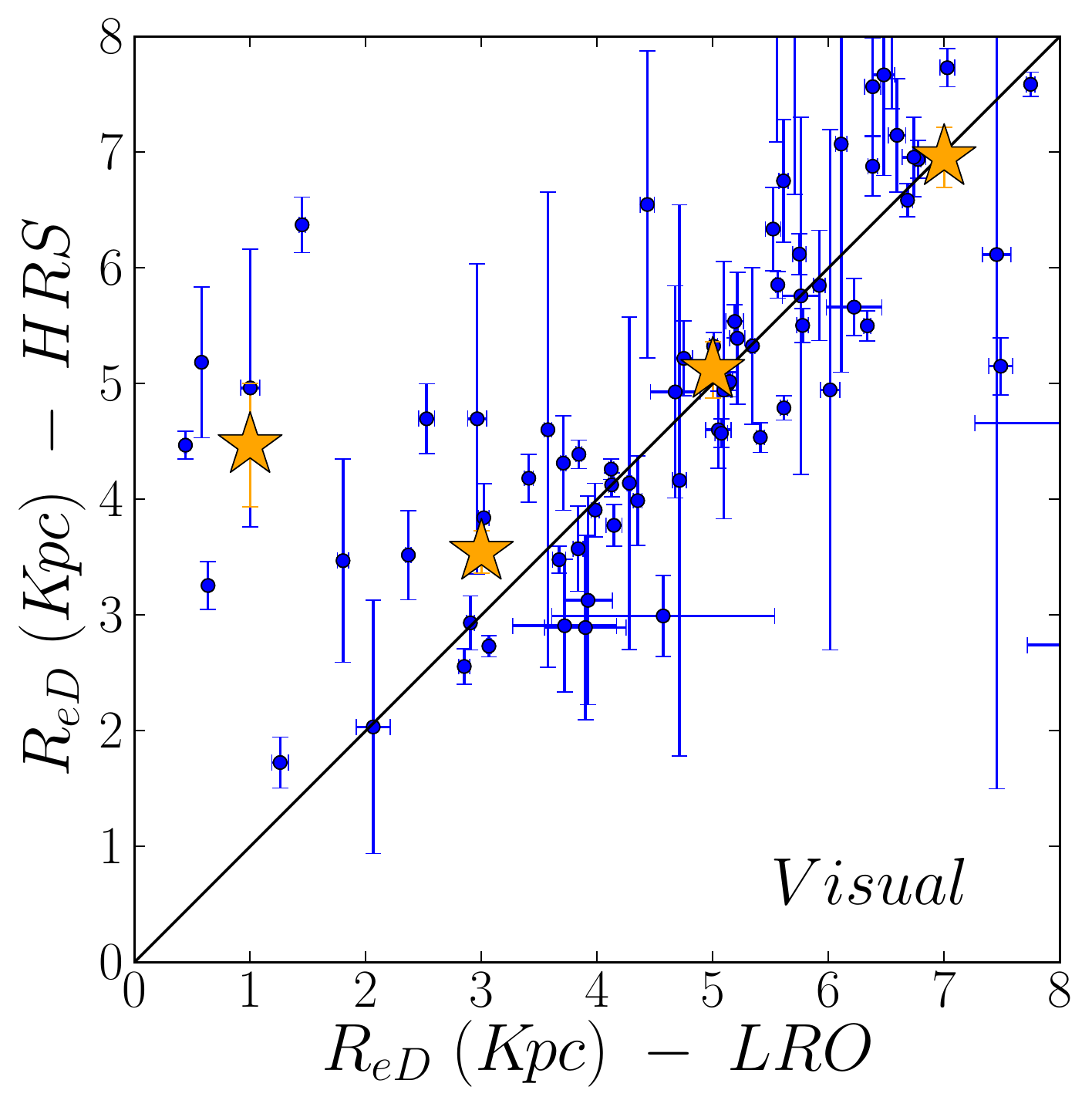}
    \includegraphics[width=0.41\linewidth]{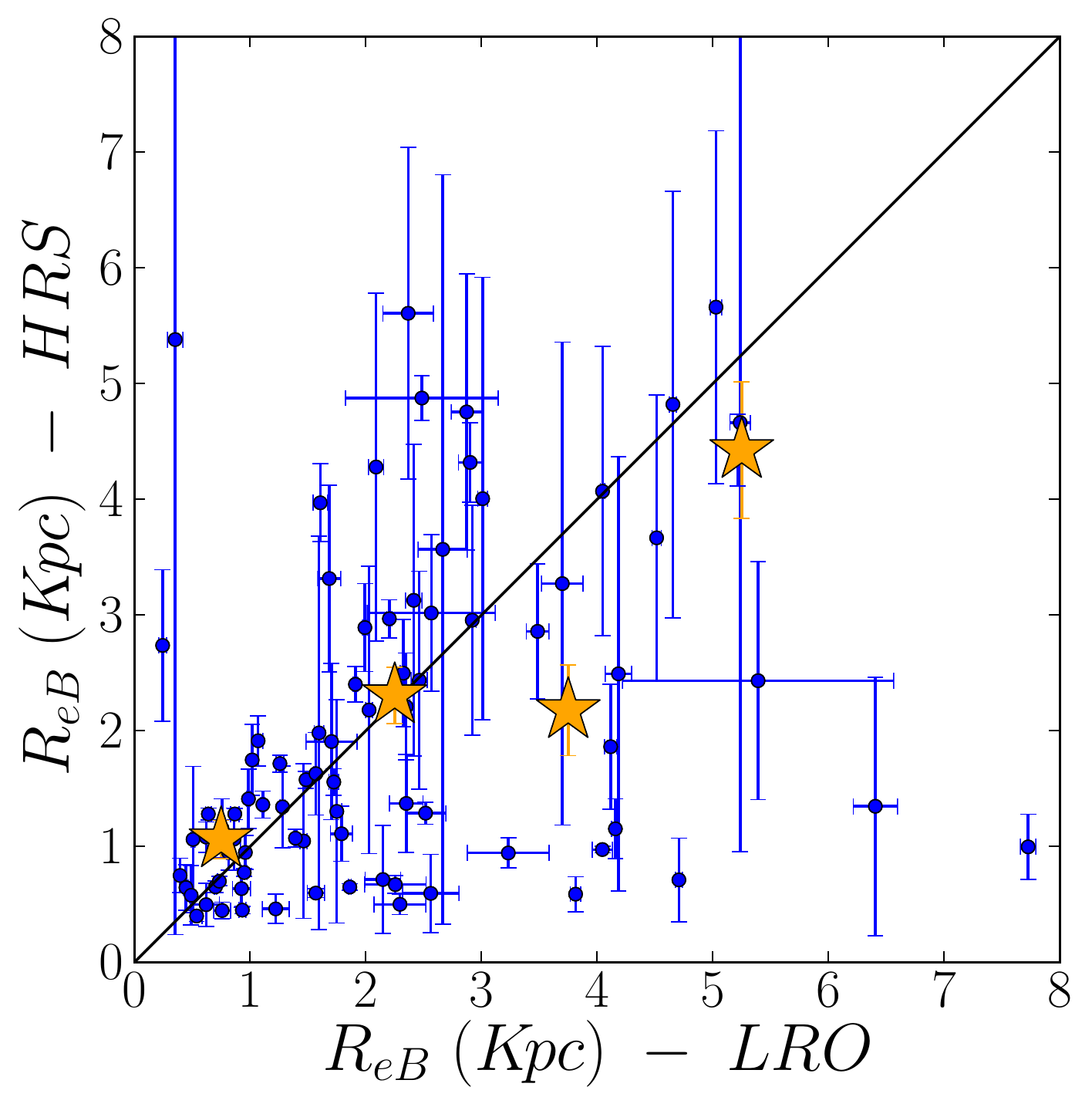}\\
    \includegraphics[width=0.41\linewidth]{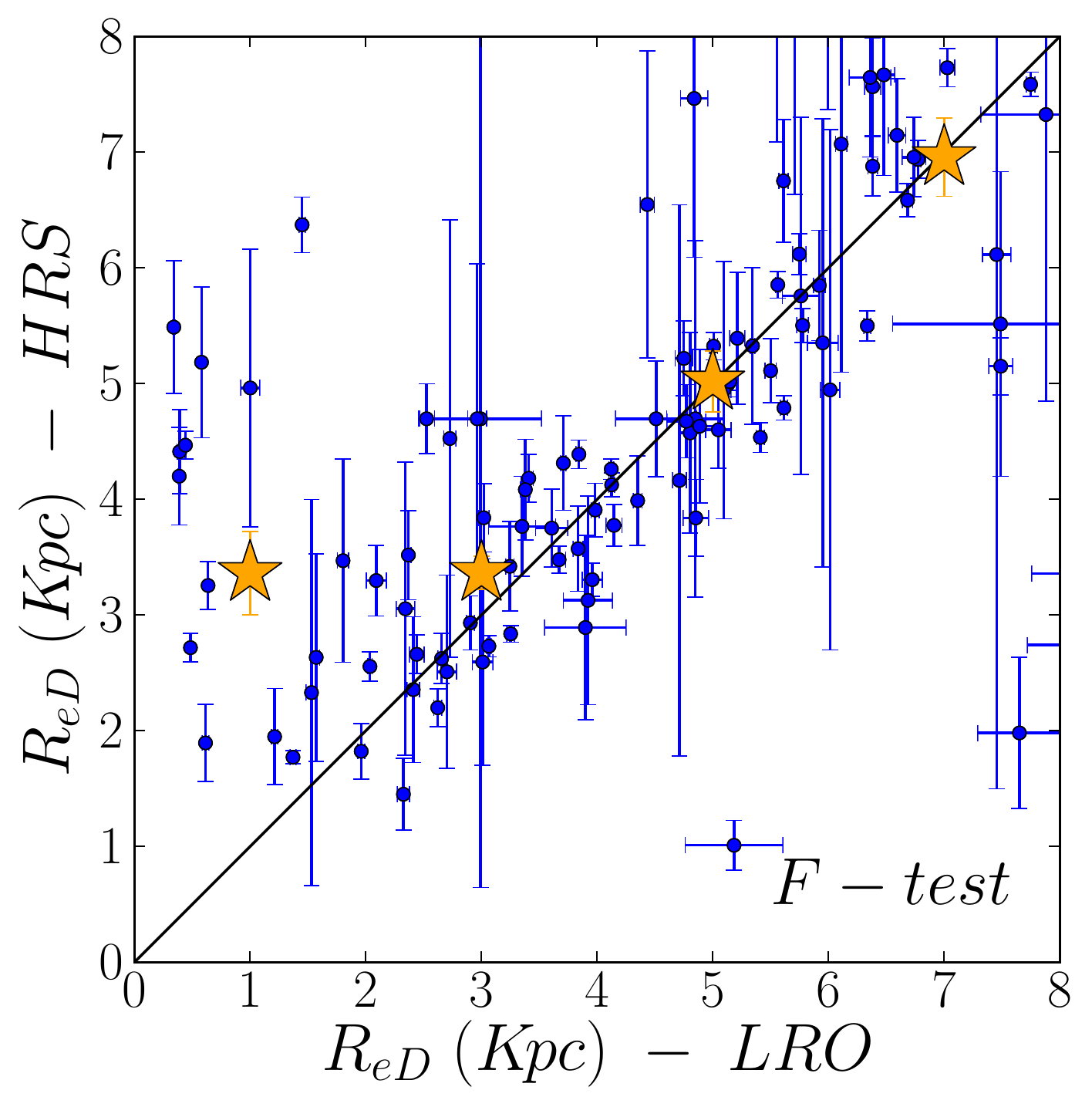}
    \includegraphics[width=0.41\linewidth]{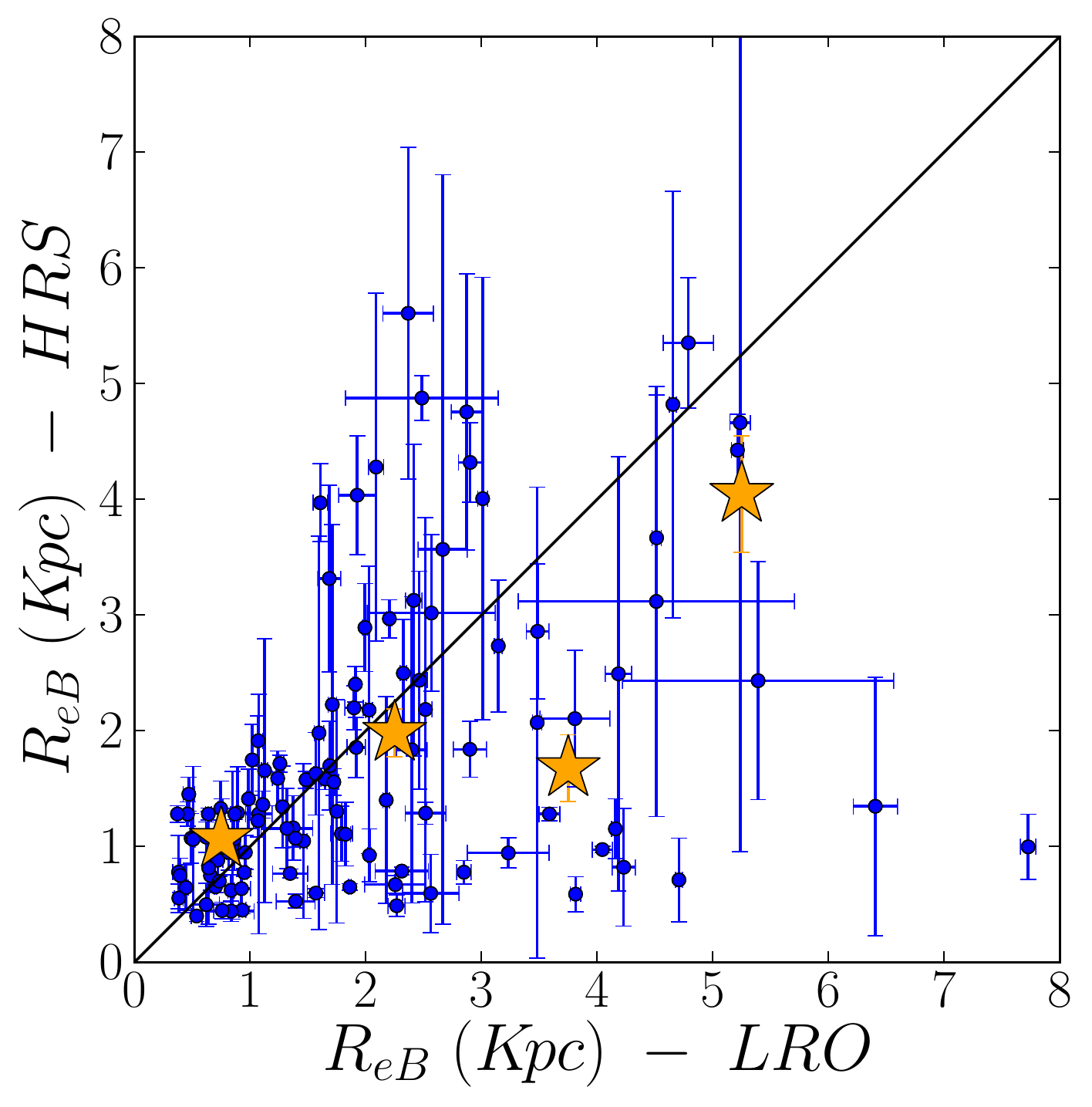}\\
    \includegraphics[width=0.41\linewidth]{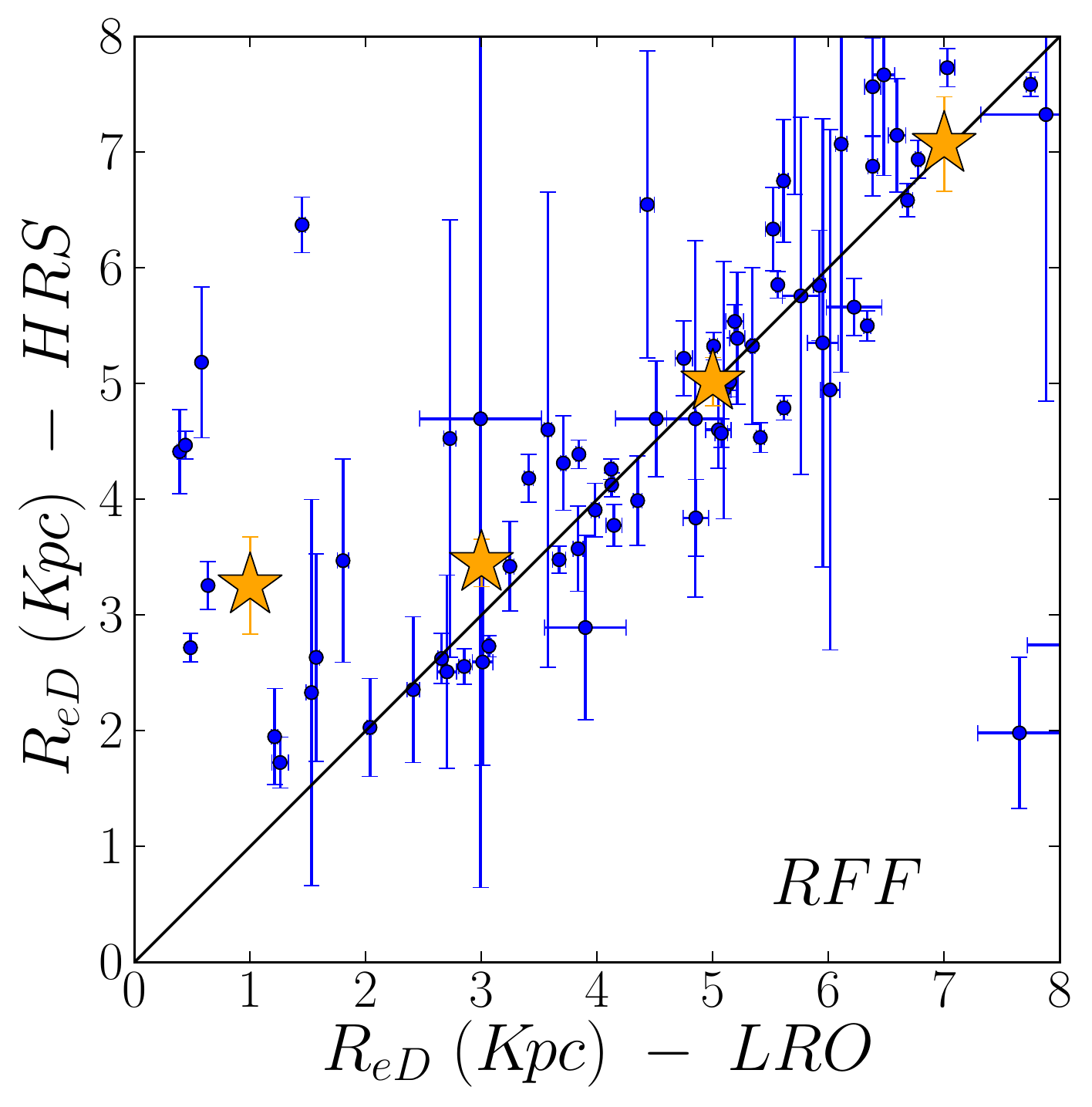}
    \includegraphics[width=0.41\linewidth]{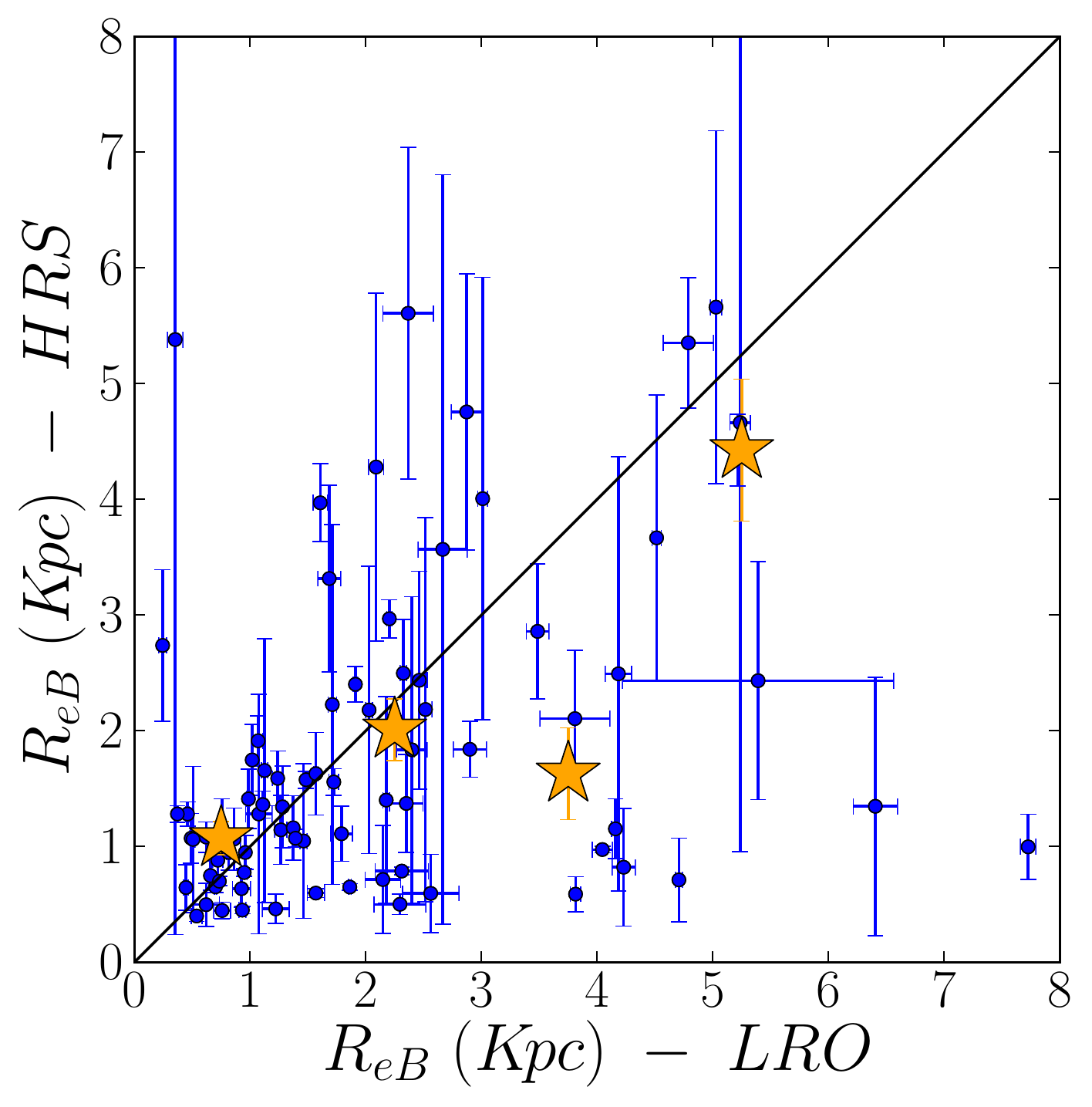}
    \caption{Effective radii of the disc-like outer (right) and bulge-like inner (left) component of high-redshift simulated  versus low-redshift original galaxies for each method (top panel: visual classification, middle panel: $F$-test and bottom panel: $RF\!F$ method). The yellow stars represent the mean value of the effective radii for the $H\!RS$ galaxies.}\label{Figure21}
\end{figure*}

An analogous comparative of the magnitude and effective radii between the $H\!RS$ and $L\!RO$ for $2$-component galaxies is shown in Figures \ref{Figure20} and \ref{Figure21}. Notice however that while in Figure \ref{Figure19} we considered the whole sample $L\!RO$ vs.\ $H\!RS$, now only the ones of $L\!RO$ classified as $2$-component (together with the corresponding ones of $H\!RS$) are taken into account. Thus we obtain for each method two symbols for the magnitude (one for each of the two components), and two for the effective radii.

Even though there is some scatter in the plots of Figure \ref{Figure21}, and the errors are in some cases large, the mean values of the effective radii in $H\!RS$ galaxies, for both components, are generally in agreement with the measures obtained for the $L\!RO$ ones. For smaller values of the outer component in the $L\!RO$, the simulations show that we would measure larger effective radii at hight redshift, which suggests that the small values of sizes we obtain for the $H\!RO$ must be real.

\begin{center}
    \begin{figure}
      \includegraphics[width=1\linewidth]{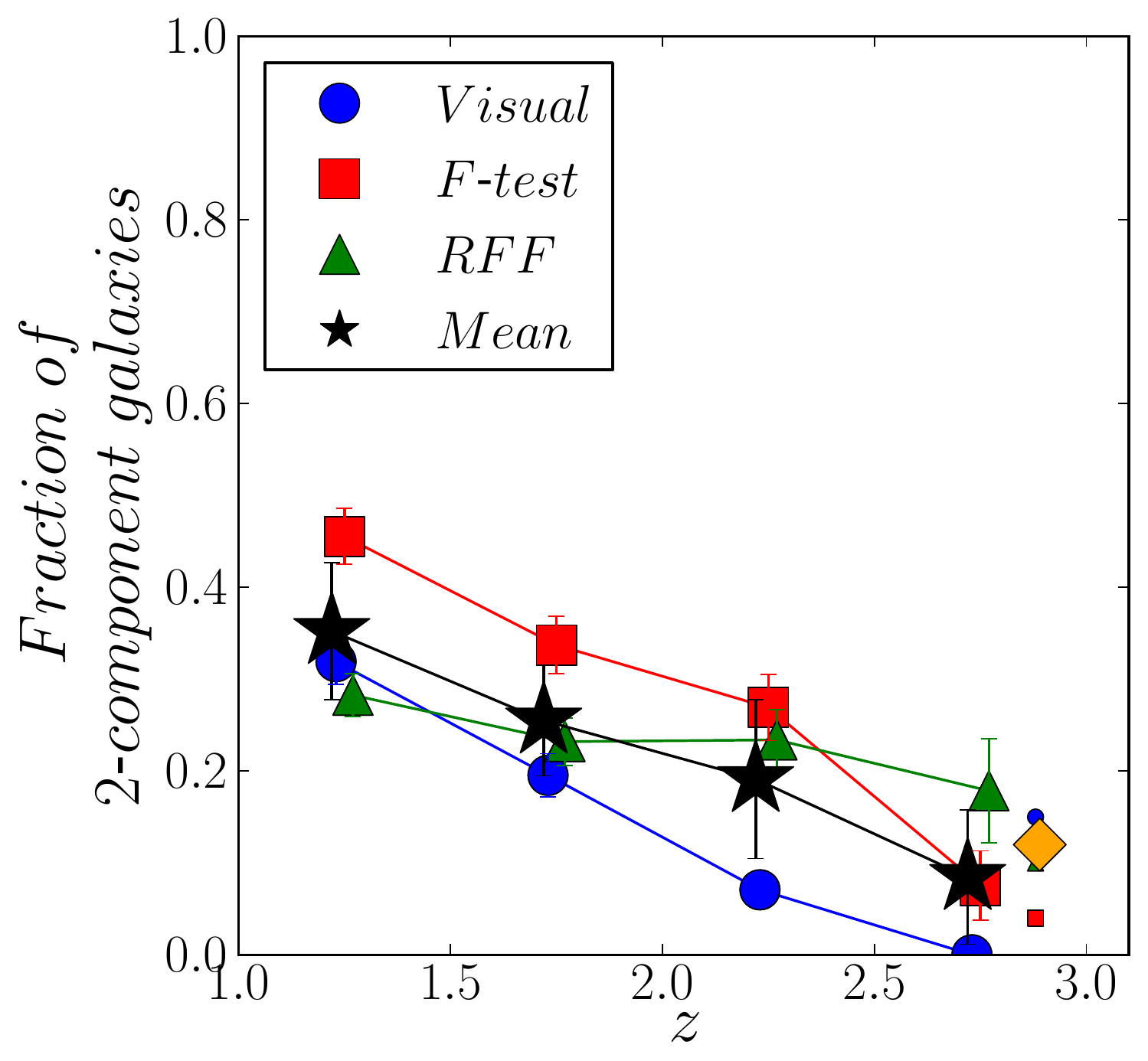}
      \caption{Comparison of the fraction of $2$-component galaxies between observations and simulations. The bigger blue circles, red squares and green triangles represent the fractions of $2$-component galaxies, from the whole sample, as function of redshift. The black stars show the mean values of the three methods (same as Figure \ref{Figure8}). The smaller blue circle, red square and green triangle represent the fractions of $2$-component $H\!RS$ galaxies for each method, and the yellow diamond is the mean between the Visual and $R\!FF$ method.}\label{Figure22}
     \end{figure}
\end{center}

Figure \ref{Figure22} also shows that the fraction of $H\!RS$ $2$-component galaxies (yellow diamond) is similar (although slightly higher) to that of the $H\!RO$ (black star at $z \sim 2.75$), which might imply that most of the evolution we see in that fraction with redshift may not be real but due to resolution and depth issues at high redshift. This shows the difficulty of carrying out a study of structure at high redshift. However, if we consider only the visual method, we observe that the evolution could be real. It is interesting to notice that while for the $H\!RO$ the visual classification fails to select any $2$-component galaxies, it selects almost $20$ per cent in the $H\!RS$.

\begin{center}
    \begin{figure}
      \includegraphics[width=1\linewidth]{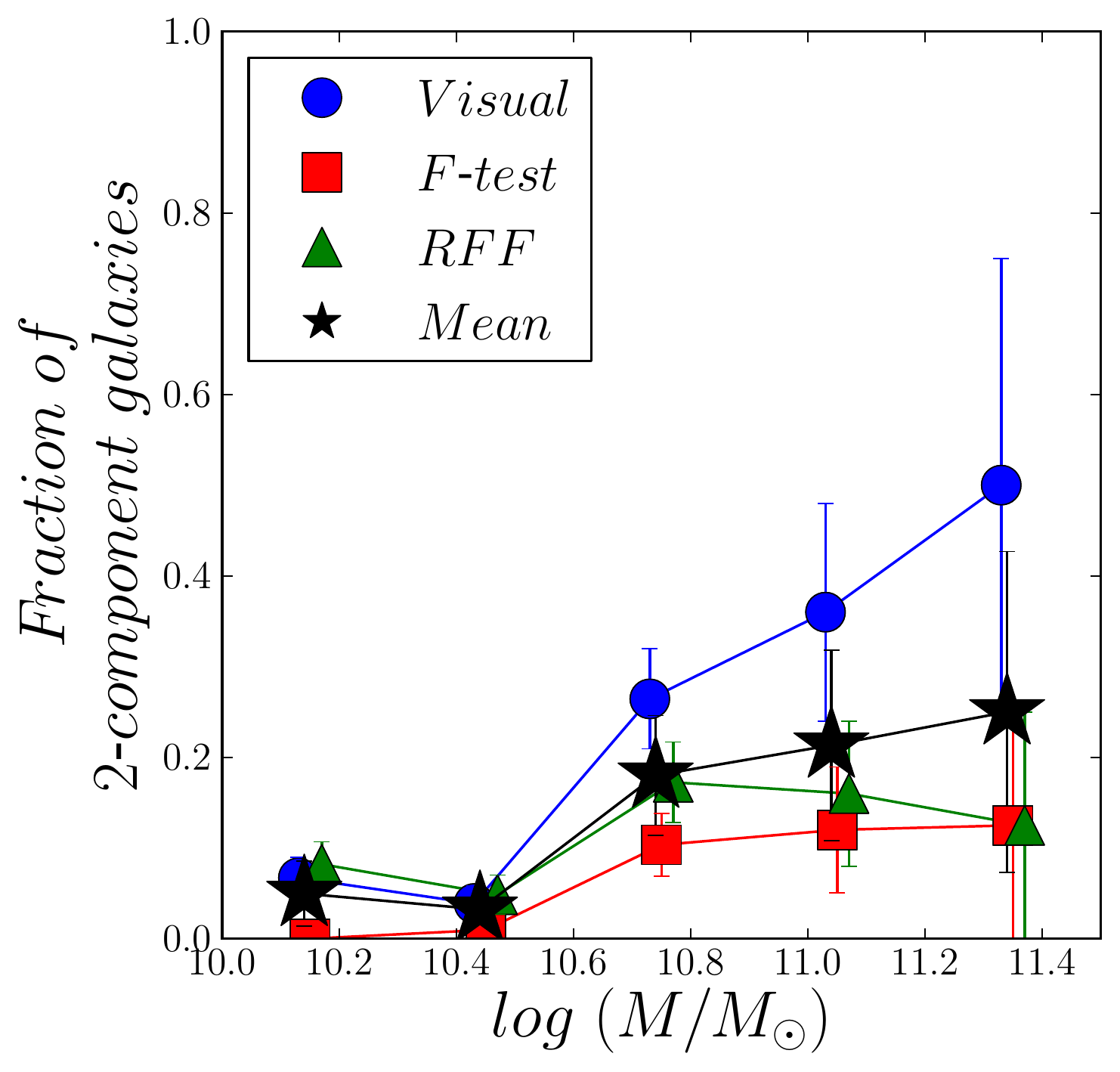}
      \caption{Fraction of $2$-component galaxies as a function of mass for the $H\!RS$ simulated galaxies. The points are coloured as in Figure \ref{Figure7}.}\label{Figure23}
     \end{figure}
\end{center}

In Figure \ref{Figure23} we plot the fraction of $2$-component $H\!RS$ galaxies as a function of mass. As in Figure \ref{Figure7} we observe that the fraction increases with higher masses in a similar manner, but is about $2$ times lower than for the original galaxies (notice that in the $H\!RO$ we detect about half of the galaxies selected as $2$-components in the $L\!RO$ sample). As we detect a larger fraction of $2$-component galaxies in the original sample than in the $H\!RS$ galaxies, this implies that the high fraction we have found at higher masses must be real. 

\end{appendices}

\bsp
\label{lastpage}
\end{document}